\documentclass[a4paper,twocolumn,longauth]{aa}
\usepackage[utf8]{inputenc}
\usepackage[varg]{txfonts}

\usepackage[]{natbib}
\bibpunct{(}{)}{;}{a}{}{,} 

\usepackage[usenames,dvipsnames]{xcolor}
\usepackage{ae,aecompl}

\usepackage{longtable} 
\usepackage{pdflscape} 
\usepackage{booktabs}
\usepackage{comment}
\usepackage{array}
\usepackage{caption}

\usepackage[]{hyperref} 
\usepackage{enumitem}
\usepackage{ulem}
\usepackage{tipa}
\usepackage{xcolor}
\definecolor{OIpink}{HTML}{CC79A7}
\definecolor{OIblue}{HTML}{0072B2}
\definecolor{OIdarkorange}{HTML}{D55E00}
\definecolor{OIlightgreen}{HTML}{009E73}
\definecolor{OIyellow}{HTML}{F0E442}

\newcommand{\logP}{$\log{P}$}
\newcommand{\Pdot}{$\dot{P}$}
\newcommand{\Ppuls}{$P_{\mathrm{puls}}$}
\newcommand{\vgamma}{$v_\gamma$}
\newcommand{\kms}{\,km\,s$^{-1}$} 
\newcommand{\ms}{\,m\,s$^{-1}$} 
\newcommand{\nfs}{$N_{\mathrm{FS}}$} 
\newcommand{\gaia}{\textit{Gaia}} %
\newcommand{\gaiarvs}{\textit{Gaia} RVs}
\newcommand{\veloce}{\texttt{VELOCE}} %
\newcommand{\coralie}{\texttt{Coralie}} %
\newcommand{\coraliev}[1]{\texttt{Coralie#1}}
\newcommand{\hermes}{\texttt{Hermes}} %

\newcommand{\vgGaiaCat}{$v_{\gamma,{\rm GDR3}}$} 
\newcommand{\vgGaiaTF}{$v_{\gamma,{\rm GDR3}}^{\star}$}

\newcommand{\ncep}{253} 
\newcommand{\ncepinclusive}{258} 
\newcommand{\ncepFS}{219} 

\newcommand{\nnoncep}{164} 
\newcommand{\nobs}{18\,225} 
\newcommand{\nobscep}{18\,225} 
\newcommand{\nobsall}{19\,386}
\newcommand{\nobsnoncep}{1161}
\newcommand{\nobscepunmodeled}{1070} 
\newcommand{\ncepunmodeled}{39} 
\newcommand{\ncepinsufficientobs}{Twenty-eight}
\newcommand{\ntypetwocep}{twelve}
\newcommand{\ntypeIIcep}{12}
\newcommand{\nrrlyrae}{4}
\newcommand{\nrrlyraestr}{four}
\newcommand{\nsbIIsbIII}{33}
\newcommand{\nextendedbaseline}{nine} 

\newcommand{\medianerror}{$0.037$}
\newcommand{\npdotall}{146} 
\newcommand{\npdotsig}{80} 
\newcommand{\nmod}{36} 
\newcommand{\nmodulatorstr}{Thirty-six}

\begin{document}

\title{VELOcities of CEpheids (VELOCE) I. High-precision radial velocities of Cepheids\thanks{Tables \ref{tab:ZPs}, \ref{tab:pdot}, \ref{app:sampletable}, \ref{app:FourierParameters} are made available in electronic form at the CDS via anonymous ftp to \url{cdsarc.u-strasbg.fr} (130.79.128.5) or via \url{https://cdsweb.u-strasbg.fr/cgi-bin/qcat?J/A+A/}}
\thanks{The catalog of radial velocity measurements and best-fit parameters is made available in FITS format via \url{https://www.zenodo.org}}} 

\author{Richard~I. Anderson\inst{1}\thanks{\email{richard.anderson@epfl.ch}} \and
Giordano Viviani\inst{1} \and 
Shreeya~S.~Shetye\inst{1} \and
Nami~Mowlavi\inst{2} \and
Laurent~Eyer\inst{2} \and
Lovro~Palaversa\inst{3} \and
Berry~Holl\inst{2}\and
Sergi~Blanco-Cuaresma\inst{2,4,5} \and  
Kateryna~Kravchenko\inst{6}\and
Micha\l~Pawlak\inst{7}\and
Mauricio~Cruz~Reyes\inst{1}\and
Saniya~Khan\inst{1}\and
Henryka~E.~Netzel\inst{1}\and
Lisa~L\"obling\inst{8}\and
P\'eter~I.~P\'apics\inst{9} \and
Andreas~Postel\inst{2}\and
Maroussia~Roelens\inst{2} \and
Zoi~T.~Spetsieri\inst{1}\and
Anne~Thoul\inst{10}\and
Ji\v{r}\'i~\v{Z}\'ak\inst{11}\and
Vivien~Bonvin\inst{1} \and
David~V.~Martin\inst{12}\and
Martin~Millon\inst{13} \and
Sophie~Saesen\inst{2}\and
Aur\'elien~Wyttenbach\inst{2} \and
Pedro~Figueira\inst{2,14}\and
Maxime~Marmier\inst{2}\and
Saskia~Prins\inst{9}\and
Gert~Raskin\inst{9} \and
Hans~van~Winckel\inst{9}
}

\authorrunning{R.I. Anderson et al.}

 \institute{
 Institute of Physics, \'Ecole Polytechnique F\'ed\'erale de Lausanne (EPFL), Observatoire de Sauverny, Chemin Pegasi 51b, 1290 Versoix, Switzerland 
 \and
 D\'epartement d'Astronomie, Universit\'e de Gen\`eve, Chemin Pegasi 51b, 1290 Versoix, Switzerland
 \and
 Ru{\dj}er Bo\v{s}kovi\'{c} Institute, Bijeni\v{c}ka cesta 54, 10000 Zagreb, Croatia
 \and
 Harvard-Smithsonian Center for Astrophysics, 60 Garden Street, Cambridge, MA 02138, USA
 \and
 Laboratoire de Recherche en Neuroimagerie, University Hospital (CHUV) and University of Lausanne (UNIL), Lausanne, Switzerland
 \and
 Max Planck Institute for extraterrestrial Physics, Giessenbachstraße 1, 85748 Garching b. M\"unchen, Germany
 \and
 Lund Observatory, Division of Astrophysics, Department of Physics, Lund University, Box 43, 221 00 Lund, Sweden
 \and
 Institute for Astronomy and Astrophysics, Kepler Center for Astro and Particle Physics, Eberhard Karls University, Sand 1, 72076 T\"ubingen, Germany
 \and
 Instituut voor Sterrenkunde, KU Leuven, Celestijnenlaan 200D bus 2401, Leuven, 3001, Belgium
 \and
 Space Sciences, Technologies and Astrophysics Research (STAR) Institute, Université de Liège, Allée du 6 Août 19C, Bat. B5C, B-4000 Liège, Belgium
 \and
 European Southern Observatory, Karl-Schwarzschild-Str. 2, 85748 Garching b. M\"unchen, Germany
 \and
 Department of Physics \& Astronomy, Tufts University, 574 Boston Avenue, Medford, MA 02155, USA
 \and
 Kavli Institute for Particle Astrophysics and Cosmology and Department of Physics, Stanford University,
Stanford, CA 94305, USA
\and Instituto de Astrof\'{i}sica e Ci\^{e}ncias do Espa\c{c}o, Universidade do Porto, CAUP, 4150-762 Porto, Portugal
 }

\date{submitted: 27 October 2023; revised: 05 February 2024; accepted: 24 February 2024}

\abstract{We present the first data release of VELOcities of CEpheids (\veloce), dedicated to measuring the high-precision radial velocities (RVs) of Galactic classical Cepheids (henceforth, Cepheids). The first data release (\veloce\ DR1) comprises \nobscep\ RV measurements of \ncepinclusive\  bona fide classical Cepheids on both hemispheres collected mainly between 2010 and 2022, along with \nobsnoncep\ observations of \nnoncep\ stars, most of which had previously been misclassified as Cepheids. The median per-observation RV uncertainty for Cepheids is \medianerror\,\kms\ and reaches $2\,$\ms\ for the brightest stars observed with \coralie. Non-variable standard stars were used to characterize RV zero-point stability and to provide a base for future cross-calibrations. We determined zero-point differences between \veloce\ and 31 literature data sets using template fitting, which we also used to investigate linear period changes of \npdotall\ Cepheids. In total, 76 spectroscopic binary Cepheids and 14 candidate binary Cepheids were identified using \veloce\ data alone, which are investigated in detail in a companion paper (\veloce~II).
\veloce~DR1 provides a number of new insights into the pulsational variability of Cepheids, most importantly:  a) the most detailed description of the Hertzsprung progression based on RVs to date; b) the identification of double-peaked bumps in the pulsation curve; and c) clear evidence that virtually all Cepheids feature spectroscopic variability signals that lead to modulated RV variability at the level of tens to hundreds of \ms\ and that cannot be satisfactorily modeled using single-periodic Fourier series. We identified \nmod\ stars exhibiting such modulated variability, of which 4 also exhibit orbital motion. Linear radius variations depend strongly on pulsation period and a steep increase in slope of the $\Delta R/p$ versus $\log{P}-$relation is found near $10$\,d. This effect, combined with significant RV amplitude differences at fixed period, challenges the existence of a tight relation between Baade-Wesselink projection factors and pulsation periods. We investigated the accuracy of RV time series measurements, \vgamma, and RV amplitudes published by \gaia's third data release (\gaia \ DR3) and determined an  offset of $0.65\pm0.11$\,\kms\ relative to \veloce. Whenever possible, we recommend adopting a single set of template correlation parameters for distinct classes of large-amplitude variable stars to avoid systematic offsets in \vgamma\ among stars belonging to the same class. 
The peak-to-peak amplitudes of \gaiarvs\ exhibit significant ($16\%$) dispersion. Potential differences of RV amplitudes require further inspection, notably in the context of projection factor calibration.}

\maketitle

\section{Introduction}\label{sec:intro}
Classical Cepheids are evolved intermediate-mass radially pulsating stars that play an important role in understanding stellar evolution and pulsations, as well as the extragalactic distance ladder. Initially, their large-amplitude radial velocity (RV) variations were frequently attributed to orbital motion. However, an exploration of the ``pulsation hypothesis'' was already underway in the 1920s  \citep[e.g.,][]{Lindemann1918,Baade1926,Tiercy1928TVul,Tiercy1928sevenCepheids}\footnote{In particular the work by Georges Tiercy at Geneva Observatory in the 1920s deserves a special mention as it has not received the attention it deserves, likely because the texts were written in French and are not yet fully accessible. Further work to render the exceptional light and RV curves of Galactic Cepheids measured by Tiercy would be warranted from a historical perspective.}, later refined by \citet{Becker1940} and \citet{Wesselink1946} to eventually develop into  ``Baade-Wesselink'' (BW) type methods for measuring distances \citep[e.g.,][and references therein]{Groenewegen2018,Gieren2018}. 

Early RV observations of Cepheids were pioneered by \citet[who already remarked on the noticeable similarity in spectra between $\delta$~Cephei and $\eta$~Aquilae, e.g., ][]{Belopolsky1894,Belopolsky1897} and \citet[measuring both the pulsation and orbital RV variability of Polaris]{Campbell1899}. The first large catalog of Cepheid velocities was prepared by \citet{Joy1937}. This  occasionally enables very long temporal baselines of RV time series being available for Cepheids, which incidentally also tell the story of technological developments, from photographic plates to physical correlation spectrometers \citet[e.g.,][]{Baranne1996,Tokovinin2014} and CCD detectors. \citet{Anderson2019polaris} presents the case of Polaris, where the very-long-baseline data available provide a very good sensitivity to the $\sim 30$\,yr orbit, while posing challenges for understanding the stability of the pulsation. However, long-term stability of RV zero-points and instrumental offsets can complicate interpretation at the level of $1\,$\kms\ when using older data sources \citep[cf. also][]{Evans2015rv}.  

The history of Cepheid RV observations in the early 1990s is also closely interwoven with the quest to identify the first extrasolar planets, since the instruments used for the latter were being extensively used to measure precision RVs Cepheids \citep[e.g.,][]{Burki+1980,Butler1992,Butler1993,Pont1994,Pont1997,Bersier1994,Bersier2002}. Since then, the discovery of extrasolar planets \citep{Mayor1995} has continued to fuel ever increasing improvements in the developments of precision RV instrumentation reaching short-term stability on the order of several tens of better than $1$\,\ms\ for very stable stars \citep[e.g.,][]{Cretignier2021}. Unfortunately, for the further understanding of Cepheids, these exciting developments shifted priorities and resulted in a significant decline in RV observations of Cepheids being collected during the 2000s. 
Large studies targeting Cepheid RVs have since primarily focused on homogenizing data from the literature. For example, \citet{Evans2015rv} homogenized RV measurements from the literature by correcting them according to time-variable zero-points. Conversely, \citet{Borgniet2019} re-measured Cepheid RVs from archival spectra using a consistent methodology aiming for BW-type applications. However, the spectra used by Borgniet et al. were collected using a large number of spectrographs with different characteristics and instrumental systematics.

Large photometric surveys, such as the Optical Gravitational Lensing Experiment \citep[OGLE]{Pietrukowicz2021} and the European Space Agency's (ESA)  \gaia\ mission \citep{GaiaMissionDR1,gaiadr3.dr3,gaiadr3.vari}, have discovered more than $3600$  Cepheids in the Milky Way based on their characteristic photometric variability.  
With the advent of large ground-based spectroscopic surveys, such as APOGEE \citep{apogee}, WEAVE \citep{weave}, and 4MOST \citep{4most}, 
thousands of Cepheids will be observed spectroscopically using a single, or usually very few, epochs. Correctly interpreting these data (e.g., to study the dynamics of the Milky Way) requires better knowledge of RV pulsation curves to determine accurate mean velocities from random phase observations. Even now, \gaia\ is collecting unprecedented RV time series using its RVS instrument \citep{gaiadr3.radvel}. The RV time series of 799 classical and type-II Cepheids have been published as part of DR3 \citep{gaiadr3.cepheid}. Since \gaia's RVS instrument is limited to a narrow wavelength range near the Ca\,II IR triplet, a detailed inspection and validation of this data set is required to understand whether and how \gaia\ time series data can be combined with literature RV measurements based on lines in the optical wavelength range \citep[e.g.,][]{Wallerstein2019}.

The \veloce\ project was initiated to provide high-quality RV time series for a significant number of Galactic Cepheid in late $2010$ \citep[for an early overview, cf.][]{AndersonPhD}. Initially, the goal of the project was to use RVs to determine cluster membership \citep{Anderson2013,CruzReyes2023} and to obtain RV data for poorly known Cepheid candidates from photometric surveys. This quickly resulted in the identification of new SB1 Cepheids \citep[e.g.,][]{2013MNRAS.430.2018S,Anderson2015} as well as the discovery of ``modulated variability,'' which can come in the form of long-term variations or cycle-to-cycle variations, and are not caused by orbital motion \citep[e.g.,][]{Anderson2014rv,Anderson2016c2c}. Hence, \citet{AndersonPhD} began systematically studying the spectroscopic variability and multiplicity of Cepheids using spectrographs capable of detecting extrasolar planets, with the goals to provide detailed pulsation curves, a systematic search for companions, and a legacy Cepheid RV catalog capable of tying together a wide range of literature sources. Since its inception, \veloce\ has already yielded several interesting interim results that were published separately, such as the detection of non-radial pulsation and modulated variability \citep{Anderson2014rv,Anderson2016c2c,Anderson2019polaris}, upper limits on parallax biases due to orbital motion \citep{Anderson2016ApJS}, the most accurate mass measurement of a Galactic Cepheids \citep{Gallenne_2018}, and studies of the BW projection factor \citep{Anderson2016vlti,Breitfelder2016}.

Here, we present the first \veloce\ data release (DR1), as well as several insights gained from the data. \veloce\ DR1 focuses on RV measurements of single-mode classical Cepheids. We plan to present additional quantities derived from the cross-correlation functions (CCFs), such as the width, asymmetry, and depth, as well as the full set of observed spectra and the computed CCFs in a future data release. Since observations for \veloce\ are still ongoing, this will include observations not featured in DR1. The systematic investigation of spectroscopic binaries is presented in a companion article submitted in parallel (Shetye et al., paper~II). With this DR1, \veloce\ is entering a new phase focused on the intensive, long-term monitoring of particularly interesting Cepheids in the search for additional signals, such as modulations and non-radial pulsation modes.

This article is structured as follows. Section\,\ref{sec:observations} describes the observations, the cross-correlation technique used for measuring RV (Sect.\,\ref{sec:obs:CCFs}), the RV drift corrections applied (Sect.\,\ref{sec:obs:pressure}), and the observations from telescopes on both hemispheres (Sect.\,\ref{sec:obs:coralie} and Sect.\,\ref{sec:obs:hermes}). To facilitate the integration of \veloce\ data with future data sets, Sect.\,\ref{sec:obs:zeropoints} provides information on the \veloce\ absolute RV zero-points and long-term stability determined using IAU RV standard stars. Sect.\,\ref{sec:catalog} presents the catalog of RV measurements and how RV curves were modeled using a Fourier series (Sect.\,\ref{sec:Fourier}). RV time series are provided also for stars that are not classical Cepheids or Cepheids with particularly complex RV curves (Sect.\,\ref{sec:beatceps}).
Section\,\ref{sec:insights} presents the discovery of a secondary bump in the Hertzsprung progression (Sect.\,\ref{sec:doublebump}), cases of modulated variability (Sect.\,\ref{sec:modulations}), and the relation between linear radius variations and \logP\ (Sect.\,\ref{sec:deltaR}).
Section\,\ref{sec:literature} compares the \veloce\ data with the literature using template fitting (Sect.\,\ref{sec:RVTF}), reports the zero-point offsets of literature catalogs (Sect.\,\ref{sec:zp}), and provides a detailed comparison of \veloce\ data with \gaia\ RVS time series published as part of \gaia\ DR3 \citep{gaiadr3.radvel} (Sect.\,\ref{sec:gaiarvs}).
Section\,\ref{sec:conclusions} summarizes this data release, which we hope will serve as a useful legacy reference for large surveys.
Additional background on \veloce\ DR1 is provided in the online Appendix, including information on the samples of Cepheids (App.\,\ref{app:sampletable}) and non-Cepheids (App.\,\ref{sec:app:classification}), and the file structure of the data set published via Zenodo (App.\,\ref{app:datastructure}).

\section{Observations}\label{sec:observations}
\begin{figure*}[t]
    \centering
    \includegraphics{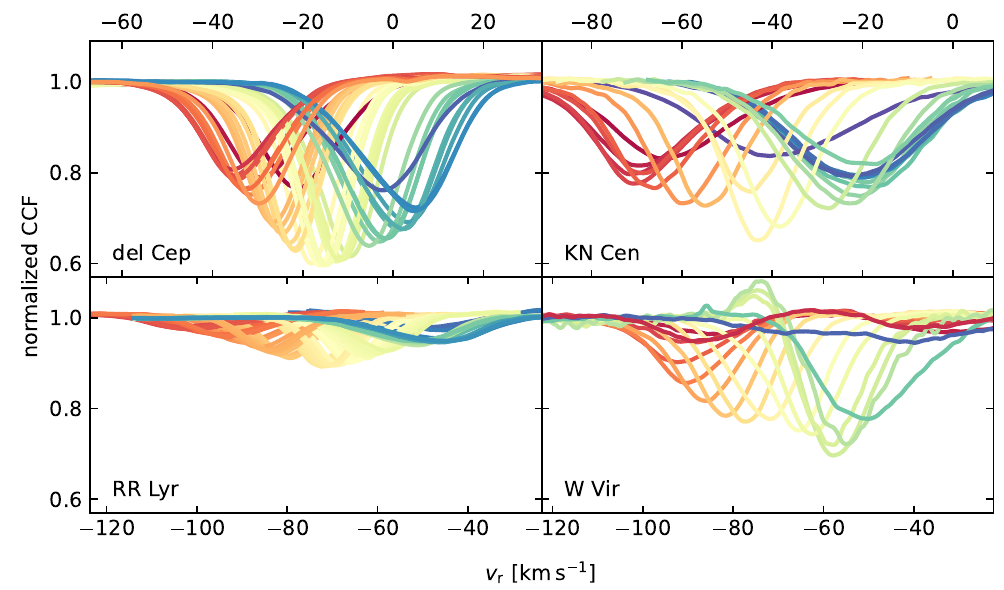}
    \caption{Example CCFs for four pulsating stars as labeled. The color traces the pulsation phase. KN\,Cen is a long-period, high-amplitude type-I Cepheid. The others are prototypes of their respective pulsating star classes. CCF shape variations follow a characteristic pattern for type-I Cepheids. RR Lyrae stars are hotter and feature shallower CCFs with very large amplitudes and line asymmetry, whereas W\,Virginis stars can exhibit signatures of shock phenomena at specific phases, including blue-shifted emission features and line splitting.}
    \label{fig:CCFs}
\end{figure*}

\veloce\ provides high-precision Cepheid RVs collected using two spectrographs on both hemispheres. 
This first data release (DR1) comprises \nobsall\ RV measurements of all stars observed as part of \veloce\ until 5 March 2022 (BJD = 2459644), including beat Cepheids, type-II Cepheids, and non-pulsators, among others. Additional, newer data, is included for \nextendedbaseline\ Cepheids where this made a difference to complete pulsational or orbital phase sampling\footnote{BG~Vel, IT~Car, MY~Pup, U~Vul, V1334~Cyg, V5567~Sgr, VY~Per, VZ~Pup, and W~Sgr}. Results from the RV curve fitting (Sect.\,\ref{sec:modeling}) are included for single-mode bona fide classical Cepheids only; non-Cepheid stars (App.\,\ref{sec:app:classification}) and double-mode Cepheid RV curves were not modeled (Sect.\,\ref{sec:unmodeled}). All data included in tabular form in this article are provided in machine-readable format at the  CDS. However, the majority of the data published here will be made available via Zenodo, as described in App.\,\ref{app:datastructure}. This data set includes the RV time series of all stars mentioned in this paper, as well as the model fits to single-mode bona fide Cepheids. Publication of the full observational data set is planned for a future data release.

Section\,\ref{sec:obs:coralie} (\coralie, south)  and Sect.\,\ref{sec:obs:hermes} (\hermes, north) present basic information on the spectrographs and telescopes used for \veloce. All RVs presented here are determined by the cross-correlation method and using Gaussian fits to the CCF (cf. Sect.\,\ref{sec:obs:CCFs}. When no simultaneous wavelength reference was available for RV drift corrections, we used a physical model of RV drift due to atmospheric pressure variations, cf. Sect.\,\ref{sec:obs:pressure}. Observations of RV standard stars are presented in Sect.\,\ref{sec:obs:zeropoints}. Section\,\ref{sec:targets} provides background on target selection for \veloce.

\subsection{Southern hemisphere observations with Coralie\label{sec:obs:coralie}}

\coralie\ is a fiber-fed high-resolution (R $\sim$ 60\,000) cross-dispersed echelle spectrograph and an improved copy of {\it ELODIE} \citep{Baranne1996}, which is famous for the discovery of the first extrasolar planet orbiting a main sequence star \citep{Mayor1995}. \coralie\ is mounted to one of two Nasmyth foci of the Swiss 1.2m Euler telescope at ESO La Silla Observatory in Chile and housed in a temperature-controlled Coud\'e room. The absolute wavelength calibration is provided by a ThAr lamp, whose spectrum is recorded during afternoon calibrations and can be reobserved as required for recalibration. 
\coralie\ was commissioned in 1998 and originally described by \citet{2001Msngr.105....1Q}. A first round of instrumental upgrades implemented in 2007 are described in \citet{2010A&A...511A..45S}. Observations performed between 2007 and 2014 are referred to as \coraliev{07} observations in the following. \coralie\ underwent a second significant upgrade in November 2014 \citep{VanMallePhD}. The most notable changes made were as follows. The original 2" optical fiber with a circular cross-section was replaced by a fiber with an octagonal cross-section that improves the stability of fiber pupil illumination, and therefore RV precision \citep{Chazelas2010,Chazelas2012,LoCurto2015}. The removal of the fiber scrambler (not necessary for hexagonal fibers) resulted in slightly reduced throughput. A Fabry-P\'erot \'etalon (FP) was introduced for simultaneous wavelength drift calibration whose absolute wavelength scale is provided by the ThAr lamp.
The altered optical path systematically impacts line shapes recorded before and after November 2014, and we therefore refer to \coralie\ observations collected after November 2014 as \coraliev{14}. An assessment of the impact of instrumental changes on the RV zero points is provided in Sect.\,\ref{sec:obs:zeropoints}.

An efficient reduction pipeline is available for \coralie. Data reduction follows standard recipes and performs pre- and overscan bias correction, flat-fielding using Tungsten (\coraliev{07}) or LED (\coraliev{14}) lamps, background modeling, and cosmic removal. The RVs are determined via cross-correlation 
\citep{Baranne1996,2002A&A...388..632P} on 2D spectra (orders not merged) using a
numerical mask designed for solar-like stars (optimized for spectral type G2).
The instrument is renowned for its stability and very
high precision of $\sim 3$\,m\,s$^{-1}$ that has enabled the discovery of hundreds of extrasolar planets 
\citep{2003ASPC..294...39P,2010A&A...511A..45S,bebop}.

\coraliev{07} observations were conducted using two different instrument modes, OBJO and OBTH. The former refers to a single-fiber observing mode, with no calibration entering the second available fiber. OBTH refers to an observing mode where the secondary fiber is illuminated by a ThAr wavelength calibration lamp for monitoring drifts of the wavelength calibration. This calibration spectrum is then interlaced between the science object's echelle orders. While initially all observations were taken in OBJO mode, we eventually shifted to OBTH mode for all observations to benefit from greater RV precision. While observing in OBTH mode, we recalibrated the wavelength solution whenever the simultaneous drift estimate exceeded $50$\ms; for OBJO observations we typically recalibrated when atmospheric pressure variations since the last wavelength calibration exceeded $\sim 0.5$\,mbar.

OBJO mode observations (\coraliev{07} only) are very sensitive to variations in atmospheric pressure, with a 1\,mbar pressure change roughly corresponding to a $80$\ms\ shift in RV. To mitigate this systematic, we applied a correction to the measured RVs using the model described in Sect.\,\ref{sec:obs:pressure}. An uncertainty of $0.015$\kms\ is added in quadrature to RVs based on OBJO observations to account for the uncertainty of the correction.

Almost all \coraliev{14} observations were conducted in OBFP mode, where the Fabry-P\'erot \'etalon provides extremely precise ($0.1 - 0.5$\,\ms) instrumental drift corrections. However, some OBTH observations were collected before OBFP became available.

\subsection{Northern hemisphere observations with Hermes\label{sec:obs:hermes}}

\hermes\ is a high-resolution echelle spectrograph, which features a resolving power of $R \sim 85,000$ and is mounted to the Flemish 1.2m Mercator telescope located on the Roque de los Muchachos Observatory on La Palma, Canary Island, Spain \citep{2011A&A...526A..69R}. \hermes\ features a wider spectral range than \coralie. However, we followed the standard \hermes\ pipeline recipe and used a correlation mask representative of a G2 spectral type that features  $1130$ metallic absorption lines in the spectral orders $55-74$ ($4781.1-6535.6$~\AA) to measure RV, as this wavelength range benefits from a better signal-to-noise ratio (S/N). 

All observations were performed in the high-resolution fiber (OBJ\_HRF) mode to achieve the best throughput and spectral resolution. The data were reduced using the dedicated reduction pipeline that carries out standard processing steps such as flat-fielding using Halogen lamps, pre- and overscan bias corrections, background modeling, order extraction, and cosmic clipping. ThAr lamps are used for the wavelength calibration at the beginning and end of the night and were used during nights to re-calibrate the wavelength solution if needed. We note that \hermes\ was upgraded with octagonal fibers to improve throughput and light scrambling properties on 25 April 2018 (G. Raskin, priv. comm. 19 November 2020) and that this change in optical path caused a systematic RV offset of $-0.075 \pm 0.007$\,\kms\ between \hermes\ RVs collected before and after the upgrade. The improved light scrambling significantly improved RV precision, with standard star\footnote{Excluding binaries HD~154417, HD~168009, and HD~42807} RVs exhibiting half the dispersion after the upgrade ($0.019$\,\kms) compared to before ($0.038$\,\kms), cf. Sect.\,\ref{sec:obs:zeropoints}. 

The OBJ\_HRF mode on \hermes\ is similar to OBJO on \coralie, and no simultaneous RV drift correction is available for this mode.
Several steps were taken to achieve maximum short-term RV precision and track long-term RV stability. To ensure short-term precision, we  monitored the nightly evolution of the ambient pressure, and re-calibrated the wavelength solution when pressure variations relative to the last calibration sequence exceeded $\sim 0.4 - 0.5$\,mbar, cf. Sect.\,\ref{sec:obs:pressure}. This  procedure achieves a short-term instrumental stability of approximately $10 - 15$\ms\ over the duration of at least 10 nights. We therefore added $0.015$\,\kms\ in quadrature to the RV uncertainties determined by the pipeline. 

We tracked the long-term stability and absolute RV scale of \hermes\ RVs using IAU RV standard stars  \citep{Udry1999a,Udry1999b}, cf. Sect.\,\ref{sec:obs:zeropoints}. A comparison of zero-points for Cepheids is shown in Sect.\,\ref{sec:zp}. Although higher-than-usual temperatures inside the temperature-controlled Coud\'e room were recorded between 22 and 26 June 2015 due to power cuts at the Observatory, no excess RV dispersion was observed among the standard stars during this BJD range ($57\,196 - 57\,200$). While we did notice that standard stars HD~168009, HD~221354, HD~144579, HD~141105, HD~154417 exhibited excess RV dispersion of $50-190$\,\ms\ between 26 June and 6 July, 2016 due to an unknown reason, we did not notice any significant issues involving the Cepheid RV data collected during these nights.

\subsection{Cross-correlation functions, Gaussian RVs, and uncertainties\label{sec:obs:CCFs}}
\begin{figure*}[t]
    \centering
    \includegraphics[width=0.8\textwidth]{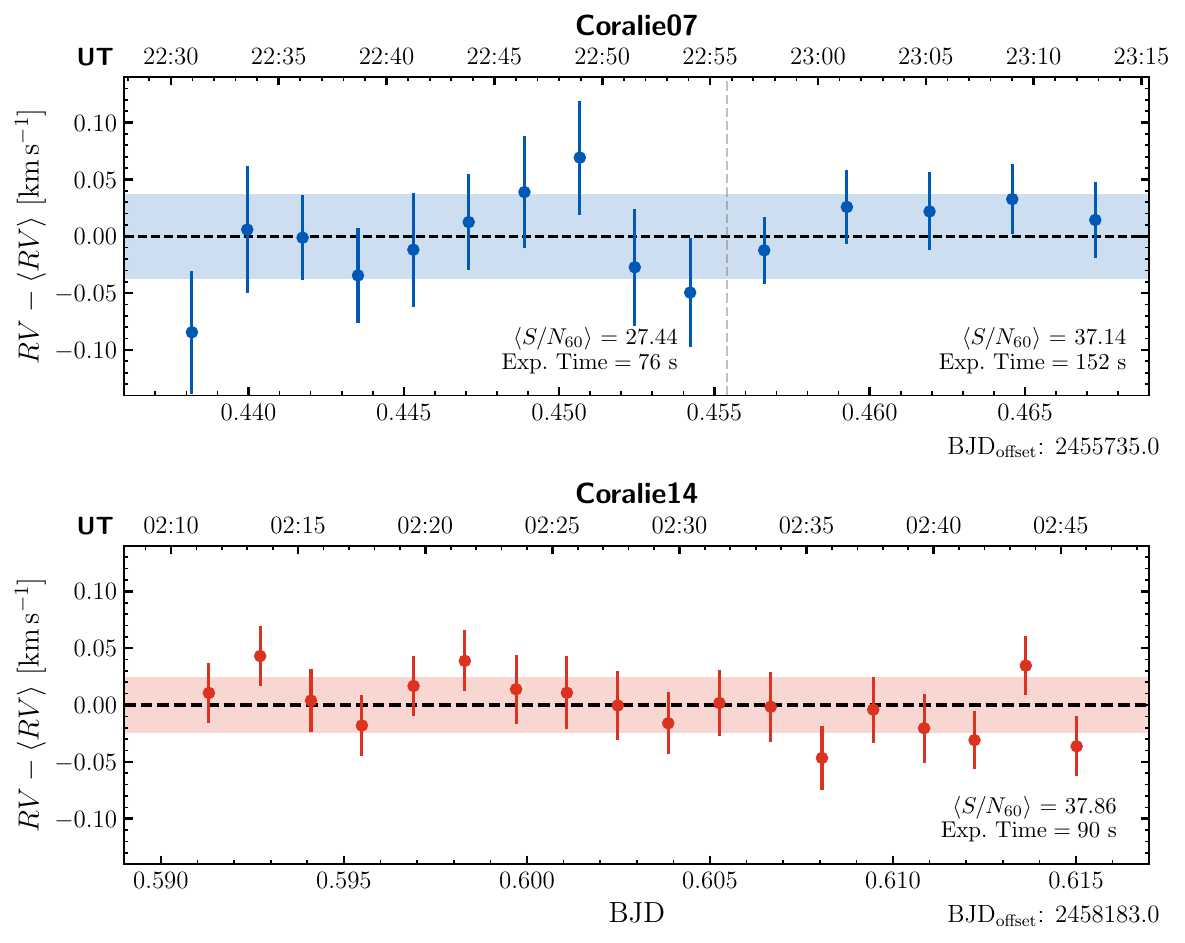}
    \caption{Short-burst (42 \& 34\,min) RV time series of the 20\,d Cepheid RZ~Vel near minimum RV collected to test the momentaneous precision of  \coraliev{07} and \coraliev{14}. Both panels show the measured RV minus the mean of the sequence. Shaded areas illustrate the dispersion of the sequences. Average S/N on the 60th echelle order are labeled for reference. Note the improved precision of \coraliev{14} over \coraliev{07}. {\it Top panel:} \coraliev{07} observations taken on 22 June 2011. Note the change of exposure time (vertical dotted line) that affects both precision and uncertainties estimated by the pipeline. {\it Bottom panel:} \coraliev{14} observations collected on 5 March 2018. 
    \label{fig:rvsequence}}
\end{figure*}

All \veloce\ RVs were determined using the cross-correlation method \citep{Baranne1996, 2002A&A...388..632P} using correlation templates made to  represent a solar-like star of roughly Solar metallicity (G2 mask). The cross-correlation technique reduces the information content of a few thousand spectral lines to a single line profile of extremely high S/N, the so-called cross-correlation function (henceforth, CCF). Example CCFs for two classical Cepheids, RR~Lyrae, and W~Virginis are  shown in Fig.\,\ref{fig:CCFs}. The key benefit of CCFs is that they enable exceptionally precise RV measurements, even from observations with a relatively low S/N. However, the increased RV precision comes at the price of a loss of information that complicates a detailed physical interpretation of CCF profiles.

All \veloce\ DR1 RVs are measured as the center position of a Gaussian profile fitted to the CCF. From this Gaussian profile, we also determine the full width at half maximum (FWHM) and the depth (contrast), which we will use in further work. 
The main advantage of Gaussian RVs is that they afford better precision than other methods commonly applied to Cepheid CCFs, such as CCF barycenters or bi-Gaussian profiles, both of which assign high weight to the lowest point of a CCF profile, where RV constraints are minimal \citep[Sect.~2.1.6 and Fig.~2.5]{AndersonPhD}. We caution that this choice notably affects the RV peak-to-peak amplitude, which would be largest in case of bi-Gaussian profiles. Moreover, bi-Gaussian profiles are more sensitive to RV curve modulation caused by line shape variations that do not follow the periodicity of the dominant mode, cf. \citet[Fig.~5]{Anderson2016c2c}. We note that RV uncertainties are estimated differently using \coralie\ and \hermes\ as explained in the following. 

The RV uncertainties of \coralie\ data are estimated by the data reduction software based on photon noise statistics and a detailed characterization of the instrument \citep{Bouchy2001}. Figure\,\ref{fig:rvsequence} illustrates that the uncertainties thus derived provide an adequate representation of the statistical precision, that is, the ability to reproduce a measurement within the determined uncertainties under nearly identical conditions. The figure shows the variation of the RV time series around the mean of the sequence for the $\sim 20$\,d Cepheid RZ~Vel near phases of maximal expansion velocity (minimum RV) during $45$\,minute intervals on 22 June 2011 (\coraliev{07}) and again on 5 March 2018 (\coraliev{14}). During each short sequence, virtually no RV variation due to pulsation is expected at pulsation phases of extremal RV, and none is indeed observed. Comparing the dispersion around the mean for each sequence of a given exposure time and the mean RV uncertainty of the same observations reveals excellent agreement; for \coraliev{07}, we find $\sigma(\mathrm{RV}-\langle{\mathrm{RV}\rangle})= 0.042$\,\kms\ for   76\,s exposures and $\sigma(\mathrm{RV}-\langle{\mathrm{RV}\rangle}) = 0.016$\,\kms\ for 5 longer 152\,s exposures, which compares to the average uncertainties of $0.048$\kms\ and $0.032$\kms, respectively. For 18 \coraliev{14} exposures of 90\,s each, we find $\sigma(\mathrm{RV}-\langle{\mathrm{RV}\rangle}) = 0.024$\,\kms\ and $\langle \sigma_{\mathrm{RV}} \rangle = 0.028$\,\kms.

RV uncertainties for \hermes\ observations are based on the covariance matrix of the Gaussian profile fit to the CCF because the instrument has not been characterized to the same RV precision as \coralie. The main shortcoming of this approach is that it does not consider photon noise as the dominant source of RV uncertainty \citep{Bouchy2001}. As a result, RV uncertainties of bright \hermes\ targets are often overestimated and in reality limited by the short-term stability set by RV drift corrections described in Sect.\,\ref{sec:obs:pressure}. This was previously shown for $\delta$~Cephei and Polaris  \citep{Anderson2015,Anderson2019polaris}. RV uncertainties derived from the Gaussian fit covariance matrix tend to be more adequate at lower S/N values ($\lesssim 30$).

Cepheids are complex stars and Gaussian fits to asymmetric CCFs yield biased estimates of the photospheric motion projected onto the line of sight. Yet, these ``Gaussian RVs'' benefit from very high precision in the differential sense that provide a powerful tool to detect additional variability signals that we collectively refer to as ``modulated variability'' \citep{Anderson2014rv}, cf. Sect.\,\ref{sec:modulations}. Hence, studies seeking to interpret Cepheid RVs in an absolute sense, e.g., to determine the Galactic rotation curve, should be warned that several complications limit the accuracy (in an absolute astrometric sense, \citealt{Lindegren2003}) of the measured RVs to the level of a few hundred \ms. These effects include a) atmospheric velocity gradients and the finite formation height of spectral lines in Cepheids that challenge the notion of ``the'' RV of the star at any given time at the level of hundreds of \ms\ to \kms\ \citep[e.g.,][]{Nardetto2007,Anderson2016c2c,Wallerstein2019}, b) spectral line asymmetry not accounted for by Gaussian profiles \citep{Nardetto2006,AndersonPhD}, c) (time-variable) gravitational redshift on the order of a few tens of \ms\ \citep{Dravins2005,Pasquini2011}, among others. The combination of such effects results in the $K-$term issue described by \citet{Nardetto2008Kterm}. We also note that amplitudes determined using other methods for RV determination (including different wavelength ranges) may yield different results, for example, if bi-Gaussian profiles are fitted to the asymmetric CCFs. It is therefore crucial to carefully evaluate whether measurements are sufficiently compatible when combining different datasets, see e.g., \citet{Anderson2019polaris}.
Further details on related issues have been presented by \citet{Anderson2018rvs}.

\subsection{RV drift corrections from atmospheric pressure changes}\label{sec:obs:pressure}

Neither \hermes\ nor \coralie\ are pressure-stabilized spectrographs. Atmospheric pressure variations during the course of a night change the diffraction index of the air within the spectrograph, causing a mismatch between the wavelength calibration and the instantaneous wavelength observed at a given pixel location on the detector. Expressed as a shift in velocity space, the displacement of the wavelength solution is usually referred to as an RV drift and measured in \ms. Atmospheric pressure variations are the leading source of RV drift in modern (temperature-, but not pressure-stabilized) spectrographs and can occur fairly rapidly on timescales of an hour or less.

\coralie\ RVs collected in OBTH or OBFP modes are corrected for instrumental drift using a simultaneous wavelength reference that is interlaced between the science orders. However, RVs measured using \coralie\ OBJO and \hermes\ OBJ\_HRF observations are subject to instrumental drift that we aimed to mitigate using a physical model of the RV drift due to ambient pressure changes.

The physical model for atmospheric drift correction was developed by \citet[Ch.~2.1.5]{AndersonPhD} and predicts the following RV drift correction for a change of atmospheric pressure, $\Delta P$, in mbar:
\begin{equation}
    \Delta v (\Delta P) \approx 2.7 \cdot 10^{-7} \cdot \Delta P \cdot c \approx 81 \cdot \Delta P [\mathrm{m\,s^{-1}\,mbar^{-1}}] \ ,
    \label{eq:RVdriftPressure}
\end{equation}
where $c$ is the speed of light in \ms.  

\begin{figure}
    \centering
    \includegraphics[width=0.48\textwidth]{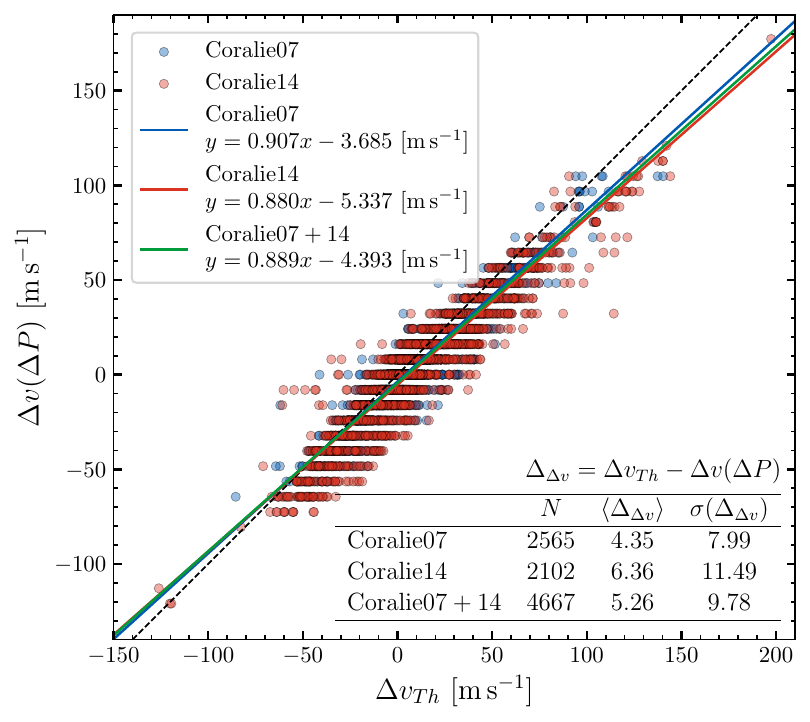}
    \caption{
    RV drift corrections determined from pressure variations (Eq.\,\ref{eq:RVdriftPressure}) against the corrections determined via 
    simultaneous drift corrections based on interlaced ThAr spectra. The discretization of the results is due to the single-decimal precision with which the atmospheric pressure is recorded in the FITS headers. The black dotted line shows the equality $\Delta v (\Delta P) = \Delta v_{Th}$, whereas the blue, red, and green lines are the best linear fits to the \coraliev{07}, \coraliev{14}, and combined \coralie\ distributions, as quantified in the legend. \label{fig:plotRVdriftPressureCorrection}}
\end{figure}

We adopted this model to correct RV drifts of  3359 \coralie\ observations collected in OBJO mode as well as of all \hermes\ (OBJ\_HRF) observations. Fig.\,\ref{fig:plotRVdriftPressureCorrection} shows a comparison of our RV drift model to 4667 \coralie\ observations collected in OBTH mode. The model adequately represents RV drifts, and we found a small global offset of $\sim 5$\,\ms\ and a $10\%$ larger RV drift estimate based on the simultaneous reference compared to the pressure model. Part of the scatter in this relation is due to the rather coarse $0.1$\,mbar resolution of ambient pressure in the data headers, as well as differences between the pressure measured by the weather station and the air pressure inside the instrument. Of course, the best results are achieved when this model is used in conjunction with intra-night wavelength recalibrations scheduled when atmospheric pressure variations exceed $\pm 0.5$\,mbar to avoid large corrections. Based on this comparison, we estimate a systematic uncertainty of approximately $10-15\,$\ms\ for this correction and conservatively added $15$\,\ms\ in quadrature to the reported RV uncertainties for \coralie\ OBJO and \hermes\ observations. Further improvements to this models would be feasible by considering the impact of the altitude of the observatory on the model as well as other instrument parameters, such as grating or CCD temperatures. However, the first-order approximation presented here is sufficient for our purposes.

\subsection{RV standard stars}\label{sec:obs:zeropoints}

\begin{table*}
\centering
\caption{List of \hermes\ RV standard stars\label{tab:hermes_stds}}
\begin{tabular}{@{}l l r r r r r r r @{}}
\toprule
Star & SpType & FWHM & HER rv & HER err & HER18 rv & HER18 err & HER - HER18 & comb. err \\
 & & \kms & \kms & \kms & \kms & \kms & \kms & \kms \\
\midrule
HD154417$^{\dagger}$ & F8.5IV & 10.883 & -16.729 & 0.009 & -16.668 & 0.005 & -0.061 & 0.011  \\
HD168009$^{\dagger}$ & G1V & 8.370 & -64.570 & 0.002 & -64.509 & 0.004 & -0.061 & 0.005 \\
HD32923  & G4V & 7.982 & 20.623 & 0.005 & 20.747 & 0.004 & -0.125 & 0.007  \\
HD197076 & G5V & 8.178 & -35.417 & 0.005 & -35.321 & 0.004 & -0.097 & 0.006  \\
HD42807$^{\dagger}$  & G5V & 9.412 & 6.067 & 0.003 & 6.148 & 0.006 & -0.080 & 0.006  \\
HD141105 & G8V & 8.225 & 1.876 & 0.003 & 1.957 & 0.003 & -0.081 & 0.004  \\
HD144579 & G8V & 7.532 & -59.445 & 0.003 & -59.372 & 0.002 & -0.074 & 0.004  \\
HD10780  & K0V & 8.243 & 2.810 & 0.007 & 2.841 & 0.004 & -0.031 & 0.008  \\
HD221354 & K0V & 8.284 & -25.107 & 0.002 & -25.049 & 0.002 & -0.058 & 0.003  \\
HD65583  & K0V & 7.523 & 14.810 & 0.037 & 14.886 & 0.006 & -0.076 & 0.038  \\
HD82106  & K3V & 9.868 & 29.803 & 0.007 & 29.880 & 0.003 & -0.076 & 0.008  \\
\midrule
\multicolumn{7}{r}{weighted average difference HER - HER18 [\kms]:} & -0.075 & 0.007 \\
\bottomrule
\end{tabular}
\tablefoot{Spectral types from SIMBAD. HER and HER18 denote \hermes\ RVs before and after the upgrade of the optical fiber on 25 April 2018, respectively. Columns labeled `rv' and `err' refer to straight means and standard errors on the mean, respectively. The last two columns show the difference before minus after the upgrade and the square summed mean errors. Low-amplitude spectroscopic binaries exhibiting orbital signals have been marked with a $^\dagger$. NB: These offsets have not yet been applied to \veloce\ data.}
\label{tab:RVstds}
\end{table*}

\begin{table*}[]
    \centering
    \caption{List of \coralie\ RV standard stars    \label{tab:coralie_stds}}
    \begin{tabular}{@{}llrrrrrrr@{}}
\toprule
Star & SpType & FWHM & COR07 rv & COR07 err & COR14 rv & COR14 err & COR07 - COR14 & comb.~err \\
 & & \kms & \ms & \ms & \ms & \ms & \ms & \ms \\
\midrule
HD1581     &    F9.5V    &  7.83        & 9315.71       & 1.34   & 9342.99      & 0.40    & -27.28        & 1.39 \\
HD65907A  &     F9.5V    &  7.55        & 14987.94      & 3.27   & 15017.66     & 0.26    & -29.72        & 3.28 \\
HD108309  &     G2V      &  8.37        & 30617.57      & 2.12   & 30642.60     & 0.33    & -25.03        & 2.14 \\
HD144585  &     G2V      &  8.52        & -14016.01     & 2.12   & $-$13995.28  & 0.32    & -20.73        & 2.14 \\
HD20794   &     G6V      &  7.67        & 87907.57      & 0.66   & 87938.41     & 0.23    & -30.84        & 0.69 \\
HD10700   &     G8V      &  8.19        & -16568.22     & 1.24   & $-$16546.5   & 0.20    & -21.72        & 1.25 \\
HD161612  &     G8V      &  8.40        & 2488.82       & 1.26   & 2509.41      & 0.50    & -20.59        & 1.35 \\
HD190248  &     G8IV     &  8.37        & -21511.53     & 3.78   & $-$21492.73  & 0.29    & -18.8 & 3.79 \\
HIP75386  &     G6III    &  8.37        & -13723.49     & 3.74   & $-$13694.49  & 3.37    & -29.00      & 5.03 \\
HIP10440  &     G9III    &  8.88        & 20615.30      & 3.95   & 20641.42     & 2.72    & -26.12        & 4.79 \\

\midrule
\multicolumn{7}{r}{weighted average difference COR07 - COR14 [\ms]:} & $-26.8$ & $0.5$ \\
\bottomrule
    \end{tabular}
    \tablefoot{\coralie\ RV standard star measurements in \ms. The finer scale compared to Tab.\,\ref{tab:hermes_stds} is warranted by the better precision due to the available simultaneous drift correction. COR07 refers to \coralie\ after instrument upgrades installed in 2007 and before upgrades installed in 2014, label henceforth as COR14. The RV zero-points and their uncertainties were determined using \texttt{DACE}\footnote{\url{https://dace.unige.ch/}} \citep{DACE}. The two last columns list the difference between COR07 minus COR14 and the square-summed errors. The weighted average of the offsets is $-26.8 \pm 0.5$\,\ms. NB: These offsets have not yet been applied to \veloce\ data.}
\end{table*}

We tracked the long-term stability and absolute RV scale of \hermes\ and \coralie\ using IAU RV standard stars  \citep{Udry1999a,Udry1999b}. In particular, we investigated whether the installations of fibers with octagonal cross-sections (2014 for \coralie, 2018 for \hermes) impacted RV zero-points and precision. Tables\,\ref{tab:hermes_stds} and \ref{tab:coralie_stds} list the spectral types, CCF full FWHM, and RVs determined before and after the two upgrades, alongside the differences and a mean offset. For \hermes, we note a factor of $2$ improvement in the dispersion of standard star RVs thanks to the upgrade (from $0.038$ to $0.019$\,\kms), as well as a systematic increase in RV by $0.075\pm0.007$\,\kms. For \coralie, the weighted average of the zero-point offsets is $-0.0267 \pm 0.0005$\,\kms. A more detailed inspection of the \coralie\ zero-point would need to account for differences in FWHM. However, this is out of scope here and will be presented elsewhere (Segransan et al., in prep.). Neither of the zero-point shifts have been applied in \veloce\ DR1 because sufficient standard star information was not available at the time when the Cepheid light curves were fitted. Future data releases will be corrected for these zero-point offsets. We note that the typical root mean square of Cepheid RV curve residuals typically exceeds $50$, if not $80$, \,\ms\ due to stellar effects (Sect.\,\ref{sec:modulations}), so that the zero-point changes do not challenge any of the conclusions presented here. However, studies targeting very detailed investigations of modulated variability should certainly account for these zero-point shifts.

\subsection{Sample selection and observational strategy\label{sec:targets}}

Observations for the \veloce\ project have been collected using \coralie\ and \hermes\ since November 2010. From 2010 to mid-2019, access to the Euler and Mercator telescopes was typically available during 10 (\hermes) and $14-15$ night (\coralie) observing runs. Data obtained during that time frame hence tends to ``clump'' within windows of about two weeks, with some exceptions made possible via time exchanges with other groups. On both telescopes, data were typically collected during two to three observing runs per year. Subsequent observing runs were spaced by at least two weeks, and up to several months apart. This mode of access to both telescopes was well suited for achieving good phase coverage for Cepheids with $P \lesssim 11-15$\,d. For longer-period Cepheids, efforts were made to complete phase coverage via time exchanges and by considering pulsation phases accessible for each star in every given observing run. Some observing runs were also specifically timed such as to coincide with pulsation phases of interest for specific very long-period Cepheids. This close monitoring of accessible pulsation phases for all sources during all observing runs made it possible to obtain very good phase sampling over relatively short periods of time so that orbital motion and other modulations could be well separated from pulsational variability and  uncertainties related to the pulsation ephemerides.

Observations for \veloce\ were involuntarily interrupted between the end May 2019 and March 2021, first due to changes in telescope operations and funding, and second due to uncertainties of the lead author's postdoctoral career path during the COVID-19 pandemic. Thankfully, time exchanges and the good will of observers, as well as KU~Leuven and Mercator staff, made it possible to obtain additional observations of $\delta$~Cephei in 2020, with crucial importance for constraining its highly eccentric orbit by probing times near the most extreme orbital velocities (cf. paper~II). 

In March 2021, \veloce\ observations resumed on both telescopes thanks to funding provided by the European Research Council. Observations collected since March 2021 follow a very different temporal sampling, which seeks to provide a more even monitoring across the year. New observations on Euler in particular can target very specific pulsation and orbital phases, and were used to fill in gaps in phase coverage and to extend baselines for many stars to improve sensitivity to orbital motion. 

The selection function of \veloce\ as a survey is not well defined, since \veloce\ was not initially conceived as a long-term project. The sample of stars was regularly modified for a variety of reasons and science interests, which notably included
(1) to densely sample pulsation periods most relevant to tracing the \citet{HertzsprungProgression} progression spectroscopically;
(2) long-period Cepheids with high-quality parallax measured using the {\it Hubble Space Telescope} ({\it HST}) or {\it Gaia} as well as Cepheids with photometric observations from {\it HST}  \citep{Anderson2016ApJS,Riess2018hstphot,Riess2018hstplx,Riess2021} owing to their importance for the extragalactic distance scale;
(3) Cepheids observed with long-baseline optical interferometry \citep[e.g.,][]{Breitfelder2016,Anderson2016vlti,Gallenne2019}, e.g., to calibrate Baade-Wesselink projection factors;
(4) spectroscopic binary Cepheids, notably ones fainter than $\sim 8-9$th magnitude, with brighter Cepheids included as fillers and to achieve orbital solutions;
(5) known or suspected spectroscopic binaries as well as new binary candidates discovered by the ongoing program;
(6) very bright Cepheids, such as $\eta$~Aql, $\ell$~Car, or $\delta$~Cep, to obtain RV curves and spectral time series of exceptional quality, e.g., to study line shape variability and atmospheric effects \citep[e.g.,][]{Anderson2016c2c};
(7) possible signatures of shock-related phenomena;
(8) monitoring Cepheids whose RV curves exhibit signatures of additional signals, such as long-term amplitude variations or cycle-to-cycle fluctuations  \citep{Anderson2014rv};
(9) Cepheids in the vicinity of open clusters \citep{Anderson2013,CruzReyes2023};
(10) spectroscopic follow-up of less well studied Cepheid candidates from the All Sky Automated Survey (ASAS) \citep{Pojmanski2002,Pojmanski2003,Pojmanski2004,Pojmanski2005,Pojmanski2005a} and other ground-based surveys, including objects located away from the Galactic plane as filler objects when no other targets were observable;
(11) a few type-II Cepheids and RR~Lyrae stars observed when no classical Cepheids were observable
(12) RV standard stars, notably on \hermes.

A large number of Cepheid candidates from photometric surveys 
were found to be non-Cepheids based on their spectral properties, notably the absence of a clear CCF peak, multiple peaks, or absence of line shape variability, cf. \citet{AndersonPhD}. Limited information concerning these targets is provided in App.\,\ref{sec:app:classification}. 

We aimed to collect at least three, better four or more, epochs of reasonably well sampled RV phase curves. Depending on the observational progress, priorities and strategies were adapted to ensure that interesting phenomena could be followed up.  The focus throughout was to obtain measurements with uncertainties $< 50$\ms\ per observation. Most observations targeted S/Ns of at least $20$ (more commonly $25-30$) per pixel near 5500\,\AA. For fainter stars, the target S/N was lowered to $\sim 10$ due to the limited collecting area, producing satisfactory results notably for detecting orbital motion given the somewhat larger RV uncertainties reaching $100$\,\ms. Higher target S/Ns were adopted for bright Cepheids of interest. For Cepheids visible to the naked eye, typical S/Ns exceed $\sim 100$ per pixel. As a result, a subset of observations is viable for more detailed spectroscopic analysis, although the S/N of the majority of observations may be too low for this purpose.

\section{A high-precision catalog of Cepheid RVs}\label{sec:catalog}

\begin{figure*}
    \centering
    \includegraphics[width=\textwidth]{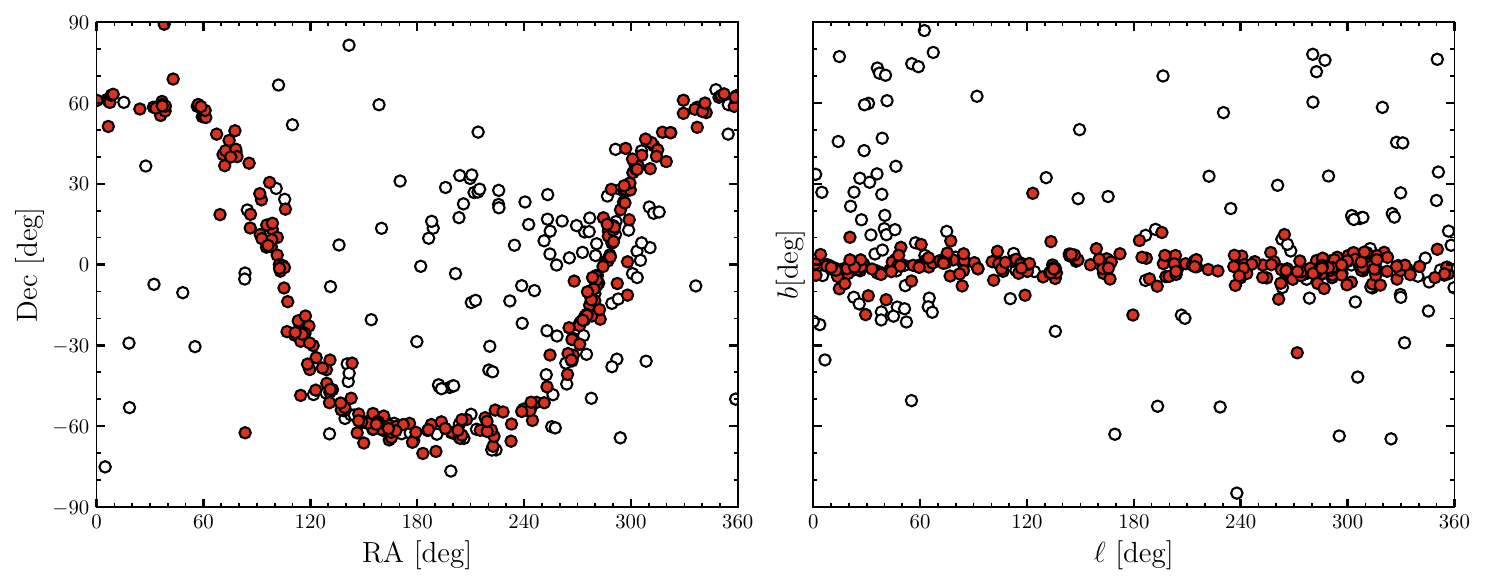}
    \caption{Sky coverage of \veloce\ targets in sky coordinates (left) and Galactic coordinates (right). Bona fide Cepheids are shown as red filled circles, stars that are not classical Cepheids are shown as open circles.}
    \label{fig:skycoverage}
\end{figure*}

\begin{figure*}
    \centering
    \includegraphics[width=\textwidth]{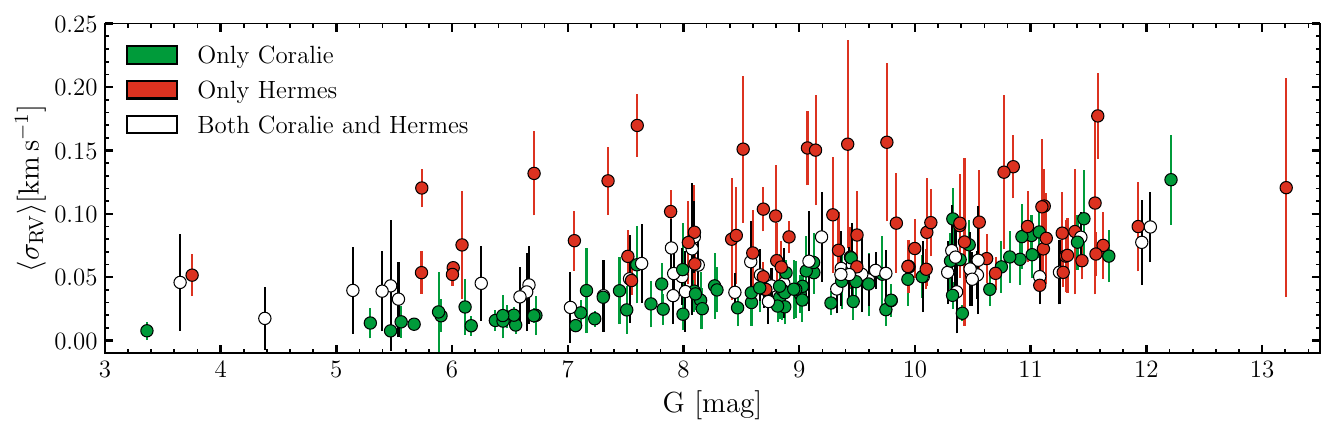}
    \caption{Mean per epoch RV uncertainty of bona fide Cepheids with at least 9 observations as a function of the mean \gaia\ DR3 $G-$band magnitude, cf. Tab.\,\ref{tab:Cepheidcatalog}. \coralie\ uncertainties (green points) are estimated differently from \hermes\ uncertainties, cf. Sect.\,\ref{sec:obs:CCFs}, and are limited by photon noise. \hermes\ uncertainties (red points) are overestimated for $G \lesssim 10$\,mag. The brightest star, Polaris, is missing due to a lack of \gaia\ photometry; the faintest target is GL~Cyg. X~Sgr is excluded from the plot due to significant line shape distortions \citep{Mathias2006,AndersonPhD} that complicate uncertainty estimation.}
    \label{fig:RVerrorMag}
\end{figure*}

\begin{figure}
    \centering
    \includegraphics[width=0.49\textwidth]{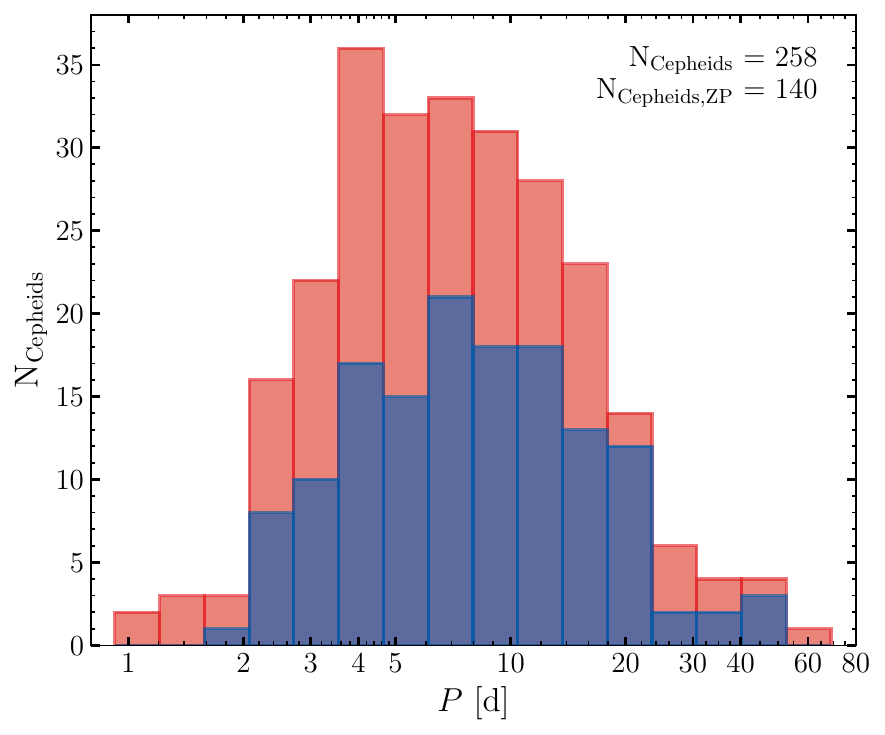}
    \caption{Distribution of pulsation periods of Cepheids in \veloce. The red histograms shows all classical Cepheids, while the blue histogram shows the stars used to determine instrumental zero-points, cf. Sect.\,\ref{sec:zp} and Sect.\,\ref{sec:gaiarvs}.}
    \label{fig:Phisto}
\end{figure}

\begin{figure}
    \centering
    \includegraphics[width=0.48\textwidth]{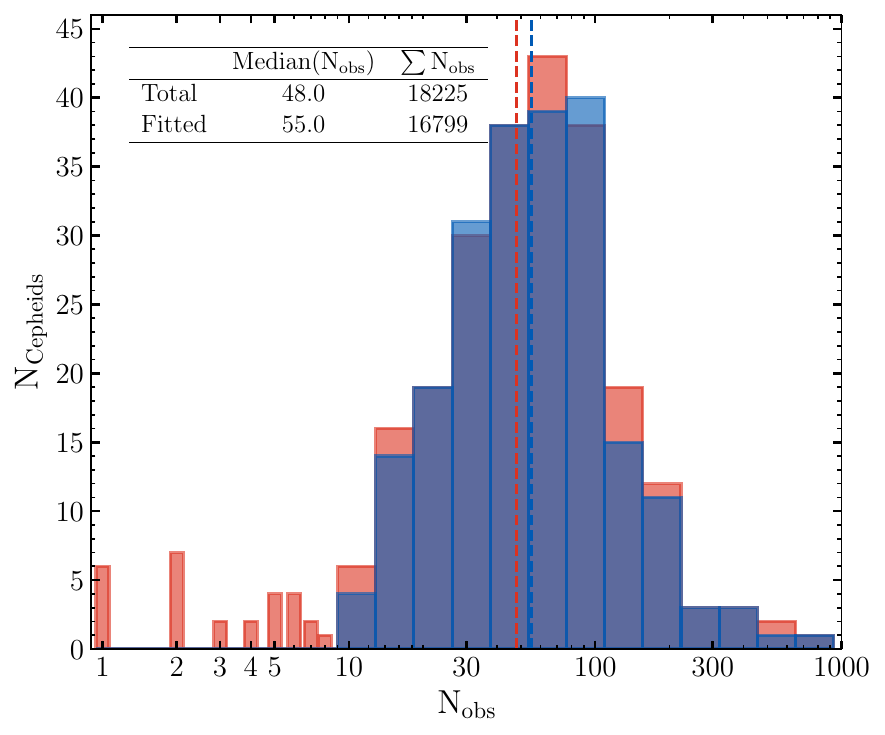}
    \caption{Distribution of the number of observations per Cepheid. The red histogram shows the available number of observations ($N_{obs}$) of all Cepheids in \veloce. The blue histogram shows the number of measurements used to fit Cepheid RV curves ($N_{\rm{obs, fitted}}$). The bin edges are identical for both histograms and logarithmically spaced. The first 8 bins of the red histogram are shown individually to improve readability; a threshold of at least 9 observations was used to fit pulsational models (Sect.\,\,\ref{sec:insufficient}). Differences between the blue and red histograms at $N_{obs} > 8$ arise due to data selection for model fitting (e.g., in the case of binaries, Sect.\,\ref{sec:adequacy}) or additional signals preventing a reasonable fit (Sect.\,\ref{sec:modulations}). The legend summarizes the median and total number of \veloce\ observations per group.}
    \label{fig:histo_nobs}
\end{figure}

\veloce\ DR1 delivers \nobsall\ RV measurements of \ncepinclusive\ bona fide MW classical Cepheids and \nnoncep\ stars that are not classical Cepheids. The targets are distributed across both hemispheres and clearly trace the Galactic disk, cf. Fig.\,\ref{fig:skycoverage}. The dominant uncertainty on the RV measurements is photon noise, so that there exists a fairly clean correlation between the best RV uncertainties and the magnitude of the source, which, however, can be altered by differences in S/N and astrophysical effects. The dependence of the mean RV uncertainty versus magnitude is illustrated in Fig.\,\ref{fig:RVerrorMag}. We note that the different estimation of RV uncertainties by the \coralie\ and \hermes\ pipelines lead to overestimated \hermes\ uncertainties, especially at bright magnitudes. An uniquely extreme example of this is the bright $7$\,d Cepheid X~Sgr (not shown), for which \hermes\ and \coralie\ uncertainties can differ by two orders of magnitude. However, this particular issue is limited to X~Sgr, where significant line deformations \citep{Mathias2006,AndersonPhD} complicate the interpretation of RVs and their uncertainties.

RV time series of stars that are not classical Cepheids are provided ``as-is'', meaning that the output of the reduction pipelines was not vetted. For example, multi-lined binaries will generally result in a single RV measurement, although it is unclear which component is being measured. The reduced spectra and CCFs of non-Cepheids are available upon request from the author. 

For \ncep\ single-mode classical Cepheids in our sample (henceforth: the Cepheid sample), we determined the pulsation-averaged velocity $v_\gamma$,  best-fit Fourier series models, and pulsation ephemerides. Figure\,\ref{fig:Phisto} shows the distribution of Cepheid pulsation periods in \veloce\ and highlights the sample used to assess zero-point differences with literature Cepheid RVs, whereas Fig.\,\ref{fig:histo_nobs} shows the distribution of the number of observations per target ($\sim 50$). Table\,\ref{tab:Cepheidcatalog} in App.\,\ref{app:sampletable} lists our preferred identifiers, coordinates at epoch J2000, \gaia\ DR3 source identifiers, the intensity-averaged \gaia\ $G-$band and average $G_{Bp} - G_{Rp}$ color, pulsation mode, a binary flag (SB1 for single-lined spectroscopic binary), the number of observations available, $N_{\mathrm{obs}}$, the number of Fourier harmonics used to fit the star, $N_{\mathrm{FS}}$, pulsation period \Ppuls, and the reference epoch $E$ used for the fit. The binary flag indicates whether we consider a given star to exhibit orbital motion based on the \veloce\ data set alone, since this affects the modeling choice for the Fourier series model. Spectroscopic binaries are investigated in detail in paper~II, including also literature data.  Table\,\ref{tab:Cepheidcatalog} also contains the names of bona fide Cepheids for which we report RV time series without Fourier modeling, e.g., due to lack of data or particularly complex signals.

Table\,\ref{tab:noncep} provides summary information for \nnoncep\ non-classical Cepheids, including \nrrlyrae\ RR~Lyrae stars, \ntypeIIcep\ type-II Cepheids, \nsbIIsbIII\ (candidate) double- or triple-lined binaries.

\begin{figure}
    \centering
    \includegraphics[width=0.48\textwidth]{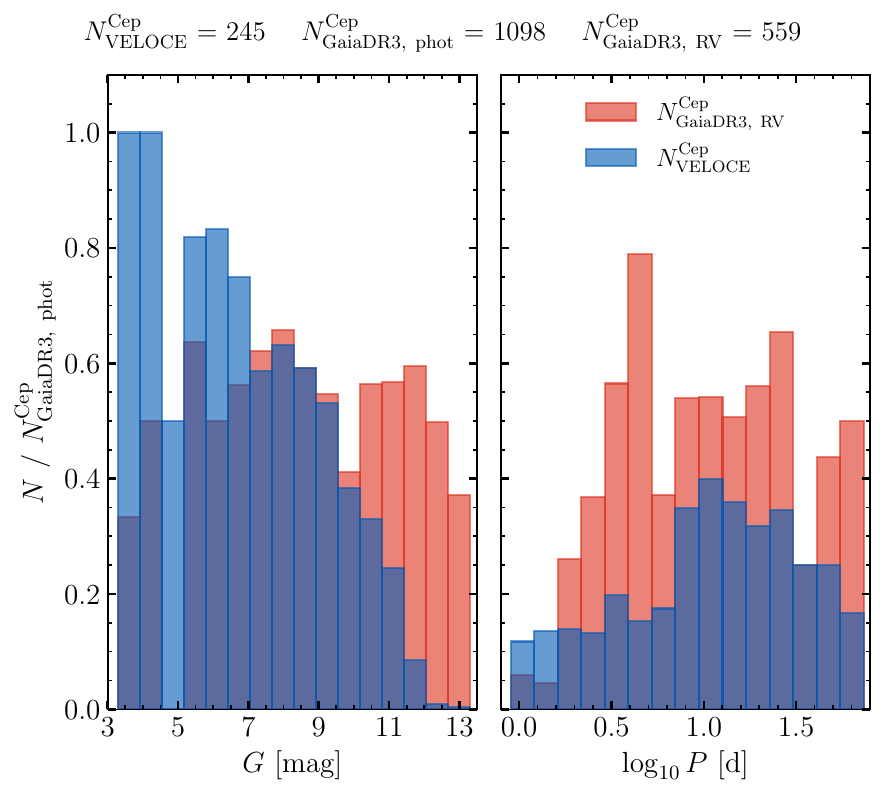}
    \caption{Distributions of the number ratios of Cepheids in \veloce\ to those in \gaia\ as a function of magnitude (left) and \Ppuls\ (right). The comparison is limited to $245$ \veloce\ Cepheids for which $G-$band magnitudes are available and to $N^{\rm Cep}_{\rm GaiaDR3,phot} = 1098$ MW Cepheids in \gaia\ with $G < 13.3$\,mag and $P > 0.9$\,d. $N^{\rm Cep}_{\rm GaiaDR3,RV} = 559$ have published RV time series measurements, cf. Sect.\,\ref{sec:gaiarvs}. Blue histograms show the fraction of \veloce\ Cepheids relative to the \gaia\ sample, $N^{\rm Cep}_{\veloce} / N^{\rm Cep}_{\rm GaiaDR3,phot}$, and red histograms show the fraction of Cepheids with \gaiarvs\ time series measurements relative to the same, $N^{\rm Cep}_{\rm GaiaDR3,RV} / N^{\rm Cep}_{\rm GaiaDR3,phot}$.
    \label{fig:frac_Ceph_GaiaRVS}}
\end{figure}

Compared to \gaia\ DR3,  \veloce\ contains $\sim 3000$ more individual RV measurements of classical Cepheids (\nobs), albeit of much fewer stars (\ncep\ vs 731). \veloce\ data thus generally sample RV curves more closely and over longer temporal baselines, similar to  \gaia's end-of-mission baseline. A detailed comparison of \veloce\ to \gaiarvs\ is provided in Sect.\,\ref{sec:gaiarvs}. Figure\,\ref{fig:frac_Ceph_GaiaRVS} illustrates the completeness of \veloce\ relative to 1098 \gaia\ DR3 MW Cepheids (from the Specific Objects Studies table \texttt{gaiadr3.vari\_cepheids}) with $P> 0.9$\,d and $G < 13.3$\,mag. The blue histograms show the fractions of Cepheids observed by \veloce\ over the \gaia\ sample as a function of magnitude (left) and \Ppuls\ (right). Red histograms show the fraction of \gaia\ DR3 Cepheids with RV time series (559 Cepheids from table \texttt{gaiadr3.vari\_epoch\_radial\_velocity})  to the same photometric sample. Unsurprisingly given the small telescope diameter, the completeness of \veloce\ is a strong function of magnitude, and only few stars fainter than $12$th magnitude were observed. 

\veloce\ data are made available in \texttt{FITS} format via Zenodo alongside supporting \texttt{python} code to facilitate data access. Appendix\,\ref{app:datastructure} provides details on the data structures containing \veloce\ data and fit results. All data tables presented in this article will additionally be made available in machine-readable form via the Centre de Donn\'ees de Strasbourg.

\subsection{Fourier series model fitting\label{sec:Fourier}\label{sec:modeling}}

\begin{figure*}[t]
    \centering
    \includegraphics[width=0.8\textwidth]{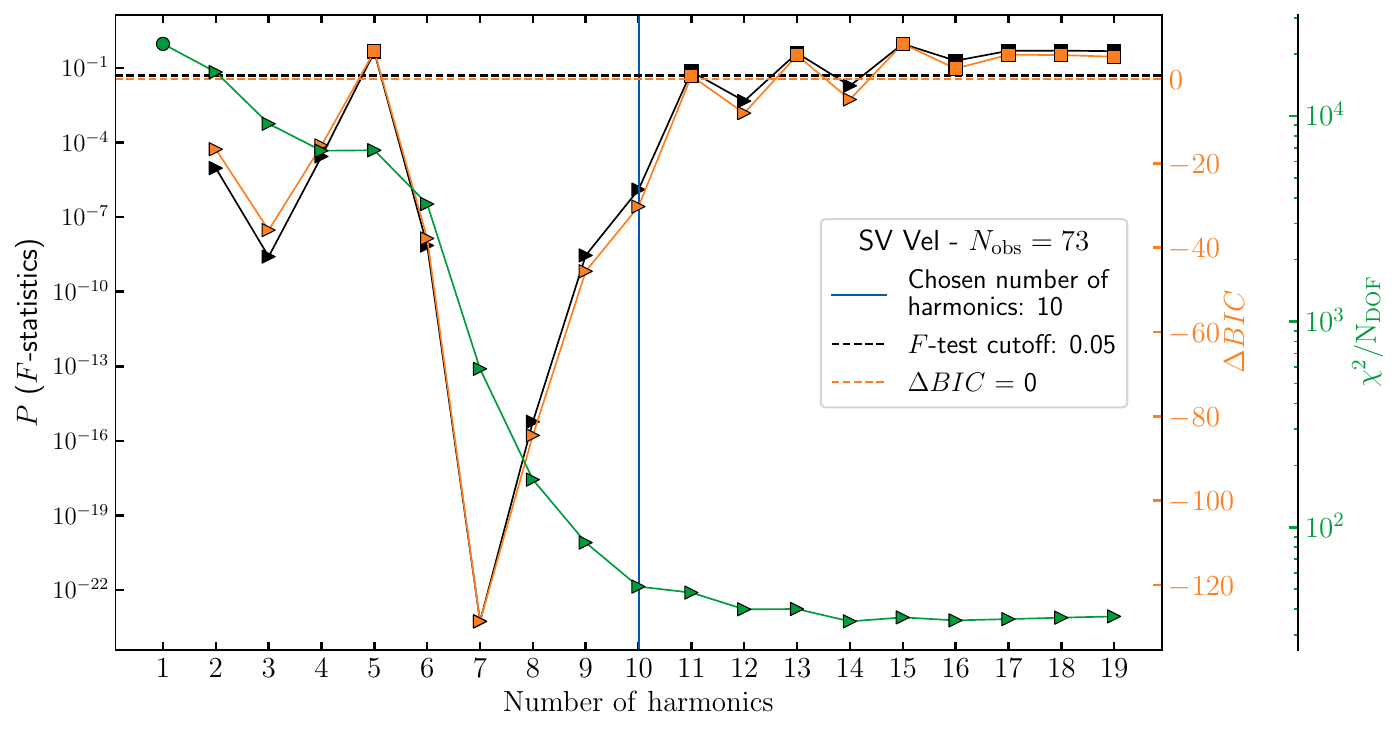}
    \caption{Illustration of the automated model selection procedure. Up to twenty harmonics were computed in a brute force approach, depending on data availability. The statistical significance of increasing the model's complexity was assessed using both F-test ($p < 0.05$) and BIC ($\Delta \rm BIC < 0$) criteria. Violations of these criteria were accepted if the two subsequent, more complex, models both indicated significant fit improvements. The improvement in $\chi^2/N_{\rm DOF}$ of the fit is shown in green.}
    \label{fig:modelselect}
\end{figure*}

We modeled the observed RV time series, $v_r(t)$, of single-mode bona fide Cepheids using Fourier series with $n$ harmonics as
\begin{equation}
v_r(t) = v_{\gamma} + \sum_{n=1,2,3,...}{a_{n} \cos{(2 n \pi \phi)} + b_{n}
  \sin{(2 n \pi \phi)}}\,,
  \label{eq:Fseries}
\end{equation}
where $v_\gamma$ denotes the pulsation averaged velocity, $t$ the midpoint of the shutter opening time in barycentric Julian Date, $\phi = (t - E) / P_{\rm{puls}}$ the pulsation phase, $P_{\rm{puls}}$ the pulsation period, and $E$ the reference epoch. Secular period changes due to stellar evolution are typically irrelevant over the timescale of the \veloce\ baseline, whereas irregular or periodic period changes appear to various degrees, but cannot be adequately modeled using the available data. However, the longer baselines covered by our template fitting analysis (Sect.\,\ref{sec:RVTF}) requires solving for phase shifts over time, which we use to measure period changes in Sect.\,\ref{sec:pdot}. Following \citet{Anderson2016c2c}, $E$ is defined such that $\phi=0$ coincides with $v_\gamma$ on the steeper, descending branch of the RV curve close to the mean observation date. This definition identifies the phase of minimum radius, which has several advantages over maximum light: a) it is most precisely measurable thanks to the steepest gradient with time; b) $\phi = 0$ corresponds to the same state of the pulsation regardless of period; c) it does not require contemporaneous photometry. For reference, $\phi = 0$ typically occurs slightly before ($\delta \phi \sim 0.1-0.2$) maximum light. The best-fit parameters are obtained by non-linear least squares. 

Cepheid RV curves differ in complexity, depending both on pulsation mode and period, and every Cepheid in \veloce\ has been observed a different number of times. The most appropriate number of harmonics, \nfs, of the Fourier series fit must therefore consider both RV curve morphology and $N_{\mathrm{obs}}$. We  adopted a sequential brute-force approach to determining the optimal model and \Ppuls\ simultaneously. First, we fitted each RV time series using $1 \le n \le min[20, (N_{\mathrm{obs}}-1)/2]$ Fourier harmonics. Since the best-fit \Ppuls\ can depend on \nfs, we performed phase dispersion minimization for every order ($N_{\rm{FS}}$) separately within typically $0.5\%$ of the literature pulsation period used as a starting value. Some short-period (overtone) Cepheids exhibit rather fast period fluctuations and thus required slightly larger flexibility for finding a best-fit \Ppuls. For all but a few Cepheids, this procedure very effectively minimized phase scatter. The epoch $E$ was then computed to match $\phi=0$ at minimum radius. Uncertainties on \Ppuls\ and $E$ were obtained by Monte Carlo simulations, and we assign systematic uncertainties on \Ppuls\ and $E$ as half the difference between the values of \Ppuls\ of the next higher and lower harmonics. Both uncertainties contribute roughly equally to the total uncertainty and are reported separately in the data files, whereas their squared sum is used in the text. 

\begin{figure}
    \centering
    \includegraphics[width=0.49\textwidth]{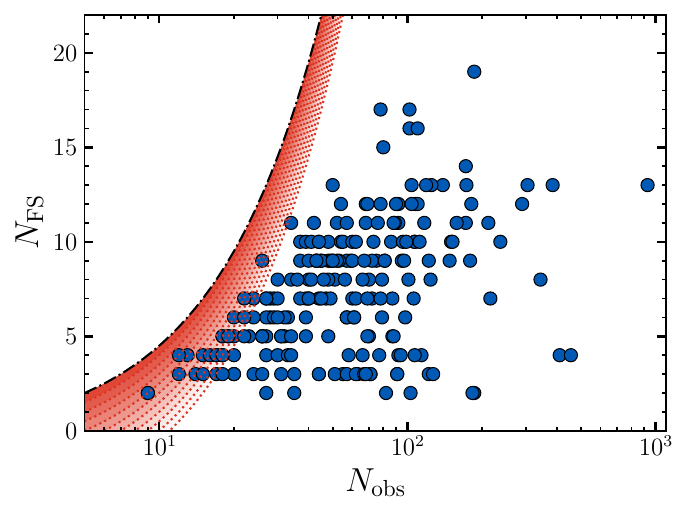}
    \caption{Number of Fourier series harmonics ($N_{\rm FS}$) used vs number of observations available ($N_{\rm obs}$). The black dash-dotted line shows the maximum number of harmonics that can be fitted for a given number of observations, $N_{\mathrm{obs}}=2 \cdot N_{\mathrm{FS}}+1$. The red dotted lines and shaded area show increasing numbers of degrees of freedom (up to $N_{\rm DOF} = 10$).}
    \label{fig:NFSvsDOF}
\end{figure}

We performed sequential model comparison using an F-test and the Bayesian Information Criterion (BIC) to select the optimal model. For the F-test, we adopt $p=0.05$ as the significance level for comparing models with $n$ versus $n+1$ harmonics. Analogously for the BIC, greater model complexity is considered acceptable if BIC decreases with increasing model complexity. We conservatively adopted as the optimal $n$ the number of harmonics above which either the F-test or the BIC criterion indicated spurious improvement. Figure\,\ref{fig:modelselect} illustrates this sequential model comparison. The final selected models were also inspected visually. Two possible exceptions to these rules were allowed: a) for seven Cepheids (AB~Cam, CH~Cas, KX~Cyg, RY~Sco, U~Sgr, WW~Mon, Z~Lac), we adopted lower $n$ after visual inspection; b) we implemented an automated rule that allowed us to skip ahead to the next higher harmonic ($n+1$) if $n<10$ and more than one of the following more complex models (e.g., $n+2$ and $n+3$) satisfied both statistical tests. Figure\,\ref{fig:NFSvsDOF} illustrates the resulting distribution of \nfs\ relative to the data available.

SB1 Cepheids exhibit temporal variations of the pulsation average velocity, so that  $v_\gamma = v_\gamma (t)$ in Eq.\,\ref{eq:Fseries}. These variations occur on timescales longer than \Ppuls\ and are discussed systematically in paper~II. Following visual inspection of Fourier series residuals, we identify Cepheids exhibiting orbital motion in the \veloce\ time series data in Tab.\,\ref{tab:Cepheidcatalog}. 

We caution that time-variable line shape distortions can lead to spurious trends in Fourier series residuals that could be mistaken for evidence of orbital motion depending on sampling \citep{Anderson2014rv,Anderson2016c2c}, and we considered this effect in setting the SB1 flag. Furthermore, some Cepheids exhibit large phase dispersion due to rapidly (possibly stochastically) varying \Ppuls, e.g., SZ Tau, and the long-period Cepheids SV~Vul and S~Vul. Period ``jitter'' reported in Cepheids \citep[e.g., V1154~Cyg, cf.][]{Derekas2017} introduces a scatter floor in the Fourier fit residuals. Expressed in root mean square, the minimum residual scatter can be as low as $\sim 60$\,\ms, although values around $100-120$\,\ms\ are more common and can reach up to several hundreds of \ms\  (Sect.\,\ref{sec:modulations}). In such cases, a simple mono-periodic Fourier series cannot provide an adequate fit to the data due to the additional signals.

To determine the pulsation periods of Cepheids exhibiting time-variable $v_\gamma(t)$ (binaries or not), we represented the variable pulsation average velocity in Eq.\,\ref{eq:Fseries} by a polynomial with coefficients $c_i$:
\begin{equation}
    v_\gamma = v_{\gamma}(t) =  v_{\gamma, E} + \sum_{i=1,...}{ c_i \cdot (t - E)^i } \ .
    \label{eq:polynomial}
\end{equation}
The polynomial allows us to represent modulated variability of any origin (aside from variable \Ppuls) while determining \Ppuls\ and $E$, and higher degree $i$ implies more complex and/or shorter-timescale modulations. This was particularly useful for dealing with high-amplitude orbital motion, especially when the orbital signal was incompletely sampled. Given the long baselines of \veloce, all polynomials trace timescales much longer than \Ppuls. The degree of any polynomials plotted are listed alongside the Fourier series coefficients in Tab.\,\ref{app:table:RVglobal}. We caution that the constant term, $v_{\gamma,E}$, is defined at the epoch $E$ when such polynomials are used, and thus, it should not be used to represent the center-of-mass velocity of the star. Further discussion of the polynomial parameters and of SB1 Cepheids is presented in paper~II.

\subsection{Fourier amplitude ratios and phase differences\label{sec:FourierRatios}}
Fourier amplitude ratios and differences were computed using the best-fit Fourier series coefficients, $a_i$ and $b_i$ (cf. Eq.\,\ref{eq:Fseries}), in order to easily visualize the Hertzsprung progression in Sect.\,\ref{sec:Hertzsprung} \citep{Simon+1981}.
The amplitude and phase of the $i$-th harmonic are defined as $A_i = \sqrt{a_i^2 +
b_i^2}$ and $\tan{\phi_i} = -b_i/a_i$. Amplitude ratios among
harmonics are defined as $R_{i1} = A_i / A_1$, and phase differences as $\phi_{i1} =
\phi_i - i\cdot\phi_1$.
Uncertainties on $A_i$ and $\phi_i$ are determined using the covariance matrix of the fit and propagated to compute uncertainties for $R_{i1}$ and $\phi_{i1}$.

\subsection{Stars exempted from detailed Fourier modeling\label{sec:unmodeled}\label{sec:adequacy}}
This data release contains \nobscepunmodeled\ observations of \ncepunmodeled\ Cepheids whose variability could not be satisfactorily modeled using a mono-periodic Fourier series due to insufficient sampling, multi-periodicity, or additional signals. We here publish the time series RV measurements for these stars without detailed Fourier modeling and include them separately in  Tab.\,\ref{tab:Cepheidcatalog}. 

An exception is the overtone Cepheid SU~Cyg \citep[paper~II]{Imbert1984}, whose high-amplitude and short-period orbital motion prevented an adequate fit using the polynomial Fourier series model using only \veloce\ data. 
Similarly, insufficient phase coverage and few measurements prevented a good Fourier series fit for ER~Aur and SU~Cas. 

The RV time series of VX~Cyg featured a noticeable gap of about $0.2$ in phase along the slowly rising RV branch, which makes up about $0.7$ in phase. During this phase, stars of this period exhibit linearly increasing RV. To obtain a clean Fourier series fit, we therefore linearly interpolated between the observed points along the rising branch and added the interpolated points with conservative $0.2\,$\kms\ uncertainties. The interpolated points are included in the data set for VX~Cyg and clearly marked in column `SOURCE' of the data tables, cf. Tab.\,\ref{tab:data_structure_table}.

\subsubsection{Cepheids with insufficient $N_{\rm obs}$ for Fourier modeling\label{sec:insufficient}}
\ncepinsufficientobs\ Cepheids featured an insufficient number of observations to adequately reproduce their variability curves: AP~Cas, ASAS~J074902$-$1906.8, ASAS~J075345$-$3658.2, ASAS~J082710$-$3825.9, ASAS~J084304$-$5117.9, ASAS~J094827$-$5801.1, ASAS~J103052$-$5903.7, ASAS~J115701$-$6218.7, ASAS~J122511$-$6120.9, ASAS~J181215$-$2029.1, ASAS~J183347$-$0448.6, ASAS~J183652$-$0907.1, ASAS~J191351+0251.3, ASAS~J192310+1351.4, BR~Vul, CG~Cas, DW~Cas, EV~Aql, GM~Cas, GSC~03996$-$00312, TV~CMa, U~Aql, V0335~Aur, V0458~Sct, V0493~Aql, V0600~Aql, V1954~Sgr, and X~Sct. While no classification was assigned for ASAS~J082710$-$3825.9 in \gaia\ DR3, we find an RV difference of $\sim 16$\,\kms\ among two measurements separated by $3$ nights (\Ppuls$=9.3$\,d), which is easily compatible with the typical peak-to-peak amplitudes of Cepheids around this period. Moreover, the star's CCFs exhibit the typical shape variation expected from a classical Cepheid.

\subsubsection{Cepheids exhibiting non-stationary variability\label{sec:nonstationary}}
Two peculiar amplitude-modulating Cepheids could not satisfactorily be modeled using a stationary Fourier series: the well-known Blazhko Cepheid V0473~Lyr \citep{Burki+1980,Molnar+2014} and the spectroscopic binary Cepheid ASAS~J103158$-$5814.7 whose Blazhko modulations and orbital motion are reported here for the first time. RS~Pup's strong cycle-to-cycle variations and period fluctuations similarly prevented an adequate fit. Additional information on such phenomena is presented in Sect.\,\ref{sec:modulations}.

\subsubsection{Double-mode (beat) Cepheids \label{sec:beatceps}}
Five double-mode (beat) Cepheids were observed: MS~Mus, TU~Cas, V1048~Cen, V1210~Cen, and Y~Car. The RV time series collected are published as part of this data release. However, they were not modeled due to the added complexity of determining two pulsation periods based on RV data and the generally short pulsation timescales of double-mode Cepheids.

\subsubsection{RR Lyrae stars and type-II Cepheids\label{sec:popII}}
\veloce\ targets classical Cepheid variables. However, \nrrlyraestr\ RR~Lyrae stars and \ntypetwocep\ type-II Cepheids were observed as backup targets at times when an insufficient number of Cepheids were visible during a particular observing run, or, in the case of the prototype RR~Lyrae, to obtain a base for comparison with classical Cepheid variability of CCFs. We did not model these stars because the number of observations available for these stars was generally lower and because type-II Cepheid RV curves exhibit more scatter than those of classical Cepheids. Further information is presented in App.\,\ref{sec:app:classification}.

Table\,\ref{tab:noncep} lists these stars and the number of RV measurements presented here alongside additional information. The RV time series for these stars are included without detailed Fourier modeling and were derived using the same procedure as the one applied above for classical Cepheids. However, we note that the G2 mask is not applicable to the hottest phases of RR~Lyrae stars, resulting in no correlation peaks (hence, no RVs) along the fast-changing descending RV branch (cf. Fig.\,\ref{fig:CCFs}). Type-II Cepheids, such as W~Vir, exhibit blue-shifted emission features indicative of shock that may further influence RV measurements (cf. Fig.\,\ref{fig:CCFs}).

\section{New insights from precision velocities}\label{sec:insights}
The combination of a large sample of Cepheids observed with unprecedented precision over a decade-long  baseline provides interesting new insights into the astrophysics of Cepheids. In the following, we highlight a) the Hertzsprung progression observed from RV data in unprecedented detail (Sect.\,\ref{sec:Hertzsprung}), b) the discovery of a secondary bump in Cepheid RV curves that also follows the Hertzsprung progression (Sect.\,\ref{sec:doublebump}), c) the ubiquity of modulated variability features, such as long-term modulations and cycle-to-cycle variations (Sect.\,\ref{sec:modulations}), and d) a puzzling dichotomy among the linear radius variations exhibited by Cepheids (Sect.\,\ref{sec:deltaR}).

\subsection{The Hertzsprung progression\label{sec:Hertzsprung}}

\begin{figure*}
    \centering
    \vspace{-3mm}
    \includegraphics[width=0.89\textwidth]{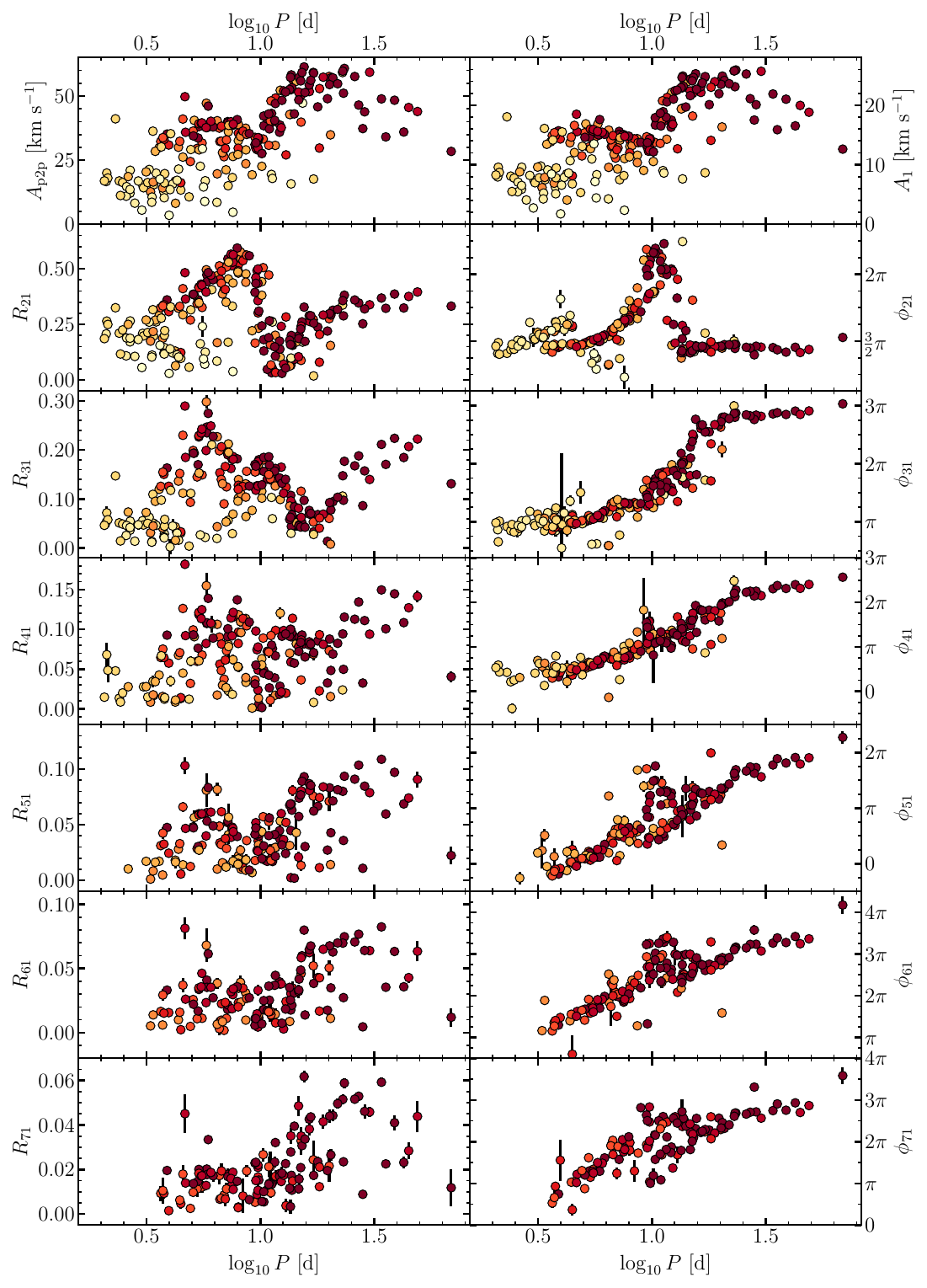}
    \caption{Fourier parameters for \ncepFS\ Cepheids in \veloce\ as a function of $\log{P}$ up to the $7th$ harmonic. Some phase differences were shifted by $\pm 2\pi$ for clarity. The plotted errorbars are usually smaller than the symbols. The number of harmonics fitted is color coded from light yellow ($N_{\rm FS}=2$) to dark red ($N_{\rm FS}=10$ is darkest shade for clarity). Overtone Cepheids are clearly apparent at low amplitudes and short periods. Excess dispersion in $\phi_{51}$ and $\phi_{61}$ near $1.0 \lesssim \log{P} \lesssim 1.2$ is a consequence of the well-known resonance between the fundamental mode and the second overtone \citep{Simon+1981,Buchler1990,Antonello+1996}. These results are further illustrated in App.\,\ref{app:FourierParameters}.}
    \label{fig:FourierParams}
\end{figure*}

The Hertzsprung progression \citep[HP]{HertzsprungProgression} refers to an apparently continuous change in the light curve shapes of Cepheids pulsating in the fundamental mode as a function of their pulsation period. It is one of the most noticeable features of Cepheids light and RV curves and provides insights into the physics of stellar pulsations. Additionally, the HP can be used to illustrate the similarity of extragalactic and nearby Cepheids to underline their physical similarity \citep{Riess2022}. Although \citet{HertzsprungProgression} already considered Fourier components in relation to the HP, the visualization of the HP using ratios of Fourier parameters has been common since \citet{Simon+1981}. 

Cepheid RV curves are known to also exhibit the HP, although the sampling in period had been rather limited until recently. Prior to \gaia\ DR3, the most detailed investigations of the RV HP were presented by \citet[based on 57~Cepheids of mostly short periods]{1990ApJ...351..606K}, \citet[using phase shifts and asymmetries rather than Fourier decomposition]{Gorynya1998Hertzsprung}, and \citet[including additional long-periods]{Anderson2016ApJS}, who remarked a group of (then: four) long-period Cepheids with particularly low RV amplitudes further discussed in Sect.\,\ref{sec:deltaR}.

The \veloce\ HP is illustrated in Fig.\,\ref{fig:FourierParams}, which shows the peak-to-peak RV amplitude, $A_{\mathrm{p2p}}$, the amplitude of the first harmonic, $A_1$, as well as the ratio of the $i$th harmonic's amplitude to the first harmonic's amplitude ($R_{i1}$) and analogous for the Fourier phase differences ($\phi_{i1}$) of several higher-order harmonics ($2 \le i \le 7$) to the first harmonic. Features known from photometric illustrations of the HP are accurately recovered, notably including the ``dip'' in $A_{\mathrm{p2p}}$ near 10 days \citep{Klagyivik+2009}, the sharp divide between overtone and fundamental mode pulsators in $R_{21}$, the flattening of $\phi_{21}$ at longer \Ppuls. Since the number of harmonics can differ among \veloce\ stars, the number of objects decreases for higher harmonics (cf. Eq.\,\ref{eq:Fseries}).

The color coding according to \nfs\ in Fig.\,\ref{fig:FourierParams} shows that stars that have been modeled using higher \nfs\ exhibit extremely clean trends, in particular among the phase ratios. Outliers tend to have a smaller number of harmonics, indicating that there were too few observations available to capture the full complexity of the RV curve. A particularly interesting case is that of the 17-day Cepheid Y~Oph, which was modeled with a very low number of harmonics (\nfs$=4$) despite a large number of available observations, and which exhibits Fourier parameters that appear as a long-period extension from overtone Cepheids, especially among $A_1$ and $R_{21}$. This is curious since the longest-period overtone MW Cepheid has a period of only $8.8$\,d \citep{Baranowski2009}.

\onecolumn
\noindent%
\begin{minipage}{\textwidth}
\makebox[\textwidth]{
    \includegraphics[width=\textwidth]{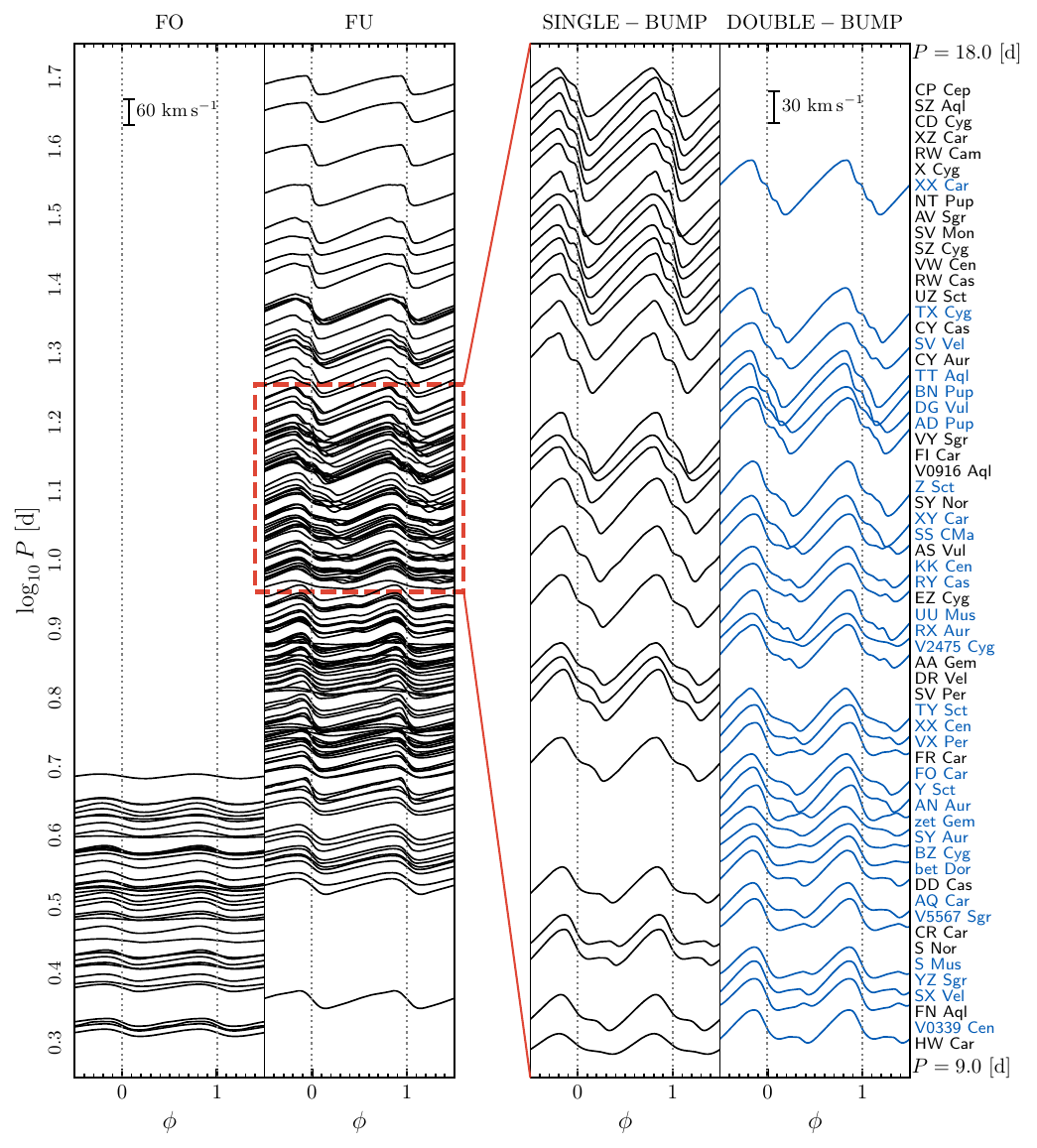}}
    \captionof{figure}{The Hertzsprung progression illustrated by the model RV curves of 208 fitted Cepheids (11 low-amplitude stars were removed for clarity). Cepheids are offset in the vertical direction according to their $\log{P_{\rm puls}}$. Errorbars near the top show the constant velocity scale. Vertical dotted lines indicate $\phi = 0$, the phase of minimum radius. {\it Left panel, left column}: Cepheids classified as first overtone pulsators in \gaia\ DR3. {\it Left panel, right column}: Cepheids classified as fundamental mode pulsators in \gaia\ DR3. Outliers with particularly small amplitudes are not shown for clarity (cf. Sect.\,\ref{sec:deltaR}). The shortest-period FU Cepheids is BP~Cir. {\it Right panels:} Close up view in the period range $9 - 18$\,days. Stars with double-peaked bump features are shown in blue on the right, stars with single-peaked bumps on the left in black. Names are color coded accordingly. Outliers Z~Lac, CH~Cas, and Y~Oph are not shown for clarity. The double-peaked bump appears in $31/61 \approx 50\%$ of \veloce\ Cepheids and may be a ubiquitous feature of Cepheid RV curves that requires very high precision and extremely dense phase sampling for detection.}
    \label{fig:RVHPfull}
\end{minipage}
\twocolumn

Further interesting features in Fig.\,\ref{fig:FourierParams} are the increased dispersion among higher-order amplitude ratios at  \logP $\gtrsim 1.15$, as well as the increased dispersion in the higher-order phase ratios that is restricted to the period range around $0.95 \lesssim $ \logP $ \lesssim 1.25$ ($9 - 18$\,d). Additional visualization of these results is provided in App.\,\ref{app:FourierParameters}.

Figure\,\ref{fig:RVHPfull} illustrates the HP using all fitted Fourier series models. The vertical axes show RV, with each star being offset according to \logP\ as marked on the right. The left panel shows the full period range, with the left column being populated by low-amplitude Cepheids, most of which pulsate in the first overtone, as well as Y~Oph and YZ~Car, two long-period Cepheids featuring atypically low RV amplitudes. The right column of the left panel shows fundamental mode Cepheids and very clearly exhibits the ``bump'' feature that appears near phase $0.75$ for a Cepheid with \logP $\approx 0.75$ and moves to lower phase with increasing period until it disappears at \logP $\gtrsim 1.55$. The bump is most noticeable when it appears on the descending part of the RV curve, that is, near minimum radius, and it can very significantly stretch the descending branch and render it much less steep. 
While the RV HP is very clearly observed, we note that some Cepheids clearly do not exhibit RV curves representative of other Cepheids at similar \logP. 

The right panel of Fig.\,\ref{fig:RVHPfull} provides a close-up view of Cepheids in the period range where this is most noticeable, around $9 - 18$\,d, and we note that this period range is also where the higher-order Fourier phase ratios exhibited excess scatter.

\subsection{A second bump in the Hertzsprung progression\label{sec:doublebump}}
\begin{figure}[!t]
    \centering
    \includegraphics[width=0.49\textwidth]{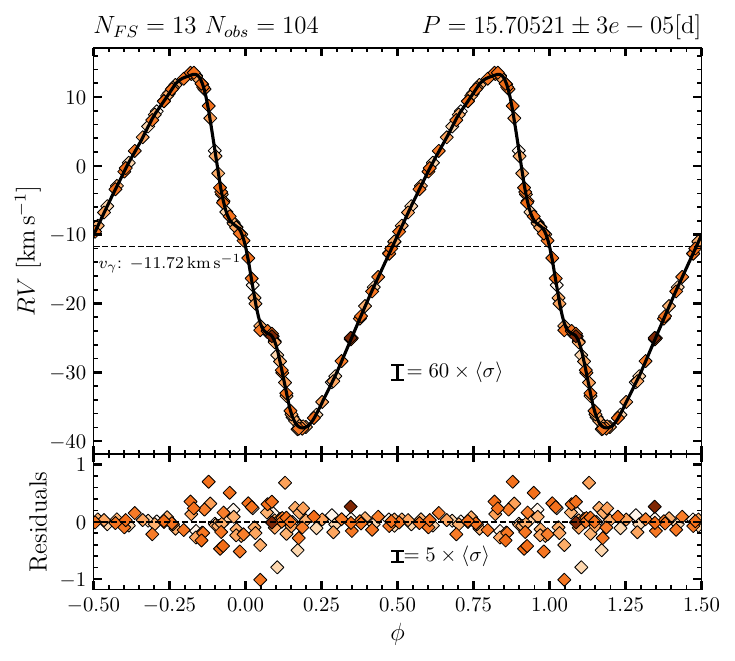}
    \caption{\veloce\ data and fit to XX~Car, which exhibits a double-bumped descending RV branch. Only \coraliev{14}\ observations were used in this Fourier series fit employing 13 harmonics. The two peaks are clearly sampled. The data quality is sufficient to show the shortcoming of the Fourier series fit in reproducing the two bumps whose slopes are slightly less steep than implied by the fit to the overall RV curve.}
    \label{fig:XXCar}
\end{figure}

Figure\,\ref{fig:RVHPfull} (right panel, right column) shows that 31 of the 61 (50\%) Cepheids in the HP period range exhibit an additional bump feature, whereas 30 Cepheids exhibit the expected single bump on the descending branch. Figure\,\ref{fig:XXCar} shows a close-up view of XX~Car to illustrate that the double bump is not caused by ringing or poor sampling. This double bump feature has not previously been reported in the literature\footnote{We previously reported this finding in September 2022 and April 2023 at the RR~Lyrae and Cepheids conference in La Palma, Spain and at IAU Symposium 376 in Budapest, Hungary.} in Cepheid light or RV curves and could be identified here thanks to the precision of \veloce\ data. One of the targets that exhibit this feature (albeit not very strongly) is the 3rd magnitude star $\beta$~Dor, which fortuitously lies within the {\it TESS} Southern Continuous Viewing Zone and therefore benefits from a particularly long-baseline high-precision photometry. Inspection of the light curve presented in \citet{Plachy2021} reveals a clear photometric double bump, although this feature was not recognized as such at the time, likely due to the lack of comparison stars exhibiting similar features. Comparison with {\it TESS} light curves for additional Cepheids will be particularly useful to shed further light on these potentially useful features of stellar pulsations. Given that the well-known bump is explained by a resonance among the fundamental mode and second overtone \citep{1990ApJ...351..606K,Antonello+1996}, it would seem likely that the double bump feature is also caused by overtone resonances, which may provide important new insights into the structure and evolution of Cepheids. 

\begin{figure}
\centering
    \includegraphics[width=0.45\textwidth]{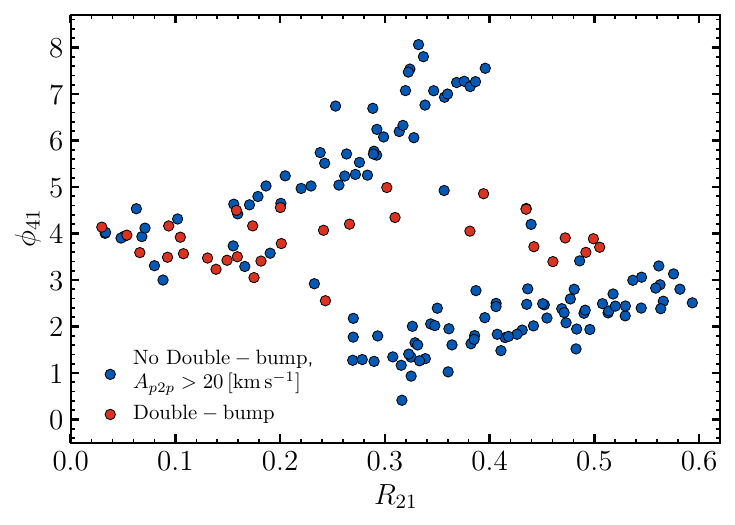}
    \includegraphics[width=0.45\textwidth]{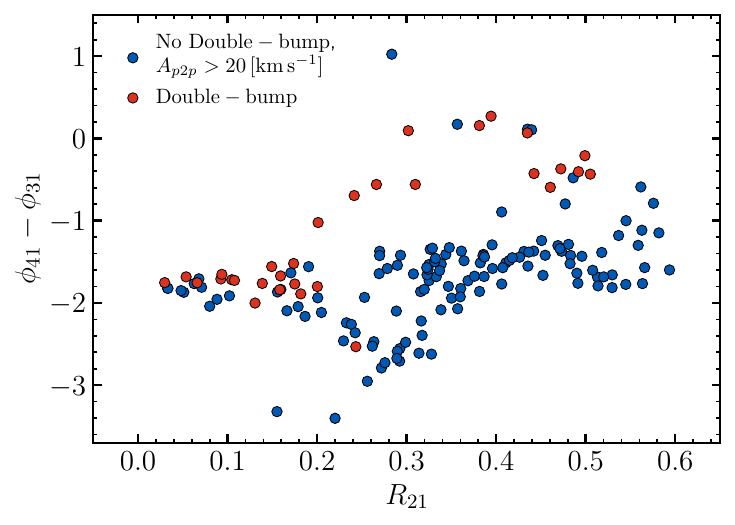}
    \caption{Cepheids exhibiting double bumps on the declining RV curve part stand out in the $\phi_{41}$ vs $R_{21}$ diagram as a ``connecting band'' between two otherwise parallel sequences. In $\phi_{41} - \phi_{31}$ vs $R_{21}$, double bump Cepheids also follow a significantly different trend. Plotting these parameters could serve to identify double bumps Cepheids in a quantitative manner rather than by visual inspection of the fits. \label{fig:doublebumps_phi41_R21}}
\end{figure}

As Fig.\,\ref{fig:doublebumps_phi41_R21} shows, double bump Cepheids have nearly constant $\phi_{41}$ across a broad range of $R_{21}$ and do not appear to follow the bulk of other Cepheids in this diagram. Additionally, double bump Cepheids follow a different trend in the $\phi_{41}-\phi_{31}$ vs $R_{21}$ diagram.

\subsection{Long-term modulations and cycle-to-cycle variations\label{sec:modulations}\label{sec:cat:modulators}}

Cepheid RV curves have been shown to exhibit signals that cannot be explained by the standard Fourier series fitting approach with a fixed period\footnote{Variable periods are discussed in Sect.\,\ref{sec:pdot} below. Suffice it here to mention that period fluctuations contribute to the rms of the RV curve, in particular near the steepest parts of the RV curve, since our baseline Fourier series model assumes a fixed \Ppuls. However, this applies primarily to fast, non-linear period changes, since secular (linear) period changes are generally too slow (mostly $<10$s/yr) to significantly change $P$ over the \veloce\ baseline.}. In an early result based on \veloce\ data, \citet{Anderson2014rv} showed the existence of at least two different categories: long-term amplitude modulations of short-period (likely overtone) Cepheids, such as V0335~Pup and QZ~Nor, and cycle-to-cycle fluctuations affecting both the RV curve shape and periods of long-period Cepheids such as $\ell$~Car and RS~Pup \citep[cf. also][]{Anderson2018rvs}. A growing body of literature has since pointed out the generally less stable behavior of overtone Cepheids, as well as the existence of additional pulsation modes, many of which are hiding underneath unstable main modes of pulsations \citep[e.g.,][]{Derekas2012,Derekas2017,Evans2015most,Poretti2015,Soszynski2015nonstandard,Smolec2016nrpSMC,Smolec2017unstable,Sueveges2018a,Sueveges2018b,Rathour2021,Csornyei2022,smolec2023}. Cycle-to-cycle variations in long-period Cepheids probe the interplay of convection with pulsations \citep{Anderson2016c2c,Anderson2016vlti} and appear to be mostly stochastic. By contrast, long-timescale modulations of overtone Cepheids may be repeating \citep{Anderson2018rvs}, or even periodic, although the physical origin is less clear. Additionally, V0473~Lyr \citep{Burki+1980,Molnar2017} exhibits large periodic amplitude modulations reminiscent of the Blazhko-effect \citep{Blazhko1907} seen in RR~Lyrae stars, whereas Polaris exhibits multi-periodic line profile variations \citep{Hatzes2000,Anderson2019polaris,Torres2023}. \citet{Moskalik2009} further reported Blazhko-like modulation in LMC double-mode Cepheids.

\begin{figure}[t]
    \centering
    \includegraphics[width=0.5\textwidth]{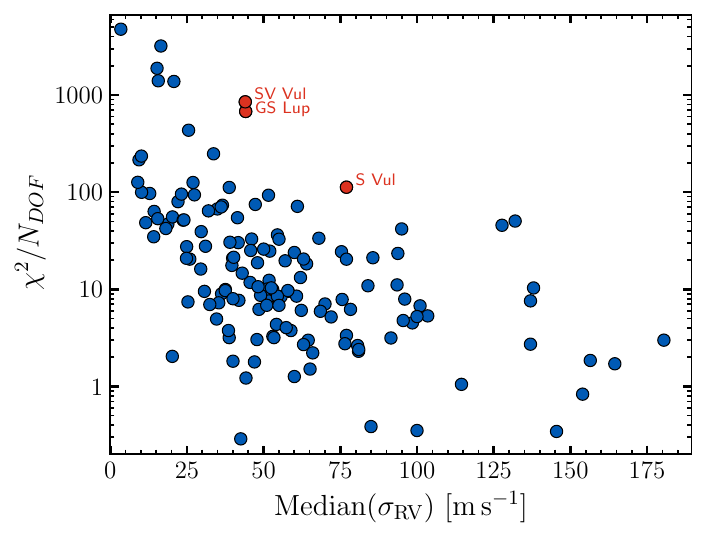}
    \caption{Normalized $\chi^2 / N_{DOF}$ of \veloce\ Fourier series fits versus the median uncertainty of the RV measurements for 151 non-SB1 Cepheids. The rapid increase of $\chi^2 / N_{DOF}$ for precise measurements with typical uncertainties better than $20-40$\,\ms\ shows the existence of additional signals that are not reproduced by Fourier series models with constant \Ppuls. The three longest-period Cepheids in the sample, GS Lup, SV Vul, and S Vul are labeled and highlighted.}
    \label{fig:modulation_error}
\end{figure}

\veloce\ data provide a powerful complement to space-based photometry for studying modulated variability thanks to high zero-point stability. While space-based photometry benefits from high precision and the ability to collect time series uninterrupted by the diurnal cycle, temporal baselines longer than a few months to a year are difficult to achieve from space for a large number of Cepheids. Unfortunately, only a single Cepheid (V1154~Cyg) was observed by {\it Kepler} \citep{Derekas2017}, and the temporal baseline of {\it TESS} \citep{TESS} is rather short outside of the continuous viewing zones. Nevertheless, {\it TESS} observations of Cepheids \citep{Plachy2021} probe up to several pulsation cycles extremely densely and without interruption. A systematic comparison between the variability signals observed by {\it TESS} and \veloce\ is therefore of interest.

For bright stars such as $\ell$~Car, the pulsational RV amplitude of order $35$\,\kms\ is $7000 - 10000$ times larger than the typical $3-5\,$\ms\ RV uncertainty achievable using 
\coralie. For comparison, a typical $V-$band amplitude of a similar Cepheid would be of order $1\,$mag, requiring a stable photometric uncertainty of $0.14$\,mmag over the course of two months to trace two full pulsation cycles in similar detail. Even considerably poorer RV precision of $50-70$\,\ms\ still corresponds to an S/N of $500-700$. Hence, \veloce\ is highly sensitive to interesting additional signals that may help to further improve the understanding of these stars.

We find that a typical RV uncertainty of $40-50$\,\ms\ will provide basic sensitivity to such additional signals, although $10-20$\,\ms\ will render them much more obvious. Figure\,\ref{fig:modulation_error} illustrates this by plotting the reduced $\chi^2/N_{\rm{DOF}}$ of the model against the median RV uncertainty, where $N_{\rm DOF}$ is the number of degrees of freedom of the fit. Most Cepheids with typical uncertainties better than $20$\,\ms\ yield $\chi^2/N_{\rm{DOF}}$ of several tens to hundreds or even thousands. Of course, inadequate temporal sampling or model complexity will contribute to higher $\chi^2$ and contributes to the scatter seen in Fig.\,\ref{fig:modulation_error}. However, the clear trend to higher reduced $\chi^2$ for better measurements is not explained by missing data and instead confirms the result by \citet{Sueveges2018b} based on OGLE photometry that modulated variability becomes a ubiquitous feature among LMC Cepheids given sufficiently precise measurements. 
Overall, we find 19 Cepheids with $\chi^2/N_{\rm{DOF}} > 100$, of which $16$ have a median error $< 40$\,\ms. 
Interestingly, some of the longest-period Cepheids in the sample, GS~Lup, SV~Vul, and S~Vul also fall into this group despite median errors of 44, 44, and 77\,\ms, respectively. This indicates that very long-period Cepheids are particularly unstable and exhibit ubiquitous cycle-to-cycle variations similar to the ones reported in $\ell~$Car and RS~Pup \citep{Anderson2014rv,Anderson2016c2c,Anderson2016vlti}.

Table\,\ref{tab:modulators} lists Cepheids exhibiting (clear or tentative) modulated variability phenomena (e.g., not orbital motion) grouped according to different types of phenomena named ad hoc according to the first object where this type of variability was seen during the course of the collection of the \veloce\ data set. The sheer number of stars exhibiting these phenomena demonstrates the power of high-precision RVs to provide significant new information in the study of Cepheids, while the diversity of phenomena implies that more than one physical effect is at play. We will specifically study this ``Modulation Zoo'' in future work, the first of which will present the  detection of non-radial modes using spectroscopic observations \citep{Netzel2024}. 

\newcolumntype{L}[1]{>{\raggedright\let\newline\\\arraybackslash\hspace{0pt}}m{#1}}
\begin{table}[]
\caption{Cepheids exhibiting modulated variability\label{tab:modulators}}
\begin{tiny}
    \begin{tabular}{@{}l|L{2.1in}@{}}
    \toprule
         Name & notes  \\
         \midrule
         \multicolumn{2}{c}{long-term RV amplitude modulation (V0335~Pup)} \\         \midrule
         ASAS~J084412$-$3528.4 & candidate  \\
         ASAS~J184741$-$0654.4 & $-$  \\
         GH~Car & candidate  \\
         MY~Pup & also SB1 (paper~II) \\
         V0335~Pup & long-term small-scale amplitude variations \citep{Anderson2014rv} \\
         V0378~Cen & candidate  \\
         V0440~Per & longest-period overtone Cepheid \citep{Baranowski2009}  \\
         \midrule
         \multicolumn{2}{c}{Blazhko-like RV amplitude modulation (V0473~Lyrae)} \\ 
          \midrule
         V0473~Lyr & large-scale amplitude variations \citep{Burki+1980,Burki1982,Molnar+2014,Molnar2017} \\
        ASAS~J103158$-$5814.7 & large amplitude modulation and SB1 (cf. paper~II) \\
         \midrule
         \multicolumn{2}{c}{long-term RV shape change (QZ~Nor)} \\         \midrule
         ASAS~J091606-5418.6 & candidate \\
         QZ~Nor & see \citet{Anderson2014rv,Anderson2018rvs} \\
         EU~Tau & candidate \\
                  \midrule
         \multicolumn{2}{c}{RV shape change, unstable \Ppuls\ (SZ~Tau)} \\         \midrule
         ASAS~J064553+1003.8 & also SB1 \\
         ASAS~J101016$-$5811.2 & also SB1 candidate \\
         R~Cru & period fluctuations, cf. Fig.\,\ref{fig:RCru}, shortest-period SB1 in MW (paper~II) \\
         SZ~Tau &  RV curve shape changes and significant period variations \\
         \midrule
         \multicolumn{2}{c}{cycle-to-cycle variations ($\ell$~Car)} \\         \midrule
         ASAS~J180342-2211.0 & candidate \\
         $\ell$~Car & cycle-to-cycle variations of RV curve shape with minor period fluctuations \citep{Anderson2016c2c} \\
         GS~Lup & $-$ \\
         KN~Cen & also SB1 (paper~II) \\
         KQ~Sco & clearest in CCF FWHM, SB1 candidate \\
         RY~Vel & $-$ \\
         SZ~Aql & candidate  \\
         SW~Vel & $-$ \\
         U~Car & $-$ \\
         V0916~Aql &  candidate  \\
         Y~Oph &  also SB1 candidate \\
         \midrule     
         \multicolumn{2}{c}{cycle-to-cycle variations, unstable \Ppuls\ (RS~Pup) } \\         \midrule
         RS~Pup & cycle-to-cycle variations of RV curve shape with long-term period fluctuations \citep{Anderson2014rv,Kervella2017rspup} \\
         RY~Sco & $> 1$\,\kms\ scatter near barely visible double bump feature \\
         S~Vul & very significant period variations, SB1 candidate \\
         SV~Vul & $-$ \\
         \midrule
         \multicolumn{2}{c}{multi-periodic line shape variations (Polaris)} \\         \midrule
         $\alpha$~UMi (Polaris) & multi-periodic non-radial pulsations \citep{Hatzes2000,Anderson2019polaris}, SB1 \\
         SZ~Cas & also c2c candidate, unstable \Ppuls \\
         \midrule
         \multicolumn{2}{c}{resolved line splitting (X~Sgr)} \\         \midrule
         BG~Cru & line splitting \citep{AndersonPhD,Usenko2014} \\
         LR~TrA & poorly resolved at $R \approx 55\,000$, SB1 \\
         X~Sgr & line splitting with cycle-to-cycle modulations \citep{Mathias2006,AndersonPhD,Kovtyukh2003} \\
    \bottomrule
    \end{tabular}
    \tablefoot{This list is incomplete; further work is ongoing to investigate this ``Modulation Zoo'' in detail. Stars are grouped by similar phenomenology.}
\end{tiny}
\end{table}

\begin{figure}
    \centering
    \includegraphics[width=0.49\textwidth]{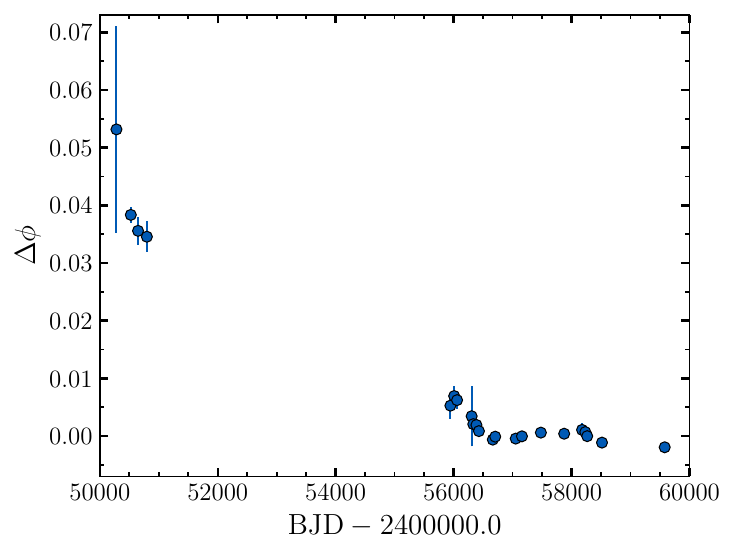}
    \caption{Phase shifts versus observation date for R~Cru determined by template fitting. The long-term change in $\Delta \phi$ is the signature of secular period change (cf. Sect.\,\ref{sec:pdot}), whereas an oscillatory pattern among the \veloce\ data is also clearly apparent. R~Cru is the Cepheid with the shortest known orbital period in the MW, $P_{\rm orb}=237.6$\,d (cf. paper~II).}
    \label{fig:RCru}
\end{figure}

A few cases merit specific mention. MY~Pup exhibits V0335~Pup-like amplitude modulations and is also an SB1 with a period of $\sim 4$\,yr (cf. paper~II). R~Cru exhibits a very noisy RV curve with apparent phase modulation over a long timescale (cf. Fig.\,\ref{fig:RCru}) in addition to being the shortest-period SB1 Cepheid in the MW with a period of $238$\,d (cf. paper~II). The presence of four superposed signals, including high-amplitude pulsations, low-amplitude orbital motion, period fluctuations, and potential further RV curve modulations renders R~Cru's RV curve particularly difficult to fit and showcases the complexity of Cepheid RV curves. The long-period Cepheid KN~Cen exhibits long-period orbital motion that appears as a long-term trend in \vgamma. At the same time, cycle-to-cycle variations similar to those seen in $\ell$~Car are clearly apparent and well-sampled in subsequent pulsation cycles. 
ASAS~J103158$-$5814.7 ($P=1.1192$\,d), one of the shortest-period Cepheids in \veloce, exhibits significant Blazhko-like modulations of the pulsational RV signal similar to V0473~Lyrae (HR~7308) \citep{Burki+1980,Molnar2017} in addition to high-amplitude orbital motion. Observations are ongoing to determine the orbit and Blazhko modulation timescale. 

\subsection{Linear radius variations\label{sec:deltaR}}

The linear radius variation, $\Delta R = p \int{v_r d\phi}$, is calculated by integration of the RV curve over the pulsational cycle and by deprojecting observed line-of-sight velocities to the speed of radial displacement as seen from the star's center using the projection factor $p$. We specify $\Delta R$ in solar radii assuming $R_\odot=695\,700.0$\,km as used in \texttt{astropy.constants} (version 5.3.4). $\Delta R$ is crucial to BW-type methods of distance determination (henceforth: BW distances), since the ratio of radial to angular diameter variation is proportional to distance. As pointed out by \citet{Anderson2014rv}, RV curve modulation (cf. Sect.\,\ref{sec:modulations}) complicates the calibration of the projection factor, $p$, and the use of BW distances for Leavitt law calibration. Additionally, $p$ is a very complex quantity that serves as a placeholder for several effects, including geometry, limb darkening, and atmospheric velocity gradients \citep[e.g.,][]{Nardetto2007,Nardetto2017deltaCepPfac}. Nevertheless, there is a strong interest in calibrating $p$ for the distance scale, notably using interferometry and \gaia\ parallaxes \citep{Breitfelder2016,Trahin2021}. However, for the time being, no clear picture has arisen as to whether $p$ should be considered a constant, or, e.g., dependent on \Ppuls.

\begin{figure*}
    \centering
    \includegraphics[width=0.49\textwidth]{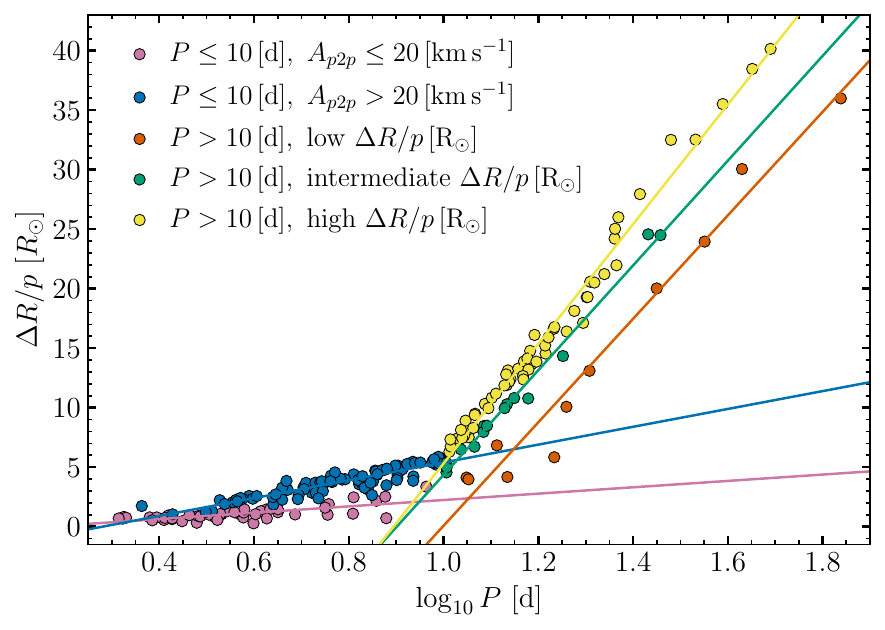}
    \includegraphics[width=0.49\textwidth]{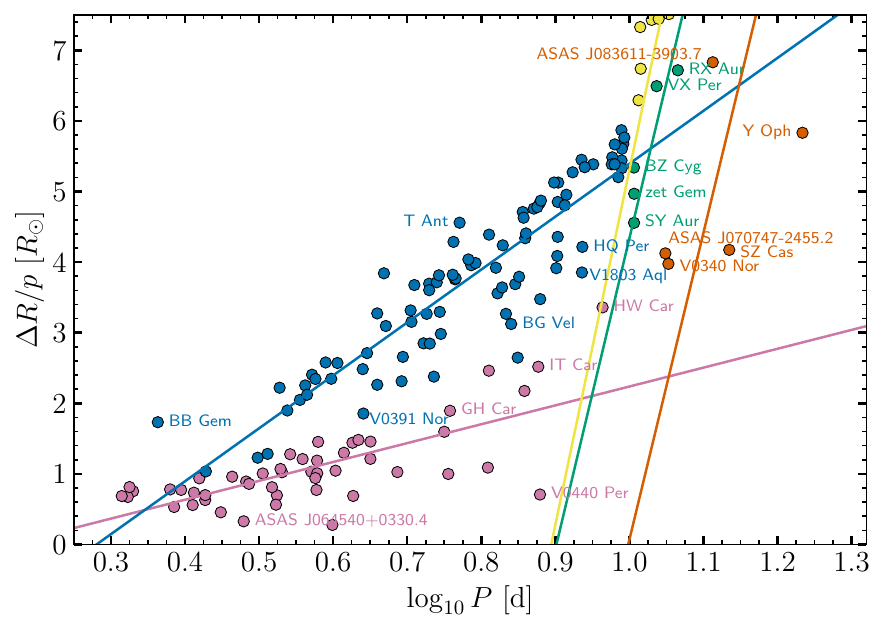}
    \caption{RV curve integrals ($= \Delta R / p$) vs \logP\ determined from the \veloce\ best fit models. Short-period Cepheids (\Ppuls $ < 10$\,d) are separated into low- and high-amplitude stars using a peak-to-peak RV threshold of $20$\,\kms\ and color coded in purple ($\le 20$\,\kms) and blue ($>20$\,\kms). Long-period ($P>10$d) FU Cepheids exhibit a strikingly ($\sim 5.6-6.6\times$) steeper trend than short-period Cepheids. We introduce an ad hoc distinction among three sequences for $P>10$d FU Cepheids into stars exhibiting large (yellow), intermediate (green), and small (orange) radius variations. The right panel shows a close-up view to more easily distinguish among short-period Cepheids, where a large number of Cepheids are found between the well-defined envelopes made up of FU and FO Cepheids, respectively. Stars near the edges are labeled to facilitate identification.}
    \label{fig:DeltaR_logP}
\end{figure*}

Here, we address the issue of determining $\Delta R$ solely from the observed RV curve by computing $\Delta R/p = \int{v_r d\phi}$ from the homogeneously measured \veloce\ data. This allows us to test whether $p$ should be constant or a function of \Ppuls\ without considering angular diameter variations, purely based on the consistency of $\Delta R/p$ at different values of \logP.

\begin{figure}
    \centering
    \includegraphics[width=0.49\textwidth]{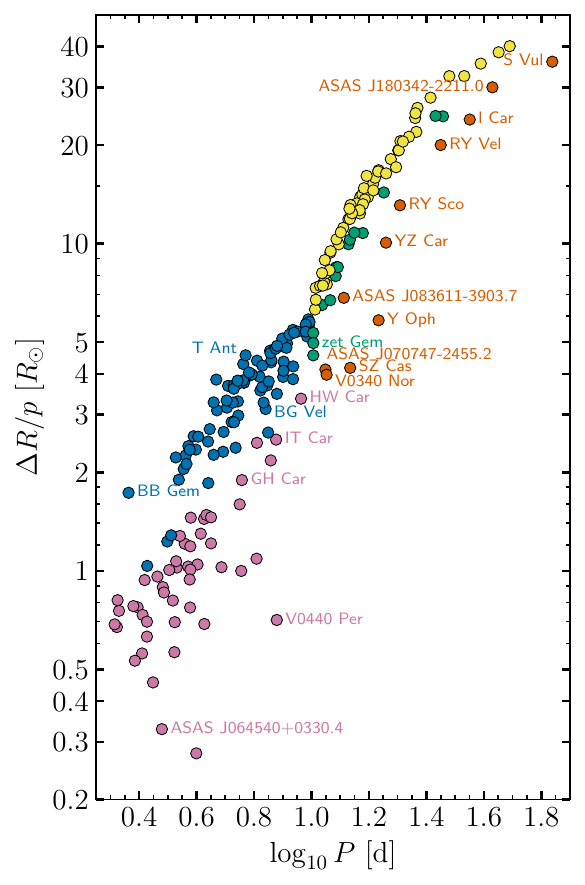}
    \caption{Double-logarithmic plot of $\Delta R / p$ vs \Ppuls. All stars on the low-$\Delta R / p$ sequence are labeled. Color coding as in Fig.\,\ref{fig:DeltaR_logP}.}
    \label{fig:deltaRpops}
\end{figure}

Figure\,\ref{fig:DeltaR_logP} shows the RV curve integrals obtained from the Fourier series fits to \veloce\ data, $\int{v_r d\phi} = \Delta R / p$, against \logP, and Fig.\,\ref{fig:deltaRpops} shows the same information with dual logarithmic axes. The most striking feature in Fig.\,\ref{fig:DeltaR_logP} is the sharp change in slope that occurs near \Ppuls $\sim 10$\,d. Assuming a constant value for $p$, this would imply that long-period (\Ppuls $> 10$\,d) Cepheids vary their size much more significantly than short-period Cepheids, and that the transition between long and short-period Cepheids is abrupt. While a break in RV amplitudes at $10.47\,$d was previously identified by \citet{Klagyivik+2009}, no reference to its significance for BW distances and $p-$factor calibration was made\footnote{\citet[their Fig.\,12]{Klagyivik+2009} considered the related, albeit less readily interpreted, quantity $A_{\rm{RV}} \times P$ as a function of $\log{P}$ to argue for a division of fundamental mode Cepheids into short- and long-period groups at a period of $10.47$\,d (\logP$=1.02$). In their search for spectroscopic binaries or identifications of pulsation modes, Klagyivik et al. came to the conclusion that light-to-RV amplitude ratios were not always reliable indicators of pulsation modes or the presence of companions, although no clear physical reason could be offered. No distinction among the long-period Cepheids was made by the authors.}.

Figure\,\ref{fig:DeltaR_logP} distinguishes five groups of Cepheids: high- ($A_{\mathrm{p2p}} > 20$\,\kms) and low-amplitude Cepheids with \Ppuls $< 10$\,d, as well as three sequences of long-period Cepheids selected in an ad hoc fashion for further investigation. Among the short-period Cepheids, the selection according to $A_{\mathrm{p2p}}$ distinguishes relatively well between stars pulsating in the fundamental mode and the first overtone, despite some imperfections. As the right panel of Fig.\,\ref{fig:DeltaR_logP} shows, there is a continuum of $\Delta R/p$ values between linear fits to the two groups of short-period Cepheids that grows increasingly wide with increasing \logP. At \Ppuls $\gtrsim 10$\,d, Cepheids with low values of $\Delta R/p$ suddenly disappear. 

We identified ad hoc three groups of long-period Cepheids using Fig.\,\ref{fig:DeltaR_logP} and marked them via orange, green and yellow symbols. In the following, we refer to these three groups as the high- (yellow, largest $\Delta R/p$), intermediate- (green, intermediate), and low-$\Delta R$ sequences (orange, smallest $\Delta R/p$). The low-$\Delta R$ sequences sequence contains $11/82 = 13.4\%$ of the Cepheids with $P > 10$\,d, the intermediate sequence $15/82 = 18.3\%$, and the high-$\Delta R$ sequence the majority of Cepheids ($56/82=68.3\%$). The names of the stars belonging to the low-$\Delta R$ sequence are annotated in Fig.\,\ref{fig:deltaRpops}. Curiously, we find Y~Oph to lie on what appears to be the extension of FO Cepheids, together with ASAS~J070747$-$2455.2, V0340~Nor, HW~Car, and others. This is potentially noteworthy, as V0440~Per was previously identified as the longest-period overtone Cepheid in the MW \citep{Baranowski2009}.

To the best of our knowledge, the three $\Delta R/p$ sequences have not previously been investigated separately. While the assignment of a specific Cepheid to each sequence (in particular the intermediate one) may be questioned, the significant outlier nature of many long-\Ppuls\ Cepheids from the (yellow) majority is obvious. Of course, the lower values of $\Delta R/p$ are a direct consequence of their lower peak-to-peak amplitudes, which are established here with high precision. In analogy to the amplitude cut applied to short-period Cepheids, we considered it useful to introduce the three sequences in order to investigate whether any  differences would appear between stars on different $\Delta R/p$ sequences. 

Interestingly, the slope of $\Delta R/p$ vs \logP\ steepens with increasing $\Delta R/p$, resulting in a growing (absolute) difference in $\Delta R/p$ among the sequences towards the longest periods. However, the {\it relative} difference between the sequences {\it decreases} towards longer periods, from nearly $50\%$ near $10\,$d to less than $25\%$ near \logP$\sim 1.7$. Assuming a strict dependence of the angular diameter variation on \logP\ would therefore require very significant differences in $p$ as a function of \logP\ that may explain the difficulties encountered in the observational calibration of $p$, and these differences would become even more noticeable if a single relation is sought for $p(\log{P})$. Interestingly, Fig.\,8 in \citet{Trahin2021} exhibits such a trend, whereby the scatter in $p$ grows towards lower \logP\ and becomes particularly high around \logP$\sim 0.8$.

\begin{figure}
\centering
    \includegraphics[width=0.45\textwidth]{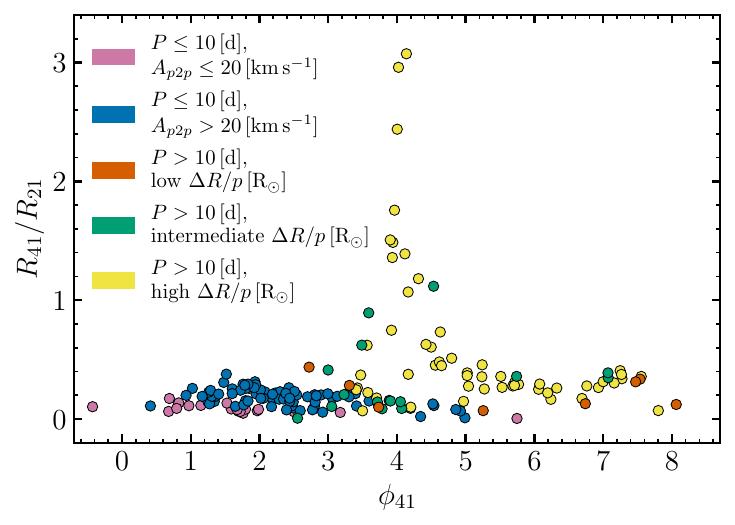}
    \caption{Ratio of amplitudes between 4th and 2nd harmonic vs $\phi_{41}$. Low $\Delta R/p$ stars have consistently low $R_{42}$ more similar to short-period Cepheids, while high $\Delta R/p$ stars exhibit a very large range in $R_{42}$ near $3.5 \lesssim \phi_{41}\lesssim 4.5$.  \label{fig:deltaRsamples_R42_phi41}}
\end{figure}

In case these may be useful, the relations fitted to the five groups are listed in Tab.\,\ref{tab:fit_deltaR}. Appendix\,\ref{app:FourierParameters}  contains figures that illustrate the Fourier parameters for all five groups, and Figure\,\ref{fig:deltaRsamples_R42_phi41} shows the ratio of the 4th to the 2nd harmonic amplitude ($R_{41}/R_{21}$) against $\phi_{41}$. An impressive discontinuity is seen among the yellow points around $3.5 \lesssim \phi_{41} \lesssim 4.5$, while the other long-period Cepheids shown in green and orange exhibit generally flat trends.

\begin{table}
\centering
\caption{Linear fits for the five subsets identified in the $\Delta R/p$ vs \logP\ relation.}\label{tab:fit_deltaR}
\begin{tabular}{@{}lcc@{}}
\toprule
Subset   &    $q$   &  $m$\\
\midrule
\textcolor{OIpink}{$\blacksquare$} $P \le 10 \mathrm{[d]}$, $A_{\mathrm{p2p}} \le 20$\,\kms &  2.238 & 2.673\\
\textcolor{OIblue}{$\blacksquare$} $P \le 10 \mathrm{[d]}$, $A_{\mathrm{p2p}} > 20$\,\kms & 5.392 & 7.494\\
\textcolor{OIdarkorange}{$\blacksquare$} $P > 10 \mathrm{[d]}$, low $\Delta R/p$ & 0.059 & 43.506\\
\textcolor{OIlightgreen}{$\blacksquare$} $P > 10 \mathrm{[d]}$, intermediate $\Delta R/p$ & 4.339 & 44.028\\
\textcolor{OIyellow}{$\blacksquare$} $P > 10 \mathrm{[d]}$, high $\Delta R/p$ & 5.306 & 50.276\\

\bottomrule
\end{tabular}
\tablefoot{$\Delta R/p = m($\logP$ - 1) + q$}

\end{table}

\begin{figure}
    \centering
    \includegraphics[width=0.49\textwidth]{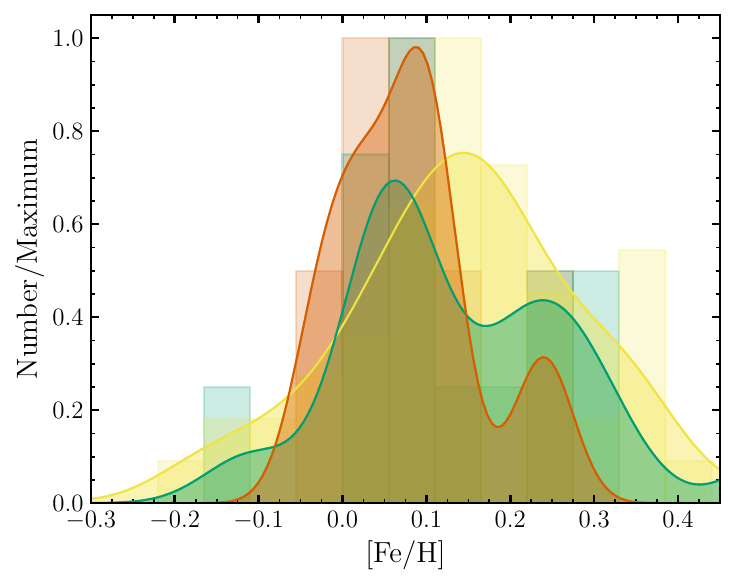}
    \caption{Iron abundances compiled by \citet{Groenewegen2018} for long-period Cepheids on the low- (orange), intermediate- (green), and high-$\Delta R/p$ (yellow) sequences introduced in Fig.\,\ref{fig:DeltaR_logP}. The binned histograms are superposed by KDE-smoothed distributions. No clear difference in metallicity is apparent between the three groups.}
    \label{fig:deltaR_meta}
\end{figure}

We considered several possible differences among the high- and low-$\Delta R$ sequences. The low-$\Delta R$ sequence contains both known (spectroscopic) binaries (YZ~Car, RY~Vel, RY~Sco) and Cepheids with no known companions (e.g., $\ell$~Car, S~Vul, Y~Oph)\footnote{Details of the SB1 nature of Cepheids in the \veloce\ sample are provided in paper~II.}. Iron abundances compiled and homogenized by \citet{Groenewegen2018} do not show a significant difference in the chemical composition of the three groups, cf. Fig.\,\ref{fig:deltaR_meta}.

\begin{figure}
    \centering
    \includegraphics[width=0.45\textwidth]{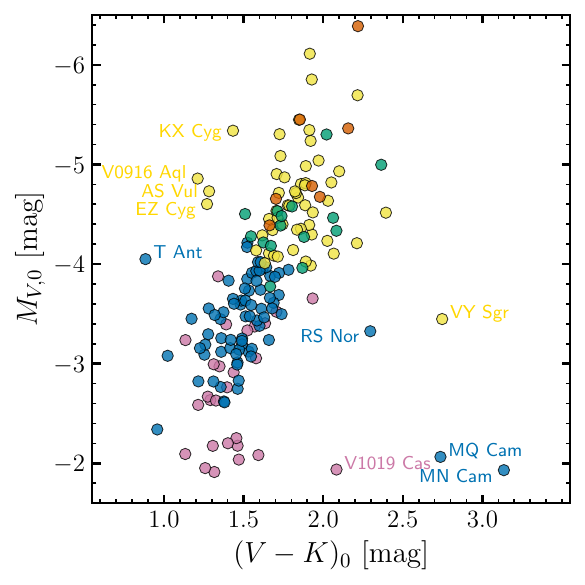}
    \caption{Color magnitude diagram of \veloce\ stars based on $V-$ and $K-$band magnitudes and color excesses compiled by \citet{Groenewegen2018}, de-reddened using the \citet{Cardelli1989} reddening law. Color coding distinguishes between long-period Cepheids (\Ppuls$>10$\,d) with low (orange), intermediate (green), and high (yellow) $\Delta R/p$, as well as short-period (\Ppuls$<10$\,d) high-RV amplitude (blue) and low-RV (purple).}
    \label{fig:deltaR_CMD}
\end{figure}

The color-magnitude diagram in Fig.\,\ref{fig:deltaR_CMD} suggests that low-$\Delta R/p$ sequence stars may prefer a location slightly closer to the hot edge of the instability strip. However, we do not consider this preference significant at this time, and we do not find a similar preference for intermediate-$\Delta R$ sequence stars, nor a preference to avoid hotter temperatures among the high-$\Delta R$ sequence Cepheids. Nonetheless, we note that the hot instability strip boundaries of overtone Cepheids are located at slightly hotter temperatures than those of fundamental mode Cepheids \citep[e.g.,][]{Anderson2016rot}. Distant outliers beyond the hot edge of the instability strip may have significant contamination by hot companion stars, although this remains unclear at the moment. Only one of the five most distant blue outliers is  an SB1 Cepheids (in order of decreasing $M_{V,0}$): T~Ant, EZ~Cyg, AS~Vul, V0916~Aql (only SB1), and KX~Cyg. The five outliers beyond the red instability strip boundaries ($M_{V,0} < -4$, in order of increasing $M_{V,0}$) are VY~Sgr (sole SB1 among red outliers), RS~Nor, MQ~Cam, V1019~Cas, and MN~Cam. The reddest long-period Cepheid in the Figure is XZ~Car (an SB1).

\begin{figure}
    \centering
    \includegraphics[width=0.49\textwidth]{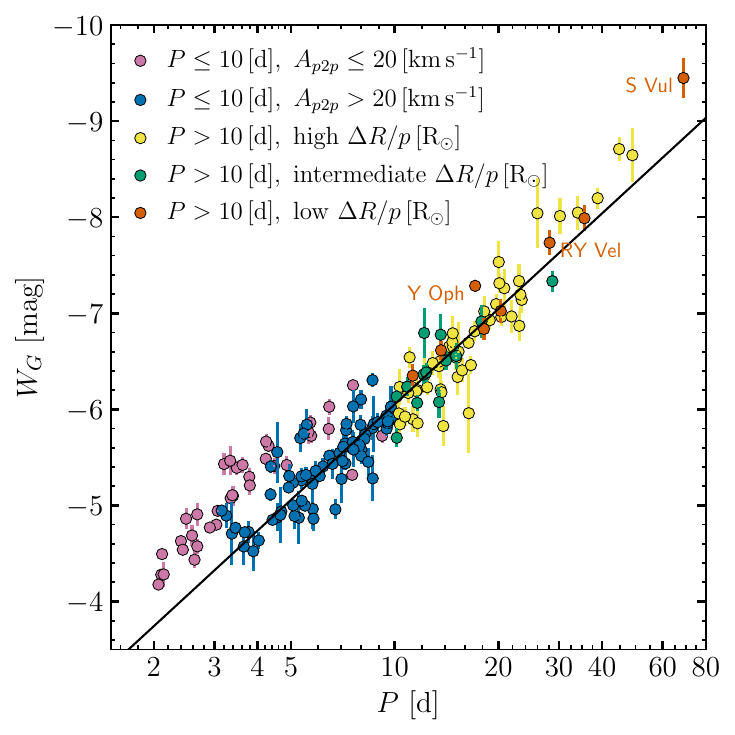}
    \caption{Comparison of the stars belonging to the five subgroups in Fig.\,\ref{fig:DeltaR_logP} shown in relation to the FU Cepheid MW Leavitt law calibrated using cluster Cepheids \citep[black solid line]{CruzReyes2023} and the reddening-free Wesenheit magnitude $W_G$. Low-$\Delta R$ Cepheids are shown in orange and are not systematic outliers from the LL. The most significant outliers are Y~Oph ($7.1\sigma$ from the LL, $\langle m_G \rangle=5.5$\,mag) and S~Vul ($3.9\sigma$, $\langle m_G \rangle=6.7$\,mag). The limitations of this comparison are described in the text.}
    \label{fig:LLoutliers}
\end{figure}

Lastly, we considered the location of stars on the three $\Delta R$ sequences in the Leavitt law (LL). Figure\,\ref{fig:LLoutliers} shows the reddening-free absolute Wesenheit magnitude \citep{Madore1982} in \gaia\ bands, $W_G = G - 1.911 \cdot( G_{Bp} - G_{Rp} )$ computed 
using (inverted) \gaia\ DR3 Cepheid parallaxes corrected for parallax systematics quantified by \citep{Lindegren2021} and the residual parallax offset of \gaia\ field Cepheids determined using cluster Cepheids\footnote{While Fig.\,\ref{fig:LLoutliers} is useful for this specific comparison, we stress that the sample of Cepheids shown here was not selected to provide a meaningful test of the LL calibration, and that it contains stars whose astrometry or photometry may not be suitable to calibrate the LL, e.g., due to blending or poor astrometry, or whose parallaxes should not be inverted to determine distance \citep[cf.][]{Luri2018parallax}. Two significant outliers with visual companions were removed, however: SV~Per, whose visual companion is seen in \coralie\ guiding camera, and SY~Nor, whose companion is barely resolved by {\it HST/WFC3} spatial scanning observations \citep{Riess2018hstphot}.} 
\citep{CruzReyes2023}. Figure\,\ref{fig:LLoutliers} shows that the low-$\Delta R$ sequence is not systematically populated by LL outliers. The notable exception is Y~Oph, which is $7.1\sigma$ from the LL \citep[second row of their Tab.\,11]{CruzReyes2023}, has very small uncertainties, and appears to be consistent with the long-period extension of the first-overtone LL dominated by low-amplitude Cepheids. This is consistent with Y~Oph following the trend of FO Cepheids in Fig.\,\ref{fig:deltaRpops}, despite its unusually long period. While none of the \gaia\ quality parameters indicate poor astrometry or photometry contaminated by a companion, we caution that its magnitude brighter than $\langle m_G \rangle=5.6$\,mag renders this star sensitive to saturation effects. Additionally, the \gaia\ parallax offsets at such bright magnitudes are not yet sufficiently characterized to ensure unbiased parallaxes for individual stars \citep{Khan2023b}. Similarly, S~Vul ($\langle m_G \rangle=6.7$\,mag) is a $\sim 3.9\sigma$ outlier from the MW LL. A few Cepheids consistent with the FO LL have $A_{\mathrm{p2p}} > 20$\,\kms, whereas a few low-amplitude Cepheids also fall on the FU Cepheid LL. Hence, a simple distinction based on RV amplitude does not allow to conclusively assign pulsation modes.  

In summary, the significant differences of $\Delta R/p$  between the three sequences are not due to companion stars, chemical composition, nor the incorrect assignment of the dominant pulsation modes, although some evidence points to the possibility that Y~Oph may be pulsating in the first overtone. This leaves interactions between pulsation modes as well as atmospheric effects as possible origins of the reduced amplitudes and differences in $\Delta R/p$. Further study is required to understand this issue, which significantly contributes to the complexity of $p-$factor calibration.

\section{A legacy reference for Cepheid velocimetry}\label{sec:literature}

We determined systematic RV (zero-point) offsets of literature Cepheid RV datasets relative to \veloce\ using a template fitting approach. The method is described in Sect.\,\ref{sec:RVTF} and literature zero-point offsets are presented in Sect.\,\ref{sec:zp}. Cepheid RV templates based on principal component analysis are currently being prepared (Viviani et al., in prep.). Sect.\,\ref{sec:gaiarvs} compares \veloce\ data to \gaia\ DR3 time series RVs.

\subsection{Template fitting of literature data\label{sec:RVTF}}
\begin{figure*}
    \centering
    \includegraphics[width=\textwidth]{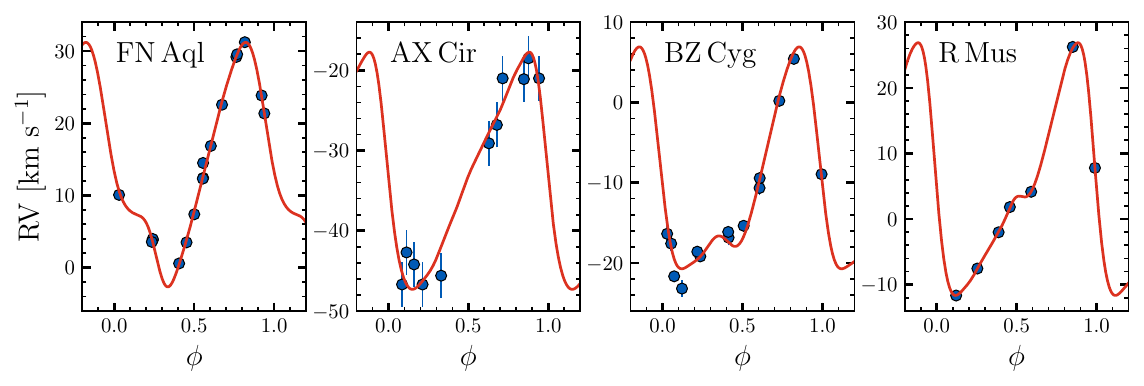}
    \caption{Example template fits to literature data of stars with \veloce\ RVs. Two parameters are fitted: the mean offset  $\Delta v_\gamma$ and the phase offset $\Delta \phi$. The RV curve shapes and amplitudes are fixed by the \veloce\ best fit models. From left to right: FN~Aql observations by \citet{Barnes2005} closely fit to the \veloce\ template; AX~Cir observations by \citet{1980SAAOC...1..257L} are sufficient to achieve a good template fit; BZ~Cyg observations by \citet{Gorynya1992} show some scatter but the template fit is reliable thanks to dense phase sampling; even just seven \veloce\ observations of R~Mus allow for a precise template fit.}   \label{fig:RVTF}
\end{figure*}

\begin{figure*}
    \centering
    \includegraphics[width=\textwidth,trim={0cm 0cm 0cm 0cm}]{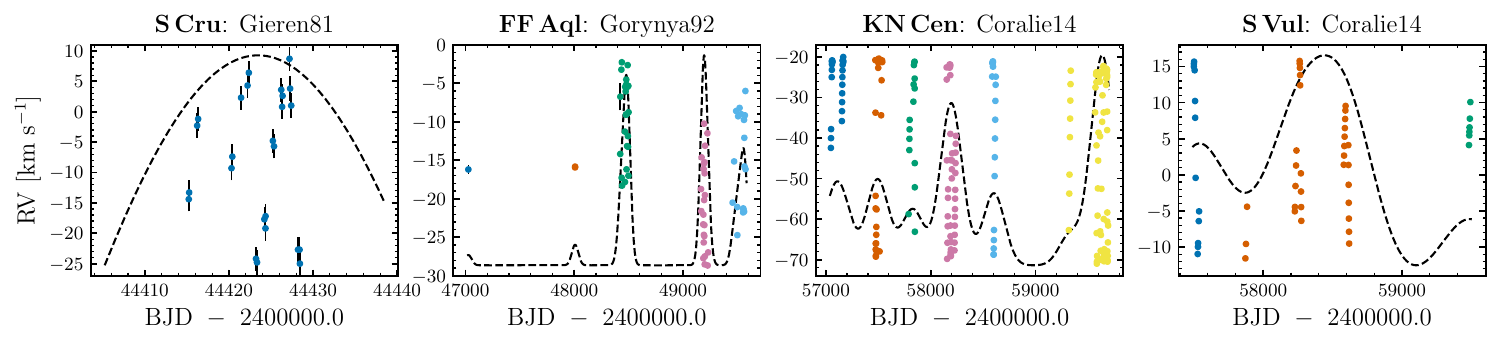}
    \caption{Example of temporal data clustering based on KDE. The solid green lines indicate the kernel density estimate, and data are clustered temporally according to local KDE minima. From left to right: Observations of S~Cru by \citet{1981ApJS...46..287G} are insufficiently sampled for temporal clustering and used all together; observations of FF~Aql by \citet{Gorynya1998} can and must be clustered during later epochs due to significant orbital motion; observations of KN~Cen from \veloce\ can be separated into several well-sampled data clusters; S~Vul observations in \veloce\ were separated into data clusters, although not all clusters could be used due to insufficient data points and rapid period fluctuations.}
    \label{fig:dataclustering}
\end{figure*}

We employed a template fitting method previously developed in \citet{Anderson2016ApJS,Anderson2019polaris} that fits our best-fit Fourier series models determined in Sect\,\ref{sec:Fourier} to literature time series of the same stars using two free parameters: the velocity offset relative to the \veloce\ template and a phase shift. Figure\,\ref{fig:RVTF} shows four examples of RV time series fitted using \veloce\ templates of the same stars.

We split the available time series into time ``clusters'' according to data availability and quality using a nearly automated procedure based on Kernel Density Estimates (KDE) as implemented in \texttt{sklearn.neighbors} \citep{scikit-learn} in order to test time-variable offsets and to improve sensitivity to orbital motion. We typically used a bandwidth of at least 30 d, or 1 pulsation cycle, whichever is longer, unless either obvious rapid orbital motion made this impossible or the data were insufficient. In cases of very good data availability, we used bandwidths as short as one pulsation cycle, for example, when fitting \veloce\ data. Figure\,\ref{fig:dataclustering} illustrates the procedure for four stars with different temporal sampling properties.

Each fitted data cluster contained at least three observations sufficiently spaced in pulsation phase in order to adequately constrain two fit parameters, $\Delta v_\gamma = v_\gamma(t) - v_{\gamma,\mathrm{T}}$ and $\Delta \phi = \phi - \phi_{\mathrm{T}}$, where subscript $_{\mathrm{T}}$ refers to the values of the \veloce\ template. Figure\,\ref{fig:RCru} shows the phase shifts determined for R~Cru, the shortest-period SB1 Cepheid in the MW (cf. paper~II). Data clusters with insufficient number of observations were automatically discarded, and references with insufficient data for KDE clustering were considered as a single time series. We inspected all resulting template fits visually and modified starting conditions for  $\Delta \phi$ (very rarely $\Delta v$) or the KDE bandwidth, if needed to obtain reasonable fits to the data.

For each clustered template fit, we recorded the reference, integer numbering of the data cluster, its mean date, its first and last observations, fit parameters $\Delta v_\gamma$ and $\Delta \phi$, as well as their uncertainties. We further recorded the time series of residuals of the data minus the template fit.

Template fitting results serve three main purposes: a) the estimation of zero-point offsets of literature data relative to \veloce, b) the detection of temporal trends in $v_\gamma$, and c) the determination of orbital solutions based on the longest available baselines. Here, we focus on element a) to facilitate the combination of \veloce\ data with other datasets and to assess the quality of \gaia\ RVS time series of classical Cepheids published as part of DR3. Elements b) and c) are presented in paper~II. Additionally, the values of $\Delta \phi$ allow for an estimation of the period changes, \Pdot, which we briefly discuss in Sect.\,\ref{sec:pdot}.

\subsection{Literature zero-point differences from template fitting} \label{sec:zp}

\begin{table}
\begin{tiny}
    \centering
    \caption{Literature zero-point offsets relative to \veloce}
    \label{tab:ZPs}
    \addtolength{\tabcolsep}{-0.4em}
    \begin{tabular}{@{}l|cc|rr@{}}
\toprule
Reference & $N_{\mathrm{stars}}$ & $N_{\mathrm{clusters}}$ & $\Delta v_{\mathrm{ZP}}$ & $e_{\Delta v_{\mathrm{ZP}}}$ \\
\midrule
\citet{Anderson2016ApJS} (Hamilton) & 9 & 25 & $-$0.039 &  0.129 \\ 
\citet{Baranowski2009} & 1 & 3 & $-$0.686 &  2.000 \\
\citet{1987ApJS...65..307B} & 9 & 16 & $-$0.916 & 0.652 \\ 
\citet{1988ApJS...66...43B} & 16 & 25 & $-$1.005 &  0.479 \\ 
\citet{Barnes2005} & 5 & 9 & $-$0.076 &  0.277 \\
\citet{Bersier1994} & 11 & 30 & $-$0.614 &  0.075 \\ %
\citet{Bersier2002} & 18 & 27 & $-$0.652 &  0.149 \\ %
\citet{Borgniet2019} & 17 & 23 & $-$0.891 &  0.118 \\ %
\citet{Caldwell2001} & 2 & 2 & $-$2.894 &  0.105 \\ %
\citet{1985SAAOC...9....5C} & 15 & 43 &  0.619 &  0.315 \\ 
\citet{1985ApJS...57..595C} & 3 & 7 & $-$0.868 &  0.695 \\ 
\citet{1981ApJS...46..287G} & 3 & 3 & $-$3.202 &  0.513 \\
\citet{1985ApJ...295..507G} & 1 & 1 &  1.560 &  2.000 \\
\citet{1989AJ.....98.1672G} & 1 & 1 & $-$2.553 &  2.000 \\
\citet{Gorynya1992,Gorynya1996,Gorynya1998} & 31 & 111 & $-$0.373 &  0.078 \\ 
\citet{Imbert1999} & 11 & 43 & $-$0.559 &  0.136 \\ 
\citet{Joy1937} & 2 & 3 & $-$2.318 &  1.790 \\
\citet{Kienzle1999} & 6 & 12 & $-$0.647 &  0.075 \\
\citet{1998MNRAS.297..825K} & 4 & 4 &  0.353 &  0.383 \\
\citet{1980SAAOC...1..257L} & 2 & 2 & $-$0.850 &  0.739 \\
\citet{1991ApJS...76..803M} & 8 & 8 &  0.127 &  0.266 \\
\citet{1992AJ....103..529M} &  6 & 6 & $-$0.826 &  0.748 \\ 
\citet{Pont1994} & 12 & 12 & -0.650 &  0.326 \\
\citet{Pont1997} (Coralie) & 6 & 8 & $-$0.525 &  0.290 \\
\citet{Pont1997} (Elodie) & 4 & 4 &  0.095 &  0.164 \\
\citet{Stibbs1955} & 2 & 2 & $-$2.308 &  1.464 \\ 
\citet{Storm2011pfactor} & 3 & 3 &  0.862 &  0.052 \\
\citet{Storm2004} & 4 & 8 & $-$0.346 &  0.431 \\
\citet{Struve1945} & 1 & 1 & $-$2.009 &  2.000 \\
\citet{1989ApJS...69..951W} & 8 & 12 & $-$0.910 &  0.561 \\
\midrule
\gaia\ DR3 RVS Cepheid time series & 62 & 121 &  0.652 &  0.105 \\
\midrule
\coraliev{07} & 35 & 75 &  0.019 &  0.019 \\
\coraliev{14} & 68 & 219 & $-$0.001 &  0.012 \\
\hermes & 62 & 207 & $-$0.002 &  0.021 \\
\bottomrule
    \end{tabular}
    \tablefoot{Zero-points are stated as $\Delta v_{\mathrm{ZP}} = v_{\gamma,\mathrm{ref}} - v_{\gamma,\mathrm{VELOCE}}$. $N_{\mathrm{stars}}$ is the number of stars used to determine $\Delta v_{\mathrm{ZP}}$, $N_{\mathrm{clusters}}$ the number of epochs used for the reference, with some stars contributing multiple epochs. Most zero-point offsets are of order $\pm 1$\kms\ or less, and we recover the zero-point change between \coraliev{07} and \coraliev{14} to within $\sim 5\,$\ms, cf. Tab.\,\ref{tab:coralie_stds}, despite this analysis being based on high-amplitude variable stars. The adopted uncertainty $e_{\Delta v_{\mathrm{ZP}}}$ is the standard median error unless only a single star was used to determine the zero-point, in which case a generous uncertainty of $2$\kms\ was adopted. The offset for \citet{Anderson2016ApJS} applies only to the Hamilton spectrograph data set. Coralie observations by \citet{Pont1997} used an instrument version prior to \coraliev{07} and feature a significantly different zero-point than \coraliev{07} and \coraliev{14}. For \gaia\ DR3 RVs, we state the result based on template fitting, see Sect.\,\ref{sec:Gaia-RVTF}.}
\end{tiny}
\end{table}

RV offsets between \veloce\ and the literature are expected due to differences in wavelength calibration, wavelength ranges used to determine RV, and the growing accuracy of RV measurements over the many decades spanned by the literature data, among others. 

\begin{figure}
    \centering
    \includegraphics[width=0.49\textwidth]{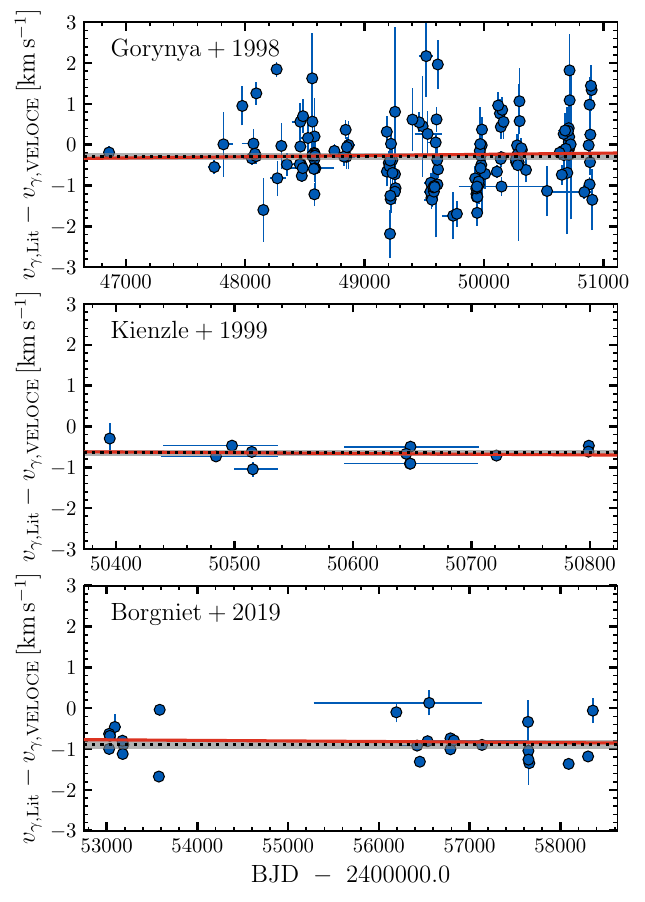}
    \caption{
    Determination of RV zero-points as a function of time for different references. From top to bottom: \citet{Gorynya1998}, the previously largest data set of classical Cepheid RV time series; \citet{Kienzle1999}; \citet{Borgniet2019}. Only stars with \veloce\ observations are used to determine zero-points. Horizontal errorbars indicate the time span of the measurements. A solid red line shows an unweighted linear fit to investigate temporal trends; the dotted black line shows the median with its standard error as a grey shaded region.}
    \label{fig:ZPvsBJD}
\end{figure}

We considered only well-observed and well-behaved\footnote{Only epochs with $\vert \Delta v_r \vert \le 4$\,\kms\ as well as  $\sigma(\Delta v_r) < 0.2$\kms\ were considered for most newer and very precise data sets. Data from \citet{1985ApJ...295..507G,1987ApJS...65..307B,1989ApJS...69..951W,Caldwell2001,1985ApJS...57..595C,1985SAAOC...9....5C,1981ApJS...46..287G,1992AJ....103..529M,1998MNRAS.297..825K,Joy1937,Stibbs1955,Struve1945} and \gaia\ DR3 were considered if $\sigma(\Delta v_r) < 3.0$\,\kms.} non-SB1 Cepheids to determine zero-point differences. Using the clustered template fitting approach, we investigated possible temporal variations in the zero-points derived from multiple sources, cf. Fig.\,\ref{fig:ZPvsBJD}.
Linear trends of $\Delta v_{\mathrm{zp}}$ were investigated for all stars from the same reference. However, such trends were generally found to be negligible.

\begin{figure}
    \centering
    \includegraphics[width=0.49\textwidth]{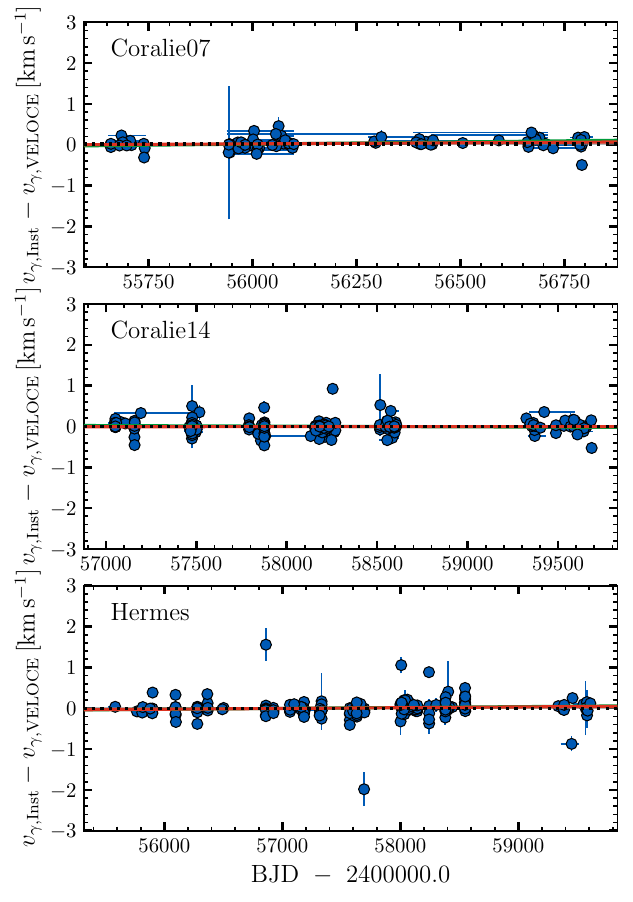}
    \caption{RV zero-point stability for the three instruments distinguished here, \coraliev{07}, \coraliev{14}, and \hermes\ relative to the overall \veloce\ zero-point set by the combined dataset of all Cepheids, which includes stars observed with both \coralie\ versions, as well as with \hermes\ and \coralie. The vertical scale and color scheme is identical to Fig.\,\ref{fig:ZPvsBJD}. Horizontal errorbars indicate the time span of the measurements. Outliers from the median are mostly due to imperfect template fits or low-amplitude signals mimicking temporal variations in $v_\gamma$. The \hermes\ zero-point change due to the intervention in 2018 is visible near BJD 58300, cf. Tab.\,\ref{tab:hermes_stds}.}
    \label{fig:VELOCEZP}
\end{figure}

Since no significant evidence for temporal zero-point variations was found, we quantified the zero-point offset relative to \veloce\ using the median $\Delta v_{\mathrm{zp}}$ and the standard error on the median for each source. Table\,\ref{tab:ZPs} lists the zero-point offsets relative to \veloce\ derived for all references considered, and Fig.\,\ref{fig:VELOCEZP} shows the zero-point stability of \coraliev{07}, \coraliev{14}, and \hermes\ derived using the same approach. We stress that a large absolute value of $\Delta v_{\mathrm{zp}}$ does not imply poor data quality.

Zero-points for \coralie\ and \hermes\ are reported in Sect.\,\ref{sec:obs:zeropoints} alongside a discussion of zero-point changes due to instrument upgrades. This information will be used in the future to further improve the precision of \veloce\ RV data and can serve to accurately tie our observations to other instruments and surveys. However, in this first data release, we considered a constant zero-point for \coralie\ and \hermes\ Cepheid RVs and worked under the hypothesis that both instruments had consistent zero-points. The fact that many Cepheids were observed with both instruments, and using each of the two instrument versions, allowed us to quantify the level of agreement of \coraliev{07}, \coraliev{14}, and \hermes\ with the overall \veloce\ DR1 average. Using this approach, we recovered the correct sign and order of magnitude of the zero-point change (20\,\ms) between \coraliev{07} and \coraliev{14} (Sect.\,\ref{sec:obs:zeropoints}), albeit using high-amplitude variable stars instead of extremely stable RV standards. Conversely, we find \hermes\ to agree extremely closely with \coraliev{14}. The \hermes\ zero-point change does not alter this picture significantly, since the majority of \hermes\ observations reported here were collected before the instrument upgrade.

\subsubsection{Literature data discarded}
Inspection of the template fits allowed us to identify individual unreliable RV observations reported in datasets from the literature. In particular, we identified two nights in the dataset by \citet{Borgniet2019} where the reported RV measurements differed by several \kms\ from expectations for 8 Cepheids. We therefore decided to remove all observations from the two nights of 24 and 25 June 2015 (reduced BJDs starting with 57198 and 57199) of that data set. We further discarded two observations for RU~Sct reported by \citet{Bersier2002} (BJD around 57635.5) and one observation of BN~Pup from the data reported by \citet{Storm2011pfactor} due to an offset of more than $20$\kms\ from the template fit.

\subsection{Period changes from RV template fitting \label{sec:pdot}}
Cepheids exhibit a variety of period change phenomena, including linear period changes most likely due to secular evolution \citep[e.g.,][]{Szabados1989,Turner2006}, oscillatory period changes \citep[e.g.,][]{Kervella2017rspup,Csornyei2022}, period ``jitter'' \citep[e.g.,][]{Derekas2017}, and possible sudden period changes \citep{Turner2005Polaris}. In particular the linear period changes are of interest for testing stellar evolutionary timescales \citep[e.g.,][]{Turner2006,Fadeyev2014,Anderson2016rot,Miller2020} as they are expected to directly trace the monotonous (secular) evolution across the instability strip and allow for an identification of the instability strip crossing numbers according to the sign and magnitude of $dP/dt = \dot{P}$ \citep[e.g.,][]{Anderson2018polaris}.

The classical approach to measuring \Pdot\ performs $O-C$ analysis \citep{Sterken2005} and compares the observed time of maximum light to the expected time of maximum light based on the known pulsation ephemerides. $O-C$ is then usually plotted against $N_{\mathrm{cyc,E}}$, the number of cycles\footnote{$N_{\mathrm{cyc,E}}$ is usually written simply as $E$ in the literature \citep{Sterken2005}. We here adopt $N_{\mathrm{cyc,E}}$ to distinguish this from our pulsation ephemerides epoch $E$ used in Sect.\,\ref{sec:Fourier} and to highlight that this refers to a number of pulsation cycles.} elapsed since the reference epoch used to determine \Ppuls, and fitted by a parabola\footnote{An erroneous period  would show as a straight line in Fig.\,\ref{fig:RCru}.} 
\begin{equation}
O-C = \frac{1}{2} \frac{dP}{dt} \cdot P_{\mathrm{puls}} \cdot N_{\mathrm{cyc,E}}^2 \ ,
\end{equation}
where \Ppuls\ denotes the period at the reference epoch, $E$. Using $O-C = \Delta \phi \cdot P$ and $N_{\mathrm{cyc,P}} = (BJD - E)/P$, we determine \Pdot\ directly from this relation for $146$ Cepheids with \veloce\ and literature RV data using the measured values of $\Delta \phi$, $P$, and $E$. In almost all instances, the small baseline yielded $\vert \Delta \phi \vert < 1$. In a few cases, sudden steps in the $\Delta \phi$ measurement sequence suggested that a full cycle had been skipped, and we added $\pm 1$ as needed to maintain a monotonous long-term period evolution.

\begin{figure*}
\centering
\includegraphics[width=1\textwidth]{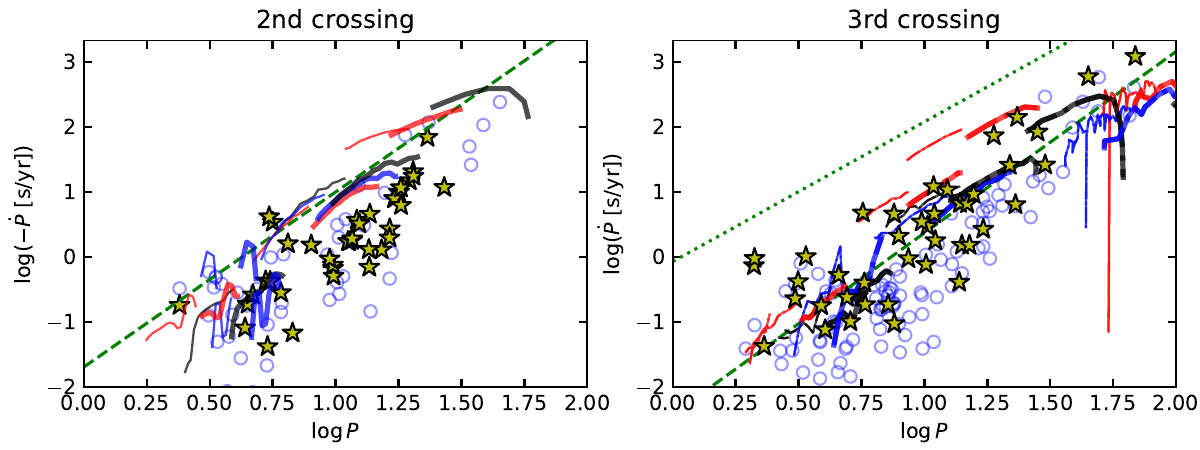}
\caption{Double-logarithmic plot of  linear period changes, \Pdot, against \Ppuls. Yellow stars show values determined using $\Delta\phi$ from our template fitting analysis if $\vert\dot{P}\vert/\sigma_{\dot{P}} \ge 3$. Blue circles were measured based on photometry by \citet{Csornyei2022}. Green straight lines are theoretical predictions from \citet[][dotted line represents first crossing]{Fadeyev2014}, blue, red, and black lines are theoretical predictions from \citet[][for first overtones (thin lines) and fundamental modes with three rotation rates, cf. their Sect.\,3.7 and Fig.\,13]{Anderson2016rot}.  {\it Left panel:} Negative \Pdot\ indicative of a second instability strip crossing. {\it Right panel:} Positive \Pdot\ indicative of a third (rarely: first) instability strip crossing. Overall, we find good agreement between our template fitting-based \Pdot\ and the literature as well as stellar models.\label{fig:pdot}}
\end{figure*}

Table\,\ref{tab:pdot} lists the values of \Pdot\ thus determined and their formal uncertainties based on the covariance matrix of the weighted fit, as well as notes for individual stars to help the interpretation of our results. Among the available sample of \npdotall, \Pdot\ is detected at $3\sigma$ significance or better for \npdotsig\ stars, $39$ ($49\%$) of which have negative \Pdot. Counting all stars does not significantly change these statistics ($53\%$ negative \Pdot).
This is rather different from large samples of photometrically measured \Pdot, such as the compilation of 196 Cepheids by \citet{Turner2006} that lists $33\%$ negative and $67\%$ positive  \Pdot\ values and the recent study by \citet{Csornyei2022}, which reports $44/141$ ($31\%$) negative and $69\%$ ($97/141$) positive \Pdot\ values. 
\begin{table}
\centering
\caption{Linear rates of period change, \Pdot\ measured by RV template fitting (extract)\label{tab:pdot}}
\begin{tabular}{@{}lrrc@{}}
\toprule
Star   &  \Pdot     &  $\sigma_{\dot{P}}$ &  Notes \\
    & (s/yr) & (s/yr) \\
\midrule
AA Gem &     -1.721 &      0.102 &   \\
AC Mon &     -0.453 &      0.294 &  $N_{\rm lit}=1$  \\
AH Vel &      0.238 &      0.091 &  non-linear \\
AK Cep &     -0.234 &      0.759 &   \\
AN Aur &      2.978 &      0.548 &  non-lin? \\
AO CMa &      0.185 &      0.014 &  $N_{\rm lit}=1$, 3 points  \\
AQ Car &     -0.669 &      0.031 &  non-linear \\
AQ Pup &     26.128 &      5.126 &  non-linear \\
AV Cir &      0.228 &      0.033 &   \\
AV Sgr &     36.292 &     46.894 &  noisy \\
\multicolumn{4}{c}{\ldots} \\
\bottomrule
\end{tabular}
\tablefoot{The full table listing all 146 Cepheids is available via the CDS. \Pdot\ is measured at a significance exceeding $3\sigma$ for 80 (shown in Fig.\,\ref{fig:pdot}. $N_{\rm lit}=1$ indicates that only one older literature epoch was available. `non-lin?' indicates possible signs for non-linear period changes, `non-linear' indicates clear evidence for this. `noisy' marks $\Delta \phi$ sequences exhibiting high scatter. `3 points' highlight stars with only few (3) epochs for determining \Pdot. `biased' indicates results where the fit result may be affected by non-linear trends or uneven weights; `biased?' indicates results where this is less clearly the case.}
\end{table}

Figure\,\ref{fig:pdot} illustrates the values of \Pdot\ thus determined, separated into negative and positive \Pdot\ for stars for which $\vert \dot{P} \vert > 3 \sigma_{\dot{P}}$. Theoretical predictions from \citet{Fadeyev2014} and \citet{Anderson2016rot}, as well as \Pdot\ values determined by \citet{Csornyei2022} using very long photometric time series are shown for comparison. Overall, we find very good agreement with both the models and the literature \Pdot\ values. Given that much fewer RV observations than photometric observations are available in the literature, this can be considered a success. A more detailed study of period changes is, however, outside the scope of this work.

Despite this promising result, we caution that the values of \Pdot\ are subject to a few important caveats and complicating factors. Figure\,\ref{fig:RCru} illustrates this for R~Cru, which exhibits both short-timescale non-linear period changes and a long-term linear trend in period. The non-linear period changes dominate over the baseline of \veloce\ data, and  the older data are required to improve sensitivity to the secular (linear) period change. Unfortunately, there are often long gaps between the sampled epochs and the older epochs are often few and not sufficient to resolve short-timescale variations at previous epochs. Fitting the long-term trend using the above parabolic model can thus occasionally produce a biased result, since a) the \veloce\ epochs will dominate the fit due to higher number of epochs and high precision and b) the effect of non-linear period changes on the older epochs cannot be assessed. Issues related to non-linear period changes can be expected to occur more frequently and severely in long-period fundamental-mode and short-period first-overtone Cepheids, whose \Ppuls\ are known to be less stable than those of short-period fundamental mode Cepheids \citep[e.g.,][]{Anderson2014rv,Sueveges2018a,Csornyei2022}. The last column of Tab.\,\ref{tab:pdot} provides further guidance for our results.

\subsection{\gaia\ DR3 RV measurements from the RVS instrument\label{sec:gaiarvs}}

\begin{figure*}
\centering
    \includegraphics[width=\textwidth]{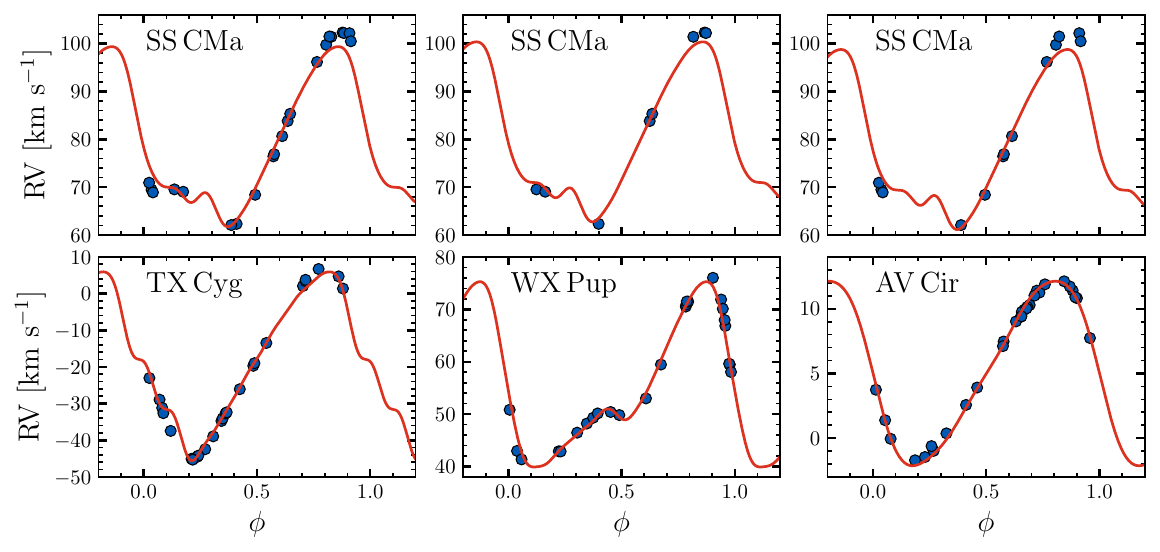}
\caption{Example \veloce\ model template fits to \gaiarvs. Top row: The full set of \gaiarvs\ for SS~CMa is fitted at once (left) or split into two epoch (center \& right). \texttt{gaiadr3.gaia\_source} does not provide the template parameters for SS~CMa despite \gaiarvs\ being available in \texttt{gaiadr3.vari\_epoch\_rv}. Bottom row: stars whose \gaiarvs\ were determined using different temperatures (from left to right:  TX~Cyg (4250K), WX~Pup (5250K), AV~Cir (6250K)).
\veloce\ uses a correlation mask created to match spectral type G2. \label{fig:gaiarvszp_examples}}
\end{figure*}

On 13 June 2022, $43\,298$ time series radial velocity measurements obtained with the \gaia\ Radial Velocity Spectrometer (henceforth: \gaiarvs) became available as part of the third \gaia\ data release \citep{gaiadr3.radvel}. Among these are $15\,171$ time series observations of $731$ classical Cepheids (stars labeled `DCEP' in column \texttt{type\_best\_classification} of \gaia\ DR3 table \texttt{gaiadr3.vari\_cepheid}). 
\gaiarvs\ are of particular interest given their great number, the large sample of stars \citep{gaiadr3.cepheid}, and the growing number of epochs being collected as the mission progresses. For Cepheids, the mean per-epoch \gaiarvs\ uncertainty is $2.0$\,\kms (median $1.4$\,\kms, standard deviation $2.1$\,\kms). The most precise \gaia\ RV observation has an uncertainty of $0.135$\,\kms, and the least precise $46.8$\,\kms. The mean number of observations per star is $21$ (\texttt{gaiadr3.vari\_cepheid} table \texttt{num\_clean\_epochs\_rv}), with a full range of $4 - 74$. The mean reported uncertainties for pulsation averaged velocities, \vgamma, according to table \texttt{gaiadr3.vari\_cepheid} is $1.9$\,\kms\ (median $1.5$\,\kms, standard deviation $2.3$\,\kms, full range $0.03 - 25$\,\kms), somewhat larger than the typical uncertainties reported by \citet[$1-1.5$\,\kms]{gaiadr3.cepheid}. Thus, \gaiarvs\ are available for a $2.8\times$ larger sample than \veloce, albeit with typically fewer ($\sim$ half) observations per target, shorter baselines, and significantly ($\sim 30\times$) lower precision. Importantly, \gaiarvs\ are fully contemporaneous with \veloce, offering a unique opportunity for direct comparisons with minimal complications due to orbital motion or variable ephemerides. 

The number of \gaia\ observations are set to grow substantially as \gaia\ completes its unprecedented survey, and the final \gaia\ data release will feature similar temporal baselines as \veloce. However, there are several important systematic differences in the way that RVs are measured here and using \gaia. The following investigates these differences, first using the RV time series (Sect.,\ref{sec:Gaia-RVTF}), followed by a comparison of the values of \vgamma\ and peak-to-peak amplitude derived by \citet{gaiadr3.cepheid} using Fourier series models (Sect.\,\ref{sec:Gaia-Fourier}).

\subsubsection{Template fitting of \gaia\ RVS time series observations\label{sec:Gaia-RVTF}}

We investigated the overall zero-point offset of \gaiarvs\ relative to \veloce\ by fitting the \gaia\ RV time series observations using \veloce\ templates, analogous to Sect.\,\ref{sec:RVTF}. Figure\,\ref{fig:gaiarvszp_examples} illustrates the examples of four stars. The top row illustrates template fits to the full \gaiarvs\ time series for SS~CMa (left), as well as template fits to two time clusters made up of the same data set (center and right). The bottom row illustrates three additional stars with different RV curve shapes and amplitudes. The largest differences are seen near the extremes of the RV curve and near bump features, which appear visibly different from the \veloce\ templates of RV curves measured using optical lines. We surmise that this effect arises primarily due to the different behavior of the lines near the Ca\,II triplet compared to the metallic line RVs reported in the Cepheid literature and in \veloce\ \citep[e.g.,][]{Wallerstein2019,Hocde2020Halpha}. 

\begin{figure}
    \centering
    \includegraphics[width=.49\textwidth]{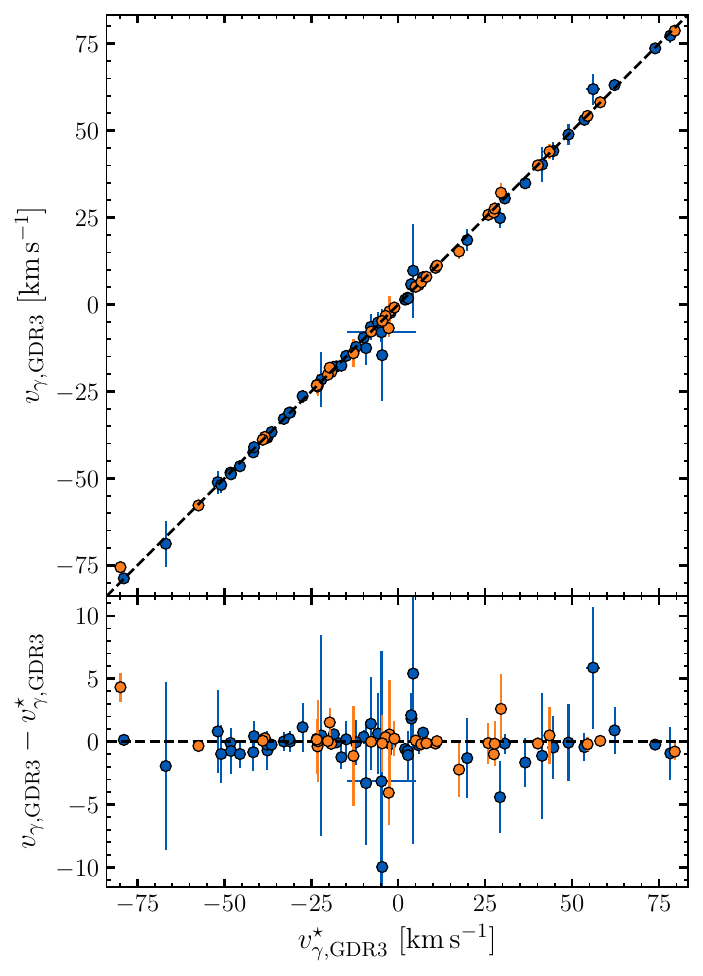}
    \caption{Comparison of \gaia\ DR3 parameter \texttt{average\_rv} from table \texttt{gaiadr3.vari\_Cepheid}, \vgGaiaCat, to the pulsation-averaged velocity derived by template fitting of the \gaia\ DR3 RV time series, \vgGaiaTF. Orange symbols highlight values based on \gaia\ RV measurements for which the most suitable template parameters were used, see Sect.\,\ref{sec:gaiarvs}.}
    \label{fig:GDR3_vgamma_consistency}
\end{figure}

Figure\,\ref{fig:GDR3_vgamma_consistency} compares the values of \vgGaiaTF, determined here by template fitting to the \gaia\ RV time series measurements, to  \vgGaiaCat, the constant term of the Fourier series fitted to \gaiarvs\ data as part of the SOS processing \citep[parameter \texttt{average\_rv} in table \texttt{gaiadr3.vari\_cepheids}]{gaiadr3.rrlyrae,gaiadr3.cepheid}. Overall, we found good agreement between the the two \vgamma\ values over the full range from $-80$\,\kms\ to $+100$\,\kms, with an rms of $1.9$\,\kms. 

\begin{figure}
\centering
\includegraphics[width=0.49\textwidth]{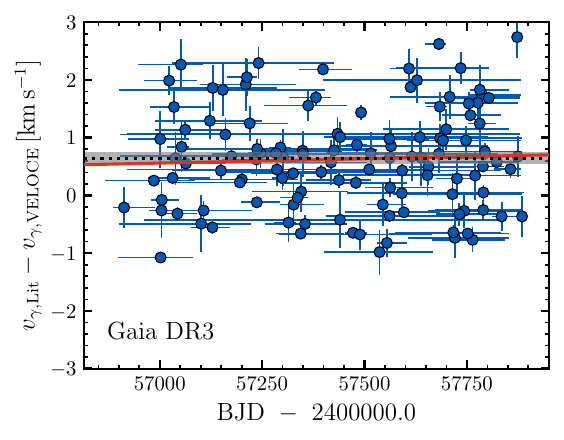}
\caption{Differences between \vgGaiaTF\ (template fits to \gaiarvs) and \veloce\ \vgamma\ of 121 time clusters of 66 non-SB1 Cepheids where $\vert \sigma_{\Delta v_\gamma} \vert \le 0.5$\kms. Horizontal errorbars indicate the duration of the time series used for the fit, vertical errorbars show the uncertainty of the velocity difference. The median difference is $0.65\pm 0.11$\kms.\label{fig:gaiarvszp}}
\end{figure}

To quantify the offset of \gaiarvs\ relative to \veloce, we measured the differences in \vgamma\ resulting from the template fit minus the \veloce\ pulsation-averaged velocity, $\Delta$\vgGaiaTF $ = $ \vgGaiaTF $- v_{\gamma,\veloce}$. We first used the full \gaiarvs\ time series at once and then repeated the procedure using KDE clustering to investigate temporal trends. The clustered approach using $121$ estimates of $62$ Cepheids with precise template fits ($\sigma_{\Delta v_\gamma}  \le 0.5$\kms) yielded a median offset of $\Delta v_\gamma = 0.65\pm 0.11$\kms\ (dispersion $0.92$\kms) and is reported in Tab.\,\ref{tab:ZPs}. Relaxing the constraint on precision to $\sigma_{\Delta v_\gamma} \le 1.0$\kms\ yielded 151 estimates of 74 Cepheids and a similar offset of  $0.67 \pm 0.12$ (dispersion $1.15$\,\kms).
Considering the entire time series for all $66$ Cepheids with $\sigma_{\Delta v_\gamma} \le 0.5$\kms\ at once yielded the consistent, albeit slightly less precise, result of $0.70\pm0.15$\kms\ ($\sigma=1.00$\kms). Including less precise template fits ($\sigma_{\Delta v_\gamma} \le 1.0$\kms)  changes this to 76 Cepheids and a median offset of $0.77 \pm 0.17$\,\kms.

Figure\,\ref{fig:gaiarvszp} illustrates the clustered results and exhibits an initially surprising degree of scatter ranging from $-1 \lesssim \Delta v_{\gamma,\rm GDR3}^\star \lesssim 3$\,\kms. This range is larger when including less precise template fits. The large scatter partially arises due to the use of different synthetic spectra for every star in the \gaia\ RVS cross-correlation pipeline \citep{gaiadr3.radvel}. For stars brighter than $g_{\rm{RVS}} \leq 12$\,mag  \citep{gaiadr3.grvsphot}, the choice of template is decided automatically in favor of the spectral template that consistently yields the highest correlation peaks. This approach is well adapted for dealing with the enormous number and diversity of objects observed by \gaia. However, when comparing RVs of a group of similar objects, such as Cepheids, it is advantageous to adopt the same correlation mask for all stars in the group and their spectra in order to ensure consistent RV zero-points. Choosing different templates either from Cepheid to Cepheid, or as a function of phase for the same Cepheid, may complicate the interpretation of the results because different correlation templates assign different weights to the same spectral lines. 

\subsubsection{Comparison of \gaia\ RVS Fourier fit results to \veloce\label{sec:Gaia-Fourier}}
\begin{figure*}[t]
    \centering
    \includegraphics[width=\textwidth]{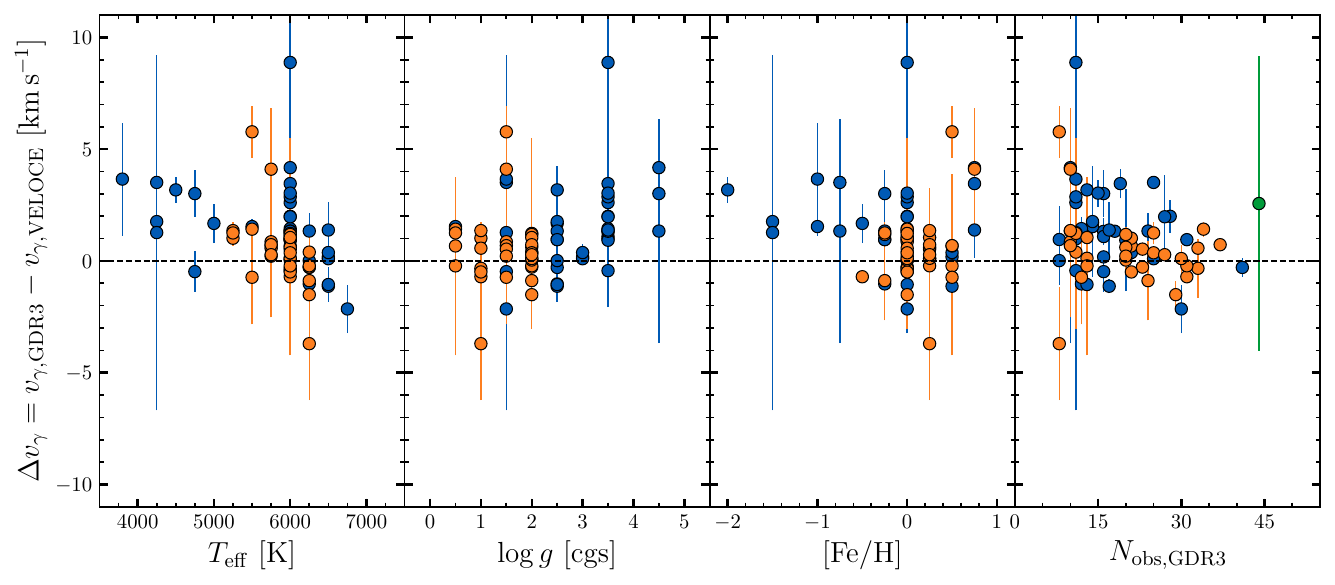}
    \caption{Difference $\Delta v_\gamma = $\vgGaiaCat$- v_{\gamma,\veloce}$ between parameter \texttt{average\_rv} from \texttt{gaiadr3.vari\_cepheid} and \vgamma\ from \veloce\ for 71 well-behaved non-SB1 Cepheids as a function of the \gaia\ correlation template parameters and the number of \gaia\ time series observations, \texttt{num\_clean\_epochs\_rv}. Orange points highlight the 33 Cepheids whose \gaiarvs\ were measured using the most suitable template parameter range $5000 \le T_{\rm eff} \le 6500$, $\log{g} \le 2$, and $-0.75 \le$ [Fe/H] $\le 0.75$. The green point in the right-most panel is RW~Cas, for which the template parameters are not known. Trends appear among all template parameters, with cooler templates or higher-$\log{g}$ templates yielding larger positive offset. RVs measured using adequate template parameters show significantly improved scatter.}
    \label{fig:GDR3_deltavgamma}
\end{figure*}

\begin{figure*}
    \centering
    \includegraphics[width=\textwidth]{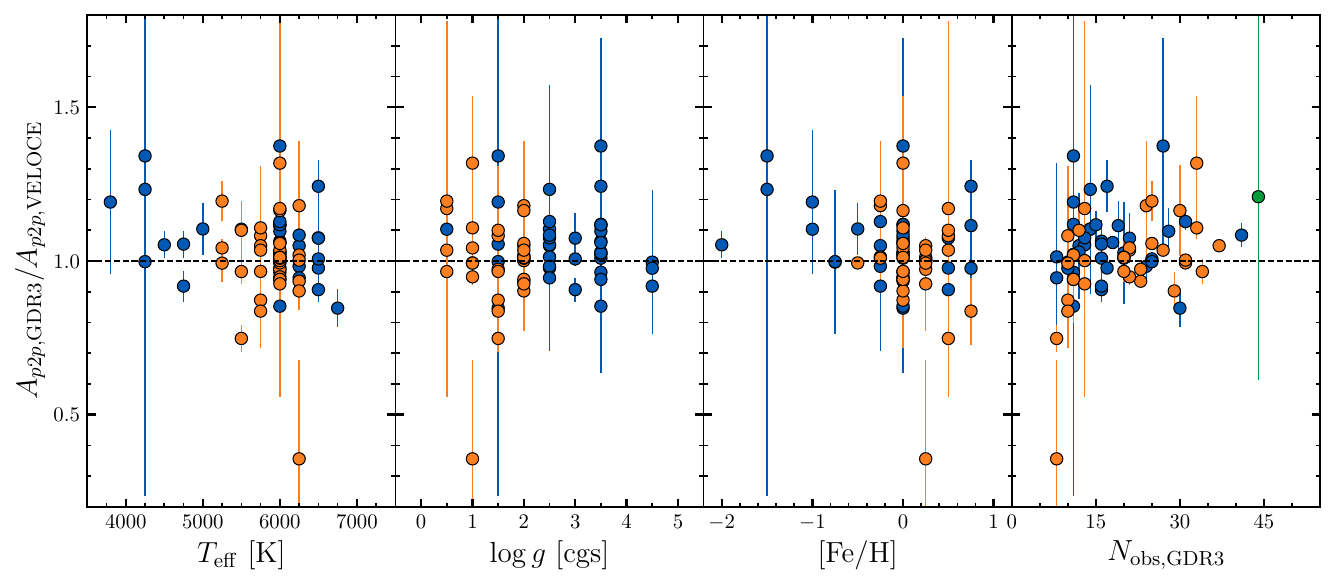}
    \caption{Ratio of the peak-to-peak RV amplitudes from \gaia\ DR3 (\texttt{peak\_to\_peak\_rv} from table \texttt{gaiadr3.vari\_cepheid}) to the \veloce\ peak-to-peak RV amplitudes as a function of the template parameters, analogous to Fig.\,\ref{fig:GDR3_deltavgamma}. The amplitude ratios exhibit large scatter. However, no clear trends emerge.}
    \label{fig:GDR3_deltaAp2p}
\end{figure*}

We considered the parameters \texttt{average\_rv} and \texttt{peak\_to\_peak\_rv} from table \texttt{gaiadr3.vari\_cepheid} to investigate the impact of template parameters on the offset between \gaia\ and \veloce. The spectral template parameters \texttt{rv\_template\_teff}, \texttt{rv\_template\_logg}, and \texttt{rv\_template\_fe\_h} were taken from table \texttt{gaiadr3.source\_id}, and are available for the majority of stars with \gaia\ RV time series measurements or \texttt{average\_rv} values, but not all (exceptions include RW~Cas, SS~CMa, and BM~Per). We considered the range $5000 < T_{\rm eff} \mathrm{[K]} < 6500$, $\log{g} \le 2.0$, and $-0.75 \le {\rm [Fe/H]} \le 0.75$ to be particularly adequate for Cepheids and distinguish \gaia\ results based on this range in the following as orange points. 

Figure\,\ref{fig:GDR3_deltavgamma} shows $\Delta$\vgGaiaCat $ = $ \vgGaiaCat $- v_{\gamma,\veloce}$ against the correlation template parameters for $71$ stars, $33$ of which were processed using the most adequate range of template parameters. Trends can be seen among all three template parameters, although their relative contributions are not immediately distinguished. The most obvious trend in Fig.\,\ref{fig:GDR3_deltavgamma} is that \vgGaiaCat\ is much more consistent with \veloce\ when the adequate template parameter range is considered. In this case, just over $\gtrsim 10$ measurements suffice to find a similar result ($\pm 2\,$\kms). When the full parameter range is considered, a larger number of observations is required. Figure\,\ref{fig:GDR3_deltavgamma} also suggests that the automated identification of the appropriate parameter range works best when a larger number of observations are available ($\gtrsim 25$).

Figure\,\ref{fig:GDR3_deltaAp2p} shows the ratio between the peak-to-peak RV amplitudes from \gaia\ and  \veloce\ as a function of the \gaia\ correlation template parameters and $N_{\mathrm{obs,GDR3}}$. Interestingly, neither the number of observations, nor the choice of template parameters appears to significantly affect this comparison. We did not find significant evidence of biased amplitudes, although there is a noticeable dispersion of $\sim 16\%$ between the amplitudes. For the time being, it remained unclear whether this is a physical effect related to the amplitudes of near-IR lines, or  the precision of \gaia\ peak-to-peak amplitudes, since the number of harmonics used to fit \gaia\ RV curves was likely low ($2-3$; G. Clementini, priv. comm.) and is not reported in the \texttt{vari\_cepheids} table. Larger number of observations in the next \gaia\ DR4 will allow for a more detailed comparison of amplitudes. In the meantime, we caution that inferences sensitive to RV amplitudes, such as BW-type distances or $p-$factor calibrations, will be limited by the dispersion seen in Fig.\ref{fig:GDR3_deltaAp2p}.

\begin{figure*}
    \centering
    \includegraphics[width=\textwidth]{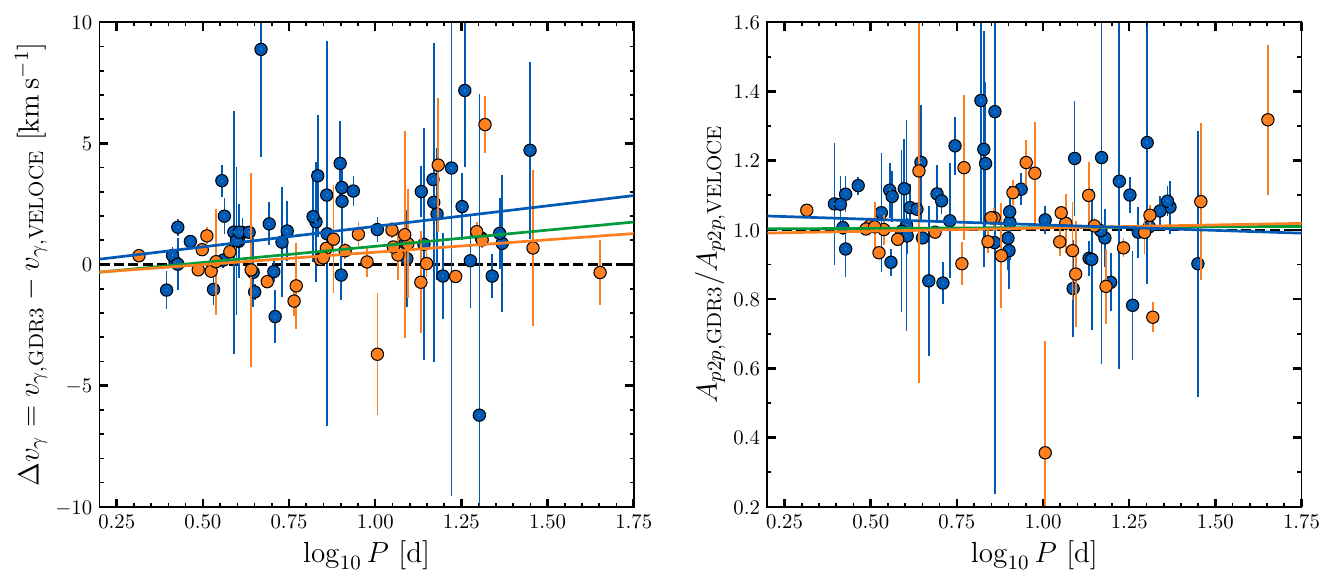}
    \caption{\vgamma\ (left) and peak-to-peak RV amplitude (right) determined by \gaia\ SOS compared to \veloce\ as a function of logarithmic pulsation period. \vgGaiaCat\ and $A_{\rm p2p,GDR3}$ are the values published in table \texttt{gaiadr3.vari\_cepheid} as parameters \texttt{average\_rv} and \texttt{peak\_to\_peak\_rv}, respectively. Orange points highlight Cepheids for which \gaia\ used the most adequate template parameters. Blue points show measurements processed using template parameters outside this range. Blue and orange solid lines show linear fits to the respective points; green solid lines show linear fits to the combined data sets. At \logP$= 1$, we find $\Delta v_\gamma = 0.51 \pm 0.13\,$\kms, and a slight trend of $1.03\pm0.29$\,\kms/\logP\ for the orange points. The blue points exhibit a much more significant offset of $1.6\pm 0.2\,$\kms\ and trend $1.69\pm0.44$\,\kms$/\log{P}$. Peak-to-peak amplitudes are generally consistent with a dispersion of approximately $16\%$ and do not exhibit a trend with \logP.}
    \label{fig:GDR3-VELOCE}
\end{figure*}

Figure\,\ref{fig:GDR3-VELOCE} illustrates the differences shown in Fig.\,\ref{fig:GDR3_deltavgamma} and \ref{fig:GDR3_deltaAp2p} as a function of \logP. At a pivot period of \Ppuls$=10$\,d (\logP$=1$), the median offset if $0.51\pm0.13$\,\kms, consistent with the offset determined using template fitting. Additionally, a trend of $1.03 \pm 0.29$\,\kms$/\log{P}$ ($3.5\sigma$) trend is found. A larger offset of $1.6\pm0.2$\,\kms\ and steeper trend of $1.69\pm0.44$\,\kms$/\log{P}$ ($3.8\sigma$) is found among the stars whose template parameters were outside the most adequate range (blue points). As seen in Fig.\,\ref{fig:GDR3_deltaAp2p}, the amplitude ratio appears very stable across the full range of periods. We tentatively note better agreement for short-period Cepheids processed with the adequate template parameters (\logP$ \lesssim 0.7$), although a cleaner comparison would required a larger number of stars.

\section{Summary and conclusions}\label{sec:conclusions}

We present the first data release (DR1) of the \veloce\ project, which comprises more than \nobscep\ high-precision radial velocity measurements of \ncep\ classical Cepheids located on both hemispheres, as well as \nobsnoncep\ observations of \nnoncep\ stars that are not classical Cepheids. This data set provides an unprecedented combination of RV precision (median uncertainty of \medianerror\,\ms), sample size, and temporal baselines of up to $11$\,yr featuring well-characterized RV zero-points based on RV standard stars. \veloce\ extends the number of Cepheid RV time series at magnitudes fainter than $G \gtrsim 10$\,mag, where previous catalogs contained relatively few observations. Systematic zero-point changes following instrumental upgrades ($\sim 26$\,\ms\ on \coralie, $< 75$\,\ms\ on \hermes) have been characterized. 

We performed Fourier decomposition of RV curves with an emphasis on model selection and to determine ephemerides from the RV data directly. Contrary to common practice, we define $\phi = 0$ at minimum radius, since this phase is most precisely measurable. The fitted Fourier series models provide the most detailed view of the \citet{HertzsprungProgression} progression of MW Cepheids to date and revealed a second bump feature on the descending RV branch of approximately half of the observed Cepheids with periods between $9 - 18$\,d. These double bumps are likely to further inform mode resonances \citep{Simon+1981,Buchler1990,Antonello+1996}, and further investigation of this feature is required to understand their potential for informing the evolution and pulsation physics of Cepheids.

Approximately $30\%$ of the Cepheids studied here exhibit changes in the pulsation-averaged velocity, $v_\gamma$, due to orbital motion. A systematic study of spectroscopic binaries is presented in paper~II of this series (Shetye et al., in prep.).

\nmodulatorstr\ Cepheids were found to exhibit various forms of modulated variability, that is, variability signals beyond the dominant pulsation mode or orbital motion (not counting double-mode Cepheids). The dichotomy of long-timescale modulations among overtone Cepheids and cycle-to-cycle variations for long-period Cepheids reported by \citet{Anderson2014rv} is confirmed. Modulated variability becomes ubiquitous for given sufficient RV precision ($20-40$\,\ms\ per epoch), and the diversity of phenomena suggests that multiple astrophysical mechanisms are at play. High-precision RV observations of Cepheids thus hold great promise to cast open an asteroseismic window for studying this advanced evolutionary stage of intermediate-mass stars in unprecedented detail. Future work will systematically study this ``modulation zoo,'' starting with the discovery of non-radial pulsation in Cepheid time series spectra \citep{Netzel2024}.  

We investigated the linear radius variations and found striking differences in the RV curve integrals ($\Delta R/p$) between Cepheids with \Ppuls\ above and below $10$\,d. We noted at least two, possibly three, distinct sequences of long-period Cepheids ($P > 10$\,d) in the $\Delta R/p$ vs. \logP\ diagram, which we conjecture are at the origin of the difficulties encountered when calibrating BW-type projection factors. Although no obvious physical explanation for these sequences was identified, we noticed mild preference for the sample of $11$ Cepheids exhibiting the smallest linear radius variations (the low-$\Delta R/p$ sequence) to reside not far from the blue edge of the instability strip (Fig.\,\ref{fig:deltaR_CMD}). 

The \veloce\ project provides a legacy dataset of high-quality Cepheid RVs that will enable further study and allow for  future datasets to be combined with existing ones from the literature. All RV time series measurements and the results from Fourier series modeling of single-mode bona fide Cepheids are made publicly available in this data release (DR1). To facilitate accurate comparisons with the literature, we have determined RV zero-point offsets relative to the \veloce\ zero-point for 31 literature sources using a template fitting approach. This also allowed us to measure secular (linear) period changes of \npdotsig\ Cepheids based on RV measurements alone. We found good agreement with photometrically measured linear period changes from the literature as well as stellar model predictions.

Comparing our template fitting results to recently published \gaia\ DR3 time series measurements, we find an average offset from \veloce\ of $0.65\pm0.12$\,\kms. Systematic offsets of a few to several \kms\ were found for stars whose RVs were measured using inadequate template parameters. The published RV amplitudes of Cepheids in \gaia\ DR3 exhibit significant ($16\%$) scatter around the closely mapped \veloce\ amplitudes.  As \gaia\ continues its unprecedented survey of the MW, the number of published RV time series measurements are set to skyrocket. To ensure consistency among the measurements for specific variable star classes, we recommend using predefined sets of correlation template parameters for classes of large-amplitude variables, such as Cepheids. The accuracy of RV peak-to-peak amplitudes is likely to improve as more epochs are being collected. Ensuring the consistency of RV amplitudes measured by \gaia's RVS instrument with angular diameter variations will be paramount to achieving accurate BW projection factor calibrations based on hundreds, if not thousands, of Cepheids.

\begin{acknowledgements} 
This work would not have been possible without a great deal of help from a large number of people. Useful with discussions with the following people are acknowledged (in alphabetical order): Gisella Clementini, Xavier Dumusque, Nancy R. Evans, Michel Grenon, Martin Groenewegen, Pierre Kervella, Christophe Lovis, Antoine M\'erand, Francesco Pepe, Vincenzo Ripepi, Damien S\'egransan, Jesper Storm, Laszlo Szabados, St\'ephane Udry.
We especially thank the support staff at Euler, Mercator, La Silla Observatory, the Observatorio del Roque de los Muchachos, at the University of Geneva, KU~Leuven, and ESO for their excellence and dedication to operating and maintaining these facilities. This notably includes Luc J.M. Weber, Ren\'e Debusson, Michel Fleury, Michel Crausaz, Gregory Lambert, Vincent and Denis M\'egevand, Gaspare Lo Curto, Emmanuela Pompei, Jesus Dalgado, Florian Merges, among others. 
We also warmly thank the following people for contributing observations on Euler and/or Mercator: Maxime Spano, Pierre Dubath, Thierry Semaan, Maria S\"uveges, Mihaly Varadi, Janis Hagelberg, Nicolas Cantale, Sarah Gebruers, Maddalena Bugatti, Amaury~M.~J. Triaud, Abigail Frost, Marion Cointepas, Monika Lendl, Dominique Naef, Matthias Fabry, S\'ebastien Peretti, Uriel Conod, Julia Seidel, Siemen Burssens, Raine Karjalainen, Azzurra Bigioli, Federica Cersullo, Ga\:el Ottoni, Thibault Merle, Olivier Verhamme, Akke Corporaal, Bruno Chazelas, Fatemeh Motalebi, Javiera Rey Cerda, Rosemary Mardling, Nicolas Lodieu, Tinne Pauwels, Lea Planquart, Steven Bloemen, Beatriz Gonzalez Merino, Oliver Turner, and Sara Mancino, among others.
While we sought to include everyone who contributed, we are likely to have forgotten some, and we apologize to those we failed to mention. Your contributions were nonetheless valuable!\\

We thank the anonymous referee for a constructive report that helped to improve the manuscript.

This research has received support from the European Research Council (ERC) under the European Union's Horizon 2020 research and innovation programme (Grant Agreement No. 947660). RIA and SK are funded by the SNSF Swiss National Science Foundation Eccellenza Professorial Fellowship (award PCEFP2\_194638).
RIA acknowledges funding and support from several sources over the 11 year duration of the project, including the Swiss National Science Foundation early Postdoc.Mobility fellowship (grant No. 155687) and the ESO fellowship program in Garching. 
AT is a Research Associate at the Belgian Scientific Research Fund (F.R.S.-F.N.R.S.). 
MP is supported by BEKKER fellowship BPN/BEK/2022/1/00106 from the Polish National Agency for Academic Exchange.
The Euler telescope is funded by the Swiss National Science Foundation (SNSF). Early \veloce\ observations ($2010-2016$) were enabled by SNSF project funding from grant Nos. 119778, 130295, and 140893.
MM acknowledges the support of the Swiss National Science Foundation (SNSF) under grant P500PT\_203114.
This research is based on observations made with the Mercator Telescope, operated on the island of La Palma by the Flemish Community, at the Spanish Observatorio del Roque de los Muchachos of the Instituto de Astrof\'isica de Canarias. {\it Hermes} is supported by the Fund for Scientific Research of Flanders (FWO), Belgium, the Research Council of K.U. Leuven, Belgium, the Fonds National de la Recherche Scientifique (F.R.S.-FNRS), Belgium, the Royal Observatory of Belgium, the Observatoire de Gene\`eve, Switzerland, and the Th\"uringer Landessternwarte, Tautenburg, Germany. \\

This research was supported by the International Space Science Institute (ISSI) in Bern, through ISSI International Team project \#490, SHoT: The Stellar Path to the Ho Tension in the Gaia, TESS, LSST and JWST Era. This research has made use of NASA's Astrophysics Data System; the SIMBAD database and the VizieR catalog access tool provided by CDS, Strasbourg; Astropy, a community-developed core Python package for Astronomy \citep{2013A&A...558A..33A,astropy:2018}.

\end{acknowledgements}

\bibliographystyle{aa} 
\bibliography{biblio}

\begin{appendix}
\section{Classical Cepheids in \veloce \label{app:sampletable}}
Here, we report the list of classical Cepheids in this first \veloce\ data release. Table\,\ref{tab:Cepheidcatalog} lists identifiers, coordinates, \gaia\ DR3 information (source ids, mean $G-$band magnitudes, mean $G_{Bp}-G_{Rp}$ color, and pulsation modes), whether the star exhibits orbital motion based on \veloce\ data alone (not considering literature information, which is included in paper~II), the number of \veloce\ observations, the number of Fourier harmonics used in the modeling, \nfs, the pulsation period \Ppuls, its uncertainty (squared sum of statistical and systematic error), and the epoch of minimum radius $E$ with its uncertainty (squared sum of statistical and systematic error). All information presented in Tab.\,\ref{tab:Cepheidcatalog} is contained within the data files provided via Zenodo in FITS format, cf. App.\,\ref{app:datastructure}.

Additionally, Tab.\,\ref{app:table:RVglobal} lists information from the Fourier modeling, including the degree of the polynomial used to determine \Ppuls\ and $E$, the number of degrees of freedom of the fit, $N_{\rm DOF}$, the $\chi^2$ of the fit (often $\gg N_{\rm DOF}$ due to unmodeled signals, cf. Sect.\,\ref{sec:modulations}), the pulsation-averaged velocity $v_\gamma$, peak-to-peak RV amplitude, the first Fourier harmonic's amplitude and phase, as well as the Fourier amplitude ratios and Fourier phase differences for higher harmonics relative to the first harmonic. The full Fourier models for every star are provided in the data tables described in App.\,\ref{app:datastructure}. Tables\,\ref{tab:Cepheidcatalog} and \ref{app:table:RVglobal} are made available in machine-readable form via the CDS.

\subsection{Corner plots of Fourier parameters\label{app:FourierParameters}}
\begin{landscape}
\begin{figure}
    \vspace{2cm}
    \includegraphics[scale=0.65]{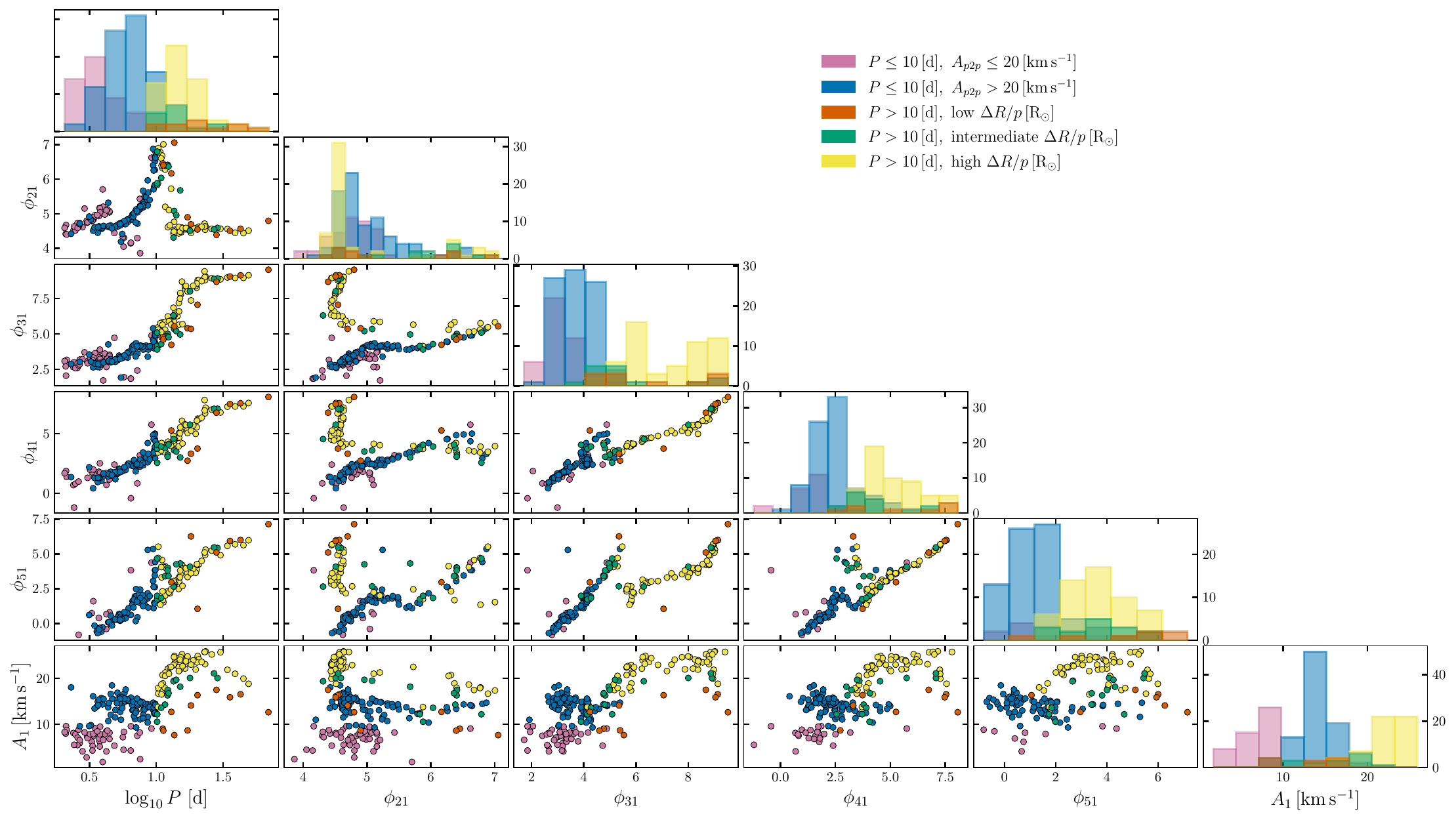}
    \caption{Fourier phase differences and the amplitude of the first harmonic shown against one another up to the 5th harmonic. The color scheme adopted in Sect.\,\ref{sec:deltaR} is applied to identify specific groups.\label{fig:corner_dphi}}
\end{figure}     
\end{landscape}

\begin{landscape}
\begin{figure}
    \vspace{2cm}
    \includegraphics[scale=0.75]{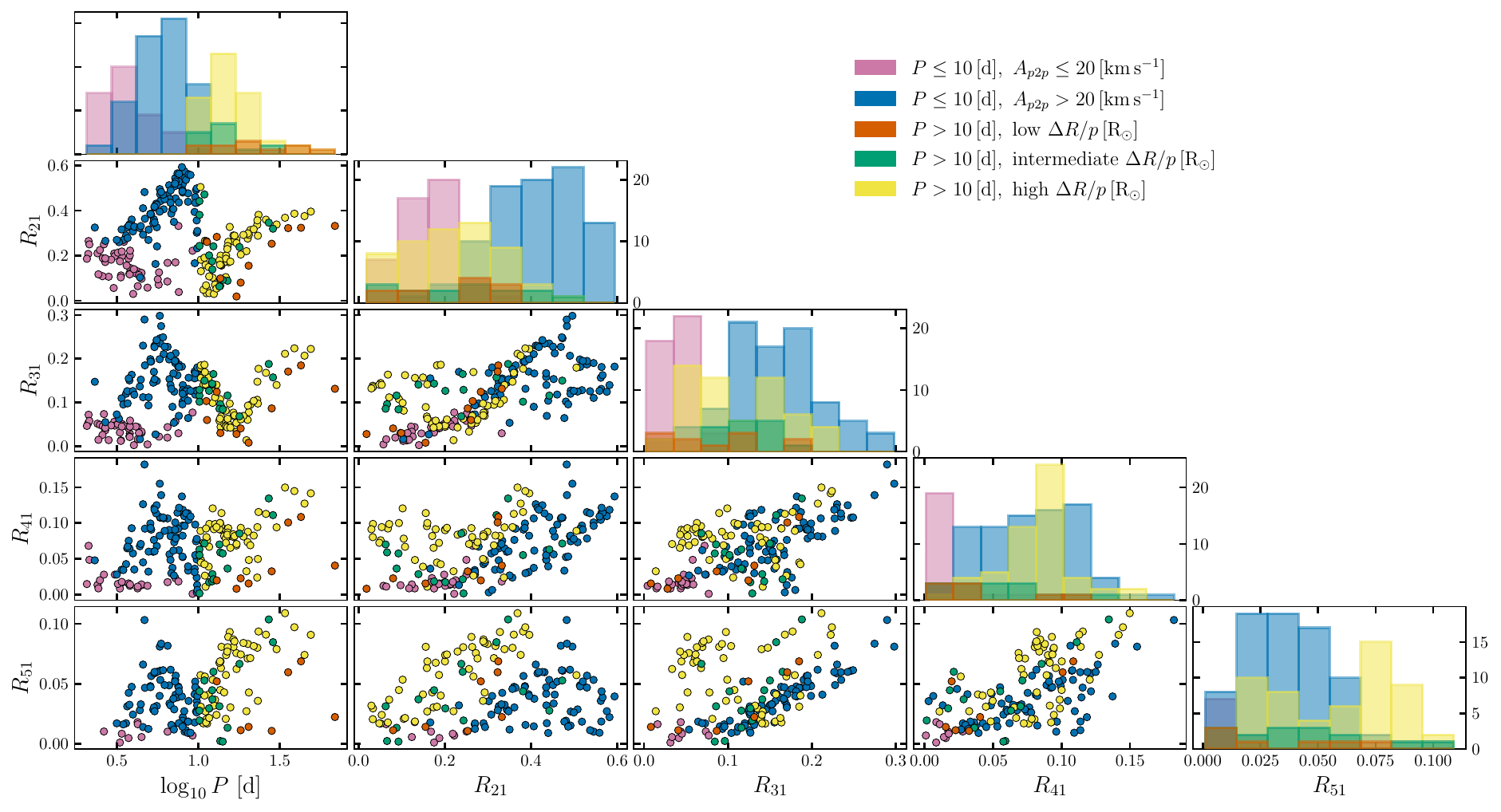}
    \caption{Fourier amplitude ratios shown against one another up to the 5th harmonic. The color scheme adopted in Sect.\,\ref{sec:deltaR} is applied to identify specific groups.\label{fig:corner_R}}
\end{figure}     
\end{landscape}

Figures\,\ref{fig:corner_dphi} and \ref{fig:corner_R} illustrate the Fourier phase differences and amplitude ratios up to the 5th harmonic for all combinations\footnote{These corner plots were inspired by \citet{Taylor2024}}. While we here do not plot uncertainties for clarity, we note that uncertainties of the Fourier parameters can be obtained from the FITS files delivered as part of this data release. The color scheme applied follows Sect.\,\ref{sec:deltaR} and is intended to help identify the specific groups of stars discussed therein. A few observations can be made at this time:
\begin{enumerate}
    \item At $P \le 10$\,d, the trend between amplitude ratio $R_{i1}$ and \logP\ increasingly disappears towards higher harmonic $i=2,3,4$. By $R_{41}$, a previously clean trend with \logP\ turns into a rather continuous distribution. 
    \item At $P \le 10$\,d, the range of $R_{21}$ clearly exceeds that of long-P Cepheids. Amplitude ratios of higher harmonics, e.g., $R_{41}$, are more similar among long-and short-period Cepheids.
    \item At $P > 10$\,d the trends of phase differences $\phi_{21}$ and $\phi_{31}$ vs \logP\ in Fig.\,\ref{fig:corner_dphi} appear ``delayed'' (follow the same trend at higher \logP) for orange and green points compared to yellow points. 
    \item At $P > 10$\,d, the orange points identified in Sect.\,\ref{sec:deltaR} have consistently lower $A_1$ than other long-period Cepheids. The amplitudes of their higher harmonics (2,3,4) are also lower than those of other long-period Cepheids, cf. left column of Fig.\,\ref{fig:corner_R}. Apparently, all harmonics contribute to the overall smaller peak-to-peak amplitude seen in low-$\Delta R/p$ stars.
    \item Low-amplitude short-period (purple) and low-$\Delta R/p$ long-period (orange) Cepheids occupy the same region in the plot of $R_{41}$ vs $R_{21}$ in Fig.\,\ref{fig:corner_R}, indicating similar ratio of $A_4/A_2$ for these stars. 
    \item High-amplitude $P>10$\,d Cepheids (yellow) feature higher values of $R_{41}$ and $R_{51}$ across the same range of $R_{21}$. 
\end{enumerate}


\onecolumn
\begin{landscape}
\begin{tiny}
\begin{longtable}{@{}lrrrrrccrrrrrr@{}}
\caption{Bona fide classical Cepheids observed as part of the \veloce\ project (excerpt)\label{tab:Cepheidcatalog}}\\
\toprule
Name & RA (J2000) & DEC (J2000) & GDR3 source\_id & $G$ & $G_{\rm Bp} - G_{\rm Rp}$ & Mode & SB1 & $N_{\mathrm{obs}}$ & $N_{\mathrm{FS}}$ & $P$ & $\sigma_{P}$ & $E$ & $\sigma_{E}$\\
 & (h:m:s) & (d:am:as) &  & (mag) & (mag) &  &  &  &  & (d) & (d) & (d) & (d)\\
\midrule
\endfirsthead
\multicolumn{4}{l}{\tablename\ \thetable\ -- \textit{Continued from previous page}} \\
\toprule
Name & RA (J2000) & DEC (J2000) & GDR3 source\_id & $G$ & $G_{\rm Bp} - G_{\rm Rp}$ & Mode & SB1 & $N_{\mathrm{obs}}$ & $N_{\mathrm{FS}}$ & $P$ & $\sigma_{P}$ & $E$ & $\sigma_{E}$\\
 & (h:m:s) & (d:am:as) &  & (mag) & (mag) &  &  &  &  & (d) & (d) & (d) & (d)\\
\midrule
\endhead
\hline \multicolumn{14}{r}{\textit{Continued on next page}} \\
\endfoot
\bottomrule
\endlastfoot
 AA Gem              & 06:06:34        & +26:19:45        & 3430067092837622272 & 9.4            & 1.4 & FU     & F     & 68                   & 12                  & 11.303705       & 1.9e-05                  & 58011.537774             & 6.0e-09                  \\
 AB Cam              & 03:46:08        & +58:47:02        & 473239154746762112  & 11.4           & 1.7 & FU     & F     & 28                   & 6                   & 5.787585        & 6.4e-05                  & 57060.317069             & 4.1e-10                  \\
 AC Mon              & 07:00:59        & -08:42:32        & 3050050207554658048 & 9.6            & 1.6 & FU     & F     & 61                   & 6                   & 8.014995        & 4.8e-06                  & 56894.022100             & 2.1e-09                  \\
 AD Pup              & 07:48:03        & -25:34:40        & 5614312705966204288 & 9.5            & 1.4 & FU     & T     & 125                  & 13                  & 13.597599       & 1.3e-05                  & 58269.802393             & 2.4e-12                  \\
 AH Vel              & 08:11:59        & -46:38:39        & 5519380081746387328 & 5.6            & 0.8 & FO     & T     & 70                   & 5                   & 4.227158        & 5.6e-06                  & 58326.793961             & 6.3e-09                  \\
 AK Cep              & 22:28:50        & +58:12:39        & 2007997408188526336 & 10.6           & 1.8 & FU     & F     & 18                   & 5                   & 7.232608        & 5.3e-03                  & 57417.918474             & 8.1e-09                  \\
 AN Aur              & 04:59:41        & +40:50:09        & 201574982848108416  & 9.9            & 1.6 & FU     & F     & 40                   & 7                   & 10.288512       & 1.1e-04                  & 58077.184797             & 5.1e-09                  \\
\multicolumn{14}{c}{\ldots}\\
\bottomrule

\end{longtable}
\tablefoot{
$G$ lists the intensity-averaged \gaia\ DR3 $G-$band magnitude (\texttt{\text{int\_average\_g}}) from table \texttt{\text{gaiadr3.vari\_cepheids}} unless preceded by a $\dagger$, in which case \texttt{\text{phot\_g\_mean\_mag}} from table \texttt{\text{gaiadr3.source}} was adopted because these stars were not included in the SOS processing \citep{gaiadr3.cepheid}. This was the case for ASAS~J082710-3825.9, BB~Gem, BB~Sgr, FM~Aql, RT~Aur, SZ~Aql, T~Mon, TT~Aql, TZ~Mon, U~Sgr, $\beta$~Dor, and $\zeta$~Gem. 
Analogously, $G_{Bp} - G_{Rp}$ is the intensity averaged \gaia\ DR3 color obtained from \texttt{\text{gaiadr3.vari\_cepheids}}, or from  \texttt{\text{gaiadr3.gaia\_source}}. 
Column `Mode' lists pulsation mode assignments according to \texttt{\text{gaiadr3.vari\_cepheids}}. Column `SB1' is `T' for `True' if the star exhibits clear signs of orbital motion either over the \veloce\ baseline or in the template fitting analysis including literature data, `U' for `Unsure' if the signal is tentative, and `F' for `False' otherwise. Note that very long-period binaries may thus be labeled as `F'. Columns $N_{\rm{obs}}$ and $N_{\rm{FS}}$ list the number of \veloce\ observations and the number of Fourier harmonics used to represent the pulsational RV variability. Columns $P$ and $\sigma_P$ specify the pulsation period and its total uncertainty. Analogously for the epoch $E$ and $\sigma_E$, which is defined so that pulsation phase $phi=0$ at $v_\gamma$ on the descending RV branch (minimum radius), cf. Sect.\,\ref{sec:Fourier}. 
Stars for which no models were fitted for various reasons are listed separately as labeled. Time-series RV measurements are published as part of this data release, even if no Fourier models were fitted. The full table is available at the CDS.
}
\end{tiny}
\end{landscape}
\twocolumn

\onecolumn
\begin{landscape}
\begin{tiny}
\begin{longtable}{@{}lrrrrrrrrrrrrrrr@{}}
\caption{Fourier parameters of single-mode Cepheids  (excerpt)\label{app:table:RVglobal}}\\
\toprule
Name & poly & N$_{\rm{DOF}}$ & $\chi^2$ & $v_\gamma$ & $A_{\rm{p2p}}$ & $A_1$ & $\phi_1$ & $R_{21}$ & $\Phi_{21}$ & $R_{31}$ & $\Phi_{31}$ & $R_{41}$ & $\Phi_{41}$ & $R_{51}$ & $\Phi_{51}$ \\
 & & & & [\kms] & [\kms] & [\kms] & [rad] &  & [rad] &  & [rad] &  & [rad] & & [rad]  \\

\midrule
\endfirsthead
\multicolumn{16}{l}{\tablename\ \thetable\ -- \textit{Continued from previous page}} \\
\toprule
Name & poly & N$_{\rm{DOF}}$ & $\chi^2$ & $v_\gamma$ & $A_{\rm{p2p}}$ & $A_1$ & $\phi_1$ & $R_{21}$ & $\Phi_{21}$ & $R_{31}$ & $\Phi_{31}$ & $R_{41}$ & $\Phi_{41}$ & $R_{51}$ & $\Phi_{51}$ \\
 & & & & [\kms] & [\kms] & [\kms] & [rad] &  & [rad] &  & [rad] &  & [rad] & & [rad] \\
\midrule
\endhead
\hline \multicolumn{16}{r}{\textit{Continued on next page}} \\
\endfoot
\bottomrule
\endlastfoot
 AA Gem              & -      & 42          & 137.5      & 19.608                        & 39.880                       & 17.355                   & -1.317              & 0.0511                      & 0.72                   & 0.1444                      & 5.81                   & 0.0759                      & 3.94                   & 0.0170                      & 1.54                   \\
 AB Cam              & -      & 14          & 999.7      & -43.980                       & 47.204                       & 16.814                   & -1.524              & 0.4957                      & 4.71                   & 0.2981                      & 3.37                   & 0.1551                      & 1.93                   & 0.0810                      & 0.38                   \\
 AC Mon              & -      & 47          & 1420.7     & 60.494                        & 38.814                       & 14.542                   & -1.355              & 0.5818                      & 5.32                   & 0.1397                      & 3.95                   & 0.1185                      & 2.80                   & 0.0223                      & 1.83                   \\
 AD Pup              & 4      & 93          & 2359.3     & 72.908                        & 52.429                       & 22.918                   & -1.337              & 0.0936                      & 4.77                   & 0.1075                      & 5.82                   & 0.1000                      & 4.16                   & 0.0449                      & 2.67                   \\
 AH Vel              & 1      & 57          & 17218.1    & 26.815                        & 16.852                       & 8.467                    & -1.544              & 0.1301                      & 5.04                   & 0.0436                      & 3.23                   & 0.0085                      & 1.69                   & 0.0181                      & 5.89                   \\
 AK Cep              & -      & 6           & 143.5      & -53.950                       & 36.077                       & 14.119                   & -1.517              & 0.4548                      & 4.72                   & 0.2124                      & 3.60                   & 0.0810                      & 2.18                   & 0.0410                      & 0.45                   \\
 AN Aur              & -      & 24          & 205.6      & 4.094                         & 33.243                       & 13.759                   & -1.230              & 0.5051                      & 5.88                   & 0.1796                      & 4.14                   & 0.0890                      & 3.70                   & 0.0573                      & 1.74                   \\
\multicolumn{16}{c}{\ldots} \\
\bottomrule
\end{longtable}
\tablefoot{Fourier parameters were determined as explained in Sect.\,\ref{sec:Fourier}, and Fourier ratios are computed as described in Sect.\,\ref{sec:FourierRatios}. Stars exhibiting long-term trends of $v_\gamma$ were modeled using polynomials, and column `poly' states its degree.
$N_{\rm{DOF}} = N_{\rm{obs}} - 1 - N_{\rm{model}}$ states the number of degrees of freedom of the fit, i.e., with $N_{\rm{model}} = 1 + 2\cdot N_{\rm{FS}} + {\rm poly}$ the number of model parameters.
$\chi^2$ has the usual meaning; high values of $\chi^2$ indicate unmodeled signals, cf. Sect.\,\ref{sec:modulations}. 
$v_\gamma$ is the constant term of the fitted Fourier series, cf. Eqs.\,\ref{eq:Fseries} and \ref{eq:polynomial}. Note that $v_\gamma$ does not correspond to the barycentric velocity for SB1 Cepheids. 
$A_{\rm{p2p}}$ is the peak-to-peak RV curve amplitude determined from the Fourier fit.
$A_1$ and $\phi_1$ are the semi-amplitude and Fourier phase of the first harmonic, respectively. 
$R_{i1}$ and $\Phi_{i1}$ are the amplitudes and phase ratios of the $i-$th harmonic relative to the first \citep{Simon+1981}. 
Uncertainties for all parameters are provided in the data files as explained in App.\,\ref{app:datastructure}. The full table is available via the CDS.
}
\end{tiny}
\end{landscape}
\twocolumn

\onecolumn
\twocolumn 

\onecolumn
\twocolumn

\section{Stars other than classical Cepheids\label{sec:app:classification}}

In an early phase (mostly between $2011-2012$), \veloce\ targeted candidate type-I Cepheids reported by different photometric surveys, including ASAS \citep{Pojmanski2002,Pojmanski2003,Pojmanski2004,Pojmanski2005,Pojmanski2005a}, NSVS \citep{NSVS}, ROTSE \citep{ROTSE}, as well as other stars previously classified as Cepheids in order to determine cluster membership \citep{Anderson2013}. A large number of these candidate Cepheids were found not to be Cepheids based on the observed CCFs or, since then, by other variability classification studies \citep[e.g.,][]{Johnson2016misclassifiedcepheids}. Additionally, a few RR Lyrae stars and type-II Cepheids were observed to fill time during observing runs. 

CCFs provide sensitive checks of variable star classifications based on four observational features that allow us to identify bona fide classical Cepheids with only a few observations \citep[Sect.\,2.4]{AndersonPhD} thanks to the enhanced S/N compared to ordinary spectra and thanks to spectroscopic features. Firstly, classical Cepheids produce noticeable correlation peaks at any phase when cross-correlated with a G2 correlation mask, since their spectral types vary between mid F to early K, cf. Fig.\,\ref{fig:CCFs}. Stars that fail to produce noticeable correlation peaks are thus outside the spectral type range of interest. Cepheid CCFs computed using the G2 masks do not usually exhibit strong rotational broadening, virtually always exhibit a single correlation peak, and virtually never exhibit emission features. However, Cepheid CCFs do generally exhibit phase-dependent displacement in velocity space, phase-dependent line broadening, phase-dependent line asymmetry, and phase-dependent line depth; these parameter are traced by RV, FWHM, BIS, and CCF contrast. Hence, stars that do not exhibit all four types of variations are very likely not classical Cepheids. 
\citet{Johnson2016misclassifiedcepheids}.

We have visually inspected the CCFs and their measured parameters of all stars observed for \veloce. Table\,\ref{tab:noncep} lists the \nnoncep\ stars that are not classical Cepheids that have been observed as part of \veloce, including \nrrlyrae\ RR Lyrae stars, \ntypeIIcep\ type-II Cepheids, and \nsbIIsbIII\ (candidate) multi-lined spectroscopic binaries. The table provides qualitative notes on whether the CCF showed a significant  correlation peak, whether this peak appeared atypical for Cepheids, whether multiple components were apparent, whether the CCF showed signs of emission features, and whether RV, FWHM, BIS, and Contrast exhibited variability. We also included a brief note concerning the CCF in the hopes that this will be useful in the future. Finally, the table includes the \gaia\ DR3 source identifier and classifiers assigned by the \gaia\ DR3 variability analysis \citep{gaiadr3.vari} and by a literature meta-study of variable stars \citep{Gavras2022}. A detailed reclassification of the stars listed in Tab.\,\ref{tab:noncep} is out of scope of this work. However, we consider that stars can be effectively shown not to be classical Cepheids if they a) exhibit no CCF peaks ($N_{RV}$), b) exhibit multiple peaks or emission features, or c) do not exhibit all four types of variability despite having been observed more than $4-5$ times. It is thus possible that a few classical Cepheids remain in Tab.\,\ref{tab:noncep} in case there were only one or two good observations available. 

The RV measurements of the stars in Tab.\,\ref{tab:noncep} is published as part of this data release. However, we stress that these RVs have not been vetted, that is, no further quality control was applied. In particular, RVs of SB2 systems need to be considered with caution because it is not always clear that the same CCF peak was measured. Further information on these stars will be published as part of the second data release and is available upon request to the lead author.

\onecolumn
\begin{landscape}
\begin{tiny}
\begin{longtable}{@{}l @{}c @{}c @{}c @{}c @{}c @{}c @{}c @{}c @{}clrll@{}}
\caption{Stars previously misclassified as Cepheid (candidates) observed as part of \veloce\ and later found out to not be classical Cepheids (excerpt) \label{tab:SB2}\label{tab:noncep}}  \\

\toprule
Name  &  N$_{\rm{RV}}$  &  Peak  &  odd?  &  N$_{\rm comp}$  &  Emi?  &  vRV  &  vFWHM  &  vAsy  &  vCont  &  CCF note  &  \gaia\ DR3 source\_id  &  GDR3 class  &  G22  \\
\hline
\endfirsthead
\multicolumn{4}{c}{\tablename\ \thetable\ -- \textit{Continued from previous page}} \\
\hline
\toprule
Name  &  N$_{\rm{RV}}$  &  Peak  &  odd?  &  N$_{\rm comp}$  &  Emi?  &  vRV  &  vFWHM  &  vAsy  &  vCont  &  CCF note  &  \gaia\ DR3 source\_id  &  GDR3 class  &  G22  \\
\hline
\endhead
\hline \multicolumn{4}{r}{\textit{Continued on next page}} \\
\endfoot
\bottomrule
\endlastfoot

\multicolumn{14}{c}{RR~Lyrae stars} \\
\midrule
ASAS J165126-2434.9  &  10  &  y  &  n  &    &    &  y  &  y  &  y  &  y  &  &  6046836528519375232  &  RR  &  RRAB   \\
GP Aqr  &  2  &  y  &  y  &    &    &  y  &  y  &  y  &  y  &  Noisy CCF &  2622261225664677120  &  RR  &  EA   \\
RR Lyr  &  146  &  y  &    &    &    &  y  &  y  &  y  &  y  &  RRAB  &  2125982599343482624  &  RR  &  RRAB   \\
V1319 Sco  &  19  &  y  &  y  &    &    &  y  &  y  &  y  &  y  &  Low contrast variable CCF and noisy  &  6246499215814759936  &  RR  &  RRC   \\
\midrule
\multicolumn{14}{c}{Type-II Cepheids} \\
\midrule
AM Vel  &  6   &  y  &  n & &  & y & y & y & y & & 5521425379537360896 & T2CEP & DCEP \\ 
AL Vir  &  15  &  y  &  n  &    &    &  y  &  y  &  y  &  y  &    &  6303152720661307648  &  CEP  &  CW   \\
\multicolumn{14}{c}{\ldots}\\
\bottomrule

\end{longtable}
\tablefoot{$N_{\rm RV}$ lists the number of RV measurements reported here; this number is $0$ if no observation produced a CCF peak. Peak specifies whether at least one observation produced a CCF peak. `odd?' specifies whether the CCFs looked significantly different than those of classical Cepheids. $N_{\rm{comp}}$ specifies if more than a single absorption component was noticed, `Emi' whether any observation exhibited signs of emission features. `vRV', `vFWHM', `vAsy' and `vCont' specify whether the object is variable in RV, line width, line asymmetry, and line depth, respectively. Column `CCF note' provides qualitative information. \gaia\ DR3 source IDs are provided for cross-matching. 
`GDR3 class' lists the best classifiers from \gaia\ DR3 classifier\_best\_class\_name, where `DSCT++' abbreviates DSCT|GDOR|SXPHE, and ``ACV++'' abbreviates ACV|CP|MCP|ROAM|ROAP|SXARI. Column `G22' lists the primary\_var\_type from \citet{Gavras2022}. `y' indicates `yes', `n' means `no', `NA' indicates `not applicable' (typically because only one observation was available). `WVir?' indicates a tentative signature of emission in the blue part of the CCF analogous to what is seen in W~Vir, cf. Fig\,\ref{fig:CCFs}. Question marks indicate tentative assessments. The full table is available via the CDS.
}
\end{tiny}
\end{landscape}
\twocolumn


\section{\veloce\ DR1 data structure\label{app:datastructure}}
The data published as part of this first \veloce\ date release (DR1) are stored in multiple FITS (Flexible Image Transport System) files. Each FITS file contains all the quantities and properties used or derived in the present paper and/or paper~II dedicated to spectroscopic binaries (Shetye et al. in prep.), and one FITS file is used per stars.

The FITS files are organized using a Header-Data Unit (HDU) list, consisting of a PrimaryHDU and multiple BinTableHDU\footnote{A detailed description of this kind of files and how to handle them is given in the \texttt{astropy.io.fits} library documentation at \url{https://docs.astropy.org/en/stable/io/fits/index.html}}. The HDU contains the header with all the metadata, whereas the BinTableHDU contain the data and results derived in tabular form. 

As mentioned above, we applied different parts of our analysis to different stars according to the available data and signal complexity. Hence, the content and number of HDUs vary among the FITS files. However, the file structure consistently follows the different phases of the analysis presented in the two papers.

The most complex file structure is used for stars that underwent all possible stages of the analysis. In this case, the structure of the HDU list is as follows: 

\begin{enumerate}[itemsep=0pt, parsep=8pt]
    \setcounter{enumi}{-1}
    \item {\bf Primary HDU} - Header containing all the metadata (see keys at \ref{tab:data_structure_header}, \ref{tab:fit_structure_header}, \ref{tab:orbit_structure_header}, \ref{tab:KP_VL_structure_header} and \ref{tab:KC_VL_structure_header})

    \item {\bf DATA - BinTableHDU} - Table containing the observational data \ref{tab:data_structure_table})

    \item {\bf FIT - BinTableHDU} - Coefficients of the model fitted to the RV data (see column names at \ref{tab:fit_structure_table})

    \item {\bf RVTF\_CLUSTER - BinTableHDU} - Results of the template fitting analysis per cluster (see column names at \ref{tab:RVTF_structure_table})

    \item {\bf ZP\_CORRECTED - BinTableHDU} - Results of the template fitting analysis per observation and zero-point offsets determined for each instrument/source (see column names at \ref{tab:ZP_structure_table})

    \item {\bf VL\_KEPLER - BinTableHDU} - Results of the fitted Keplerian orbit after combining \veloce\ and literature data (see column names at \ref{tab:KP_VL_structure_table})

    \item {\bf VL\_CIRCULAR - BinTableHDU} - Results of the fitted circular orbit after combining \veloce\ and literature data (see column names at \ref{tab:KC_VL_structure_table})
    
\end{enumerate}

The first table, {\bf DATA - BinTableHDU}, is published for all stars for which RVs were measured from CCFs, including stars not modeled and non-Cepheids. The subsequent BinTableHDUs are published only if the corresponding analysis was carried out for a given star. However, the order of the binary tables follows the stream of the analysis so that the order and hierarchy of the HDU are always respected.

All barycentric Julian dates (BJDs) are reported minus $2,400,000$ for brevity.

\begin{table*}[h!]
\tiny
\begin{center}
\caption{General keys present in the FITS file headers\label{tab:data_structure_header}}
\begin{tabular}{|p{0.30\textwidth} | p{0.6\textwidth}|}
 \hline
 \multicolumn{2}{| c |}{\bf{\normalsize{Header - PrimaryHDU}}}\\
 \hline
 \hline
 \bf{KEY}               & \bf{DESCRIPTION}\\
\hline
 NAME                         & Name of the star \\ \hline
 GAIA\_DR3\_SOURCE\_ID        & Gaia DR3 source\_id of the target \\ \hline
 RA                           & Right Ascension  in h:m:s (J2000)\\ \hline
 DEC                          & Declination  in d:am:as (J2000)\\ \hline
 USED\_SOURCES                & List of instruments used to take the RV measurements  \\ \hline
 OBSERVED\_WITH\_CORALIE07      & Flag - The star was observed with the \coralie\ instrument BEFORE the 2014 upgrade \\ \hline
 OBSERVED\_WITH\_CORALIE14    & Flag - The star was observed with the \coralie\ instrument AFTER the 2014 upgrade\\ \hline
 OBSERVED\_WITH\_HERMES       & Flag - The star was observed with the \hermes\ instrument \\ \hline
 N\_RV                   & Number of RV measurements \\ \hline
 N\_RV\_CORALIE07         & Number of RV measurements taken with the \coralie\ instrument BEFORE the 2014 upgrade \\ \hline
 N\_RV\_CORALIE14        & Number of RV measurements taken with the \coralie\ instrument AFTER the 2014 upgrade \\ \hline
 N\_RV\_HERMES           & Number of RV measurements taken with the \hermes\ instrument   \\ \hline
 BJD\_MIN                     & Minimum value of BJD (Barycentric Julian Date) among the RV measurements for the target \\ \hline
 BJD\_MAX                     & Maximum value of BJD (Barycentric Julian Date) among the RV measurements for the target \\ \hline
 BASELINE                     & Baseline of the RV measurements (BJD\_MAX - BJD\_MIN) $\mathrm{[d]}$ \\ \hline
 BOOL\_CLASSICAL\_CEPHEID & Flag - Set to \texttt{True} for targets identified as bona-fide classical Cepheids, set to \texttt{False} for the rest.  \\ \hline
 BOOL\_DOUBLE\_BUMP & Flag - Target show a double bump feature (see Sect.\,\ref{sec:doublebump})\\ \hline
 BOOL\_SB1           & Flag - Star is identified as an SB1\\ \hline
 BOOL\_FIT\_AVAILABLE         & Flag - A model of the pulsation curve was fitted and is available \\ \hline
 BOOL\_INTERPOLATED                 & Flag - If \texttt{True}, interpolated points were used to determine the best parameters that fit the pulsation curve (only VX~Cyg, see Sect.\,\ref{sec:adequacy}) \\ \hline
 BOOL\_TEMPLATE\_FITTING\_ANALYSIS & Flag - The template fitting analysis was performed   \\ \hline
 BOOL\_KEPLERIAN\_VL          & Flag - \veloce\ and Literature data were combined to estimate orbital parameters\\ \hline
 BOOL\_CIRCULAR\_VL           & Flag - \veloce\ and Literature data were combined to estimate orbital parameters assuming a circular orbit\\ \hline
\end{tabular}
\end{center}
\tablefoot{The word `Flag' indicates a Boolean variable. The word `Optional' indicates that the key is present only if it is relevant for the specific science case.}
\end{table*}

\begin{table*}[h!]
\tiny
\begin{center}
\caption{Additional FITS header keys when a Fourier series model was fitted to \veloce\ data\label{tab:fit_structure_header}}
\begin{tabular}{|p{0.30\textwidth} | p{0.6\textwidth}|}
 \hline
 \multicolumn{2}{| c |}{\bf{\normalsize{Optional Header Fit - Added to PrimaryHDU}}}\\
 \hline
 \hline
 \bf{KEY}               & \bf{DESCRIPTION}\\
\hline
 P                   & Pulsation Period \Ppuls\ $\mathrm{[d]}$\\ \hline
 P\_ERR\_SYST        & Systematic uncertainty on \Ppuls\ related to the choice of the number of harmonics ($N_{\rm FS}$) $\mathrm{[d]}$\\ \hline
 P\_ERR\_STAT        & Statistical uncertainty on \Ppuls\ determined via Monte-Carlo (fixed number of harmonics $N_{\rm FS}$) $\mathrm{[d]}$\\ \hline
 P\_ERR              & Total uncertainty on the pulsation period \Ppuls\ (combination of P\_ERR\_SYST and P\_ERR\_STAT) $\mathrm{[d]}$ \\ \hline
 E                   & Epoch ($\mathrm{BJD} - 2400000.0$) at which $\phi = 0$ (determined using Fourier fit) \\ \hline
 E\_ERR\_SYST        & Systematic uncertainty on E related to the choice of the number of harmonics ($N_{\rm FS}$) \\ \hline
 E\_ERR\_STAT        & Statistical uncertainty on E determined via Monte-Carlo (fixed number of harmonics $N_{\rm FS}$) \\ \hline
 E\_ERR              & Total uncertainty on the epoch E (combination of E\_ERR\_SYST and E\_ERR\_STAT) \\ \hline
 N\_DOF              & Degrees of freedom of the fit \\ \hline
 CHISQ               & Chi-squared statistics \\ \hline
 NORMALIZED\_CHISQ   & Normalized chi-squared \\ \hline
 STDDEV              & Standard deviation of the residuals (weighted) [\kms] \\ \hline
 STDMER              & Standard error of the mean of the residuals (STDDEV/sqrt(N\_DOF)) [\kms] \\ \hline
 N\_TERMS\_FS        & Number of harmonics (terms) used for the Fourier Series model ($N_{\rm FS}$ in the text) \\ \hline
 BOOL\_POLYNOMIAL    & Flag - True if a polynomial was used as a nuisance parameter for determining the best-fit Fourier Series model and the pulsation period\\ \hline
 BOOL\_KEPLERIAN     & Flag - True if a Keplerian orbit was used to represent orbital motion, see \autoref{tab:orbit_structure_header} for the best-fit parameters of the Keplerian model.  \\ \hline
 VGAMMA\_0              & Systemic heliocentric velocity [\kms] (When no polynomial is used for the fit, VGAMMA\_0 is equal to the pulsation averaged RV which is determined by the constant term of the fit (\vgamma in Eq.\,\ref{eq:Fseries}). In case a polynomial was also adopted for the fit, either no value is reported or we estimate the VGAMMA\_0 from the fit that combined the Fourier series with a Keplerian or from the Keplerian fit over \veloce\ and literature data (V+L).)\\ \hline
 VGAMMA\_0\_ERR         & Uncertainty on VGAMMA\_0 [\kms] \\ \hline
 DELTA\_RADIUS\_OVER\_P      & Difference in the star's radius between its most contracted and expanded configuration $\Delta R /p$ $\mathrm{[R_\odot]}$ \\ \hline
 INTERCEPT\_PHASE    & Phase at which the ascending branch cross the mean value of the Fourier series \\ \hline
 PEAK2PEAK           & Peak-to-peak amplitude ($A_{p2p}$) of the pulsation [\kms]\\ \hline
 MIN\_RV             & Minimum RV during the pulsation cycle [\kms] \\ \hline
 MIN\_RV\_PHASE      & Phase of minimum RV \\ \hline
 MAX\_RV             & Maximum RV during the pulsation cycle [\kms]\\ \hline
 MAX\_RV\_PHASE      & Phase of maximum RV \\ \hline
 R$n$1               & Amplitude ratio between the $n$-th and the first harmonics (A$n$/ A1) (up to $n=7$)\\ \hline
 R$n$1\_ERR         & Uncertainty on the amplitude ratio between the $n$-th and the first harmonics\\ \hline
 PHI$n$1             & Phase difference between the $n$-th and the first harmonics (PHI$n$ - $n$ PHI1) (up to $n=7$)  \\ \hline
 PHI$n$1\_ERR       & Uncertainty on the phase difference between the $n$-th and the first harmonics  \\ \hline
 POLYDEG             & Optional - Degree of the polynomial used to improve the fit the pulsation curve, if applied (key exists only if BOOL\_POLYNOMIAL is True) \\ \hline

\end{tabular}
\end{center}
\tablefoot{Only present if BOOL\_FIT\_AVAILABLE in \autoref{tab:data_structure_header} is True. Fourier amplitude ratios, R$n1$, and phase differences, PHI$n1$ are reported up to the 7th order via the integer $2 \le n \le 7$. Higher-order Fourier parameters can be computed from the Fourier series fit coefficients, cf. Tab.\,\ref{tab:fit_structure_table}.}
\end{table*}

\begin{table*}[h!]
\tiny
\begin{center}
\caption{Additional FITS header keys if a Keplerian orbit was fitted to  \veloce\ data\label{tab:orbit_structure_header}}
\begin{tabular}{|p{0.30\textwidth} | p{0.6\textwidth}|}
 \hline
 \multicolumn{2}{| c |}{\bf{\normalsize{Optional Header Keplerian Orbit - Added to PrimaryHDU}}}\\
 \hline
 \hline
 \bf{KEY}               & \bf{DESCRIPTION}\\
\hline
 P\_ORB            & Orbital period $\mathrm{[d]}$ \\\hline
 P\_ORB\_ERR       & Uncertainty on the orbital period $\mathrm{[d]}$ \\\hline
 ECCENTRICITY      & Eccentricity of the orbit \\\hline
 ECCENTRICITY\_ERR & Uncertainty on the eccentricity of the orbit \\\hline
 T0                & Time of periastron passage  ($\mathrm{BJD} - 2400000.0$) \\\hline
 T0\_ERR           & Uncertainty on the time of periastron passage \\\hline
 ARGPERI           & Argument of periastron $\omega$  $\mathrm{[deg]}$ \\\hline
 ARGPERI\_ERR      & Uncertainty on the argument of periastron $\mathrm{[deg]}$ \\\hline
 K                 & Semi-amplitude of the velocity curve (orbital signal) $\mathrm{[km\,s^{-1}]}$  \\\hline
 K\_ERR            & Uncertainty on the semi-amplitude of the orbital signal $\mathrm{[km\,s^{-1}]}$        \\\hline
 A\_SINI           & Projected semi-major axis of the orbit $\mathrm{[au]}$  \\\hline
 A\_SINI\_ERR      & Uncertainty on the projected semi-major axis $\mathrm{[au]}$  \\\hline
 F\_MASS           & Mass function of the binary system $\mathrm{[M_\odot]}$                      \\\hline
 F\_MASS\_ERR      & Uncertainty on the mass function $\mathrm{[M_\odot]}$                            \\\hline
 
\end{tabular}
\end{center}
\tablefoot{These keys are present only if BOOL\_KEPLERIAN of \autoref{tab:fit_structure_header} is True.}
\end{table*}

\begin{table*}[h!]
\tiny
\begin{center}
\caption{Additional FITS header keys if both \veloce\ and literature data were used to fit a Keplerian orbit using the MCMC method\label{tab:KP_VL_structure_header}}
\begin{tabular}{|p{0.30\textwidth} | p{0.6\textwidth}|}
 \hline
 \multicolumn{2}{| c |}{\bf{\normalsize{Optional Header Keplerian Orbit V+L - Added to PrimaryHDU}}}\\
 \hline
 \hline
 \bf{KEY}               & \bf{DESCRIPTION}\\
\hline
KP\_VL\_N\_ITERATIONS       & Number of iterations of the MCMC for the VL Keplerian model \\ \hline
KP\_VL\_N\_WALKERS          & Number of walkers used in the MCMC for the VL Keplerian model \\ \hline
KP\_VL\_ACCEPTANCE\_RATIO   & Mean of the acceptance fraction for each walker of the MCMC chain for the VL Keplerian model \\ \hline
KP\_VL\_N\_CHAIN            & Number of independent samples MCMC \\ \hline
KP\_VL\_N\_DOF              & Degrees of freedom of the VL Keplerian fit \\ \hline
KP\_VL\_CHISQ               & Chi-squared statistics \\ \hline
KP\_VL\_NORMALIZED\_CHISQ   & Normalized chi-squared statistics \\ \hline
KP\_VL\_STDDEV              & Standard deviation of the residuals (weighted)[\kms] \\  \hline
KP\_VL\_STDMER              & Standard error of the mean of the residuals (STDDEV/sqrt(N\_DOF))[\kms] \\  \hline
KP\_VL\_PORB                & Orbital period of the VL Keplerian model (mean) $\mathrm{[d]}$\\  \hline
KP\_VL\_PORB\_ERR           & Uncertainty on the orbital period of the VL Keplerian model (standard deviation) $\mathrm{[d]}$ \\  \hline
KP\_VL\_PORB\_Q50           & 50th percentile of the orbital period $\mathrm{[d]}$ \\  \hline
KP\_VL\_PORB\_Q16           & 16th percentile of the orbital period $\mathrm{[d]}$ \\  \hline
KP\_VL\_PORB\_Q84           & 84th percentile of the orbital period $\mathrm{[d]}$ \\  \hline
KP\_VL\_K                   & Semi-amplitude of the radial velocity curve of the VL Keplerian model (mean) [\kms] \\  \hline
KP\_VL\_K\_ERR              & Uncertainty on the semi-amplitude of the radial velocity curve of the VL Keplerian model (standard deviation) [\kms] \\  \hline
KP\_VL\_K\_Q50              & 50th percentile of the semi-amplitude of the radial velocity curve [\kms] \\  \hline
KP\_VL\_K\_Q16              & 16th percentile of the semi-amplitude of the radial velocity curve [\kms] \\  \hline
KP\_VL\_K\_Q84              & 84th percentile of the semi-amplitude of the radial velocity curve [\kms] \\  \hline
KP\_VL\_ECCENTRICITY        & Eccentricity of the orbit for the VL Keplerian model \\ \hline
KP\_VL\_ECCENTRICITY\_ERR   & Uncertainty on the eccentricity of the orbit for the VL Keplerian model \\ \hline
KP\_VL\_ARGPERI             & Argument of periastron $\omega$ for the VL Keplerian model $\mathrm{[deg]}$ \\  \hline
KP\_VL\_ARGPERI\_ERR        & Uncertainty on the argument of periastron $\omega$ $\mathrm{[deg]}$ \\  \hline
KP\_VL\_T0                  & Time of periastron passage for the VL Keplerian model ($\mathrm{BJD} - 2400000.0$) \\  \hline
KP\_VL\_T0\_ERR             & Uncertainty on the time of periastron passage for the VL Keplerian model \\  \hline
KP\_VL\_A\_SINI             & Projected semi-major axis of the orbit for the VL Keplerian model $\mathrm{[au]}$ \\  \hline
KP\_VL\_A\_SINI\_ERR        & Uncertainty on the projected semi-major axis of the orbit for the VL Keplerian model $\mathrm{[au]}$ \\  \hline
KP\_VL\_F\_MASS             & Mass function of the binary system for the VL Keplerian model $\mathrm{[M_\odot]}$ \\  \hline
KP\_VL\_F\_MASS\_ERR        & Uncertainty on the mass function $\mathrm{[M_\odot]}$ \\  \hline
\end{tabular}
\end{center}
\tablefoot{Only present if BOOL\_KEPLERIAN\_VL is True, cf. \autoref{tab:data_structure_header}. The prefix `KP\_VL' identifies orbital solutions involing both \veloce\ and Literature (VL) data from those based purely on \veloce\ data, cf. \autoref{tab:orbit_structure_header}.}
\end{table*}

\begin{table*}[h!]
\tiny
\begin{center}
\caption{Additional FITS header keys if both \veloce\ and literature data were used to fit a circular orbit using the MCMC method\label{tab:KC_VL_structure_header}}
\begin{tabular}{|p{0.30\textwidth} | p{0.6\textwidth}|}
 \hline
 \multicolumn{2}{| c |}{\bf{\normalsize{Optional Header Circular Orbit V+L - Added to PrimaryHDU}}}\\
 \hline
 \hline
 \bf{KEY}               & \bf{DESCRIPTION}\\
\hline
KC\_VL\_N\_ITERATIONS & Number of iterations of the MCMC for the VL circular model \\ \hline
KC\_VL\_N\_WALKERS & Number of walkers for the VL circular model \\ \hline
KC\_VL\_ACCEPTANCE\_RATIO & Mean of the acceptance fraction for each walker of the MCMC chain for the VL circular model \\ \hline
KC\_VL\_N\_CHAIN & Number of independent samples MCMC \\ \hline
KC\_VL\_N\_DOF & Degrees of freedom of the VL circular fit \\ \hline
KC\_VL\_CHISQ & Chi-squared statistics \\ \hline
KC\_VL\_NORMALIZED\_CHISQ & Normalized chi-squared statistics \\ \hline
KC\_VL\_STDDEV & Standard deviation of the residuals (weighted) [\kms] \\  \hline
KC\_VL\_STDMER & Standard error of the mean of the residuals (STDDEV/sqrt(N\_DOF)) [\kms] \\  \hline
KC\_VL\_PORB & Orbital period of the VL circular model (mean) $\mathrm{[d]}$ \\  \hline
KC\_VL\_PORB\_ERR & Uncertainty on the orbital period of the VL circular model (standard deviation) $\mathrm{[d]}$ \\  \hline
KC\_VL\_PORB\_Q50 & 50th percentile of the orbital period $\mathrm{[d]}$ \\  \hline
KC\_VL\_PORB\_Q16 & 16th percentile of the orbital period $\mathrm{[d]}$ \\  \hline
KC\_VL\_PORB\_Q84 & 84th percentile of the orbital period $\mathrm{[d]}$ \\  \hline
KC\_VL\_K & Semi-amplitude of the radial velocity curve of the VL circular model (mean) [\kms] \\  \hline
KC\_VL\_K\_ERR & Uncertainty on the semi-amplitude of the radial velocity curve of the VL circular model (standard deviation) [\kms] \\  \hline
KC\_VL\_K\_Q50 & 50th percentile of the semi-amplitude of the radial velocity curve [\kms] \\  \hline
KC\_VL\_K\_Q16 & 16th percentile of the semi-amplitude of the radial velocity curve [\kms] \\  \hline
KC\_VL\_K\_Q84 & 84th percentile of the semi-amplitude of the radial velocity curve [\kms] \\  \hline
KC\_VL\_A\_SINI             & Projected semi-major axis of the orbit for the VL circular model $\mathrm{[au]}$ \\  \hline
KC\_VL\_A\_SINI\_ERR        & Uncertainty on the projected semi-major axis of the orbit for the VL circular model $\mathrm{[au]}$ \\  \hline
KC\_VL\_F\_MASS             & Mass function of the binary system for the VL circular model $\mathrm{[M_\odot]}$ \\  \hline
KC\_VL\_F\_MASS\_ERR        & Uncertainty on the mass function $\mathrm{[M_\odot]}$ \\  \hline

\end{tabular}
\end{center}
\tablefoot{Only present if BOOL\_CIRCULAR\_VL is True, cf. \autoref{tab:data_structure_header}. The prefix `KC\_VL' identifies all keys corresponding to a circular orbital solution involing both \veloce\ and Literature (VL) data from those based purely on \veloce\ data, cf. \autoref{tab:orbit_structure_header}.}
\end{table*}

\begin{table*}[h!]
\tiny
\begin{center}
\caption{Content of the first BinTableHDU containing the \veloce\ radial velocity measurements\label{tab:data_structure_table}}
\begin{tabular}{|p{0.30\textwidth} | p{0.6\textwidth}|}
 \hline
 \multicolumn{2}{| c |}{\bf{\normalsize{ DATA - BinTableHDU}}}\\
 \hline
 \hline
 \bf{COLUMN}               & \bf{DESCRIPTION}\\                                               
\hline
 RV        & Radial velocities measurements [\kms] \\ \hline
 RV\_ERR    & Uncertainties on the RV measurements [\kms] \\ \hline
 BJD       & Barycentric Julian Date (2400000.0 was subtracted to the values)\\ \hline
 SOURCE    & Instrument used to take the RV measurement\\ \hline
 UNIQUE\_ID & Unique identifier of the observation \\ \hline
 SN\_60     & Signal-to-Noise ratio per pixel near $6000$\AA \\ \hline
 MASK      & Boolean array that flags if a data point was used to fit the model. TRUE indicates that the data point was used, FALSE otherwise. \\ 
\hline
\end{tabular}
\end{center}
\end{table*}

\begin{table*}[h!]
\tiny
\begin{center}
\caption{Content of the BinTableHDU containing the model fitted to the RV curve\label{tab:fit_structure_table}}
\begin{tabular}{|p{0.30\textwidth} | p{0.6\textwidth}|}
 \hline
 \multicolumn{2}{| c |}{\bf{\normalsize{FIT - Optional BinTableHDU}}}\\
 \hline
 \hline
 \bf{COLUMN}               & \bf{DESCRIPTION}\\                                                
\hline
 COEFF            & Coefficients of the fitted model             \\ \hline
 COEFF\_COV\_MATRIX & Covariance matrix of the fitted coefficients \\ \hline
 COEFF\_ERR        & Uncertainties on the fitted coefficients     \\ \hline

\end{tabular}
\end{center}
\caption{The coefficients are arranged in correspondence with their appearance in Eq.\,\ref{eq:Fseries}. In cases where a polynomial was also fitted to the data, the $c_i$ coefficients (see Eq.\,\ref{eq:polynomial}) are added after the Fourier series. Note this table is present only if BOOL\_FIT\_AVAILABLE in \autoref{tab:data_structure_header} is set to True.}

\end{table*}

\begin{table*}[h!]
\tiny
\begin{center}
\caption{Per-cluster results of the template fitting analysis\label{tab:RVTF_structure_table}}
\begin{tabular}{|p{0.30\textwidth} | p{0.6\textwidth}|}
 \hline
 \multicolumn{2}{| c |}{\bf{\normalsize{RVTF\_CLUSTER- Optional BinTableHDU}}}\\
 \hline
 \hline
 \bf{COLUMN}               & \bf{DESCRIPTION}\\
\hline
 REFERENCE                & Name of the instrument (for \veloce\ measurement) or source in the literature\\\hline
 BIBCODE                   & Bibcode of the reference (Set to NULL if reference is an instrument)\\\hline
 CLUSTER\_ID                  & Identification number of the (KDE) data cluster \\\hline
 N\_OBS\_CLUSTERED         & Number of RV measurements in the cluster\\\hline
 MEAN\_BJD                  & Mean BJD of the RV measurements in the cluster \\\hline
 MIN\_BJD                   & Minimum BJD of the RV measurements in the cluster\\\hline
 MAX\_BJD                   & Maximum BJD of the RV measurements in the cluster\\\hline
 VGAMMA\_CLUSTER           & Radial velocity offset of the pulsation for the cluster (\vgamma) [\kms]\\\hline
 VGAMMA\_CLUSTER\_ERR       & Uncertainty on VGAMMA\_CLUSTER  [\kms]\\\hline
 DELTA\_PHI                 & Phase difference between the cluster and the initial template model \\\hline
 DELTA\_PHI\_ERR             & Uncertainty on DELTA\_PHI \\\hline

\end{tabular}
\end{center}
\tablefoot{Only present if BOOL\_TEMPLATE\_FITTING\_ANALYSIS is True, cf. \autoref{tab:data_structure_header}). Each line corresponds to a temporal data cluster composed of multiple observations.}

\end{table*}

\begin{table*}[h!]
\tiny
\begin{center}
\caption{Epoch results of the template fitting analysis\label{tab:ZP_structure_table}}
\begin{tabular}{|p{0.30\textwidth} | p{0.6\textwidth}|}
 \hline
 \multicolumn{2}{| c |}{\bf{\normalsize{ZP\_CORRECTED - Optional BinTableHDU}}}\\
 \hline
 \hline
 \bf{COLUMN}               & \bf{DESCRIPTION}\\                                                
\hline
 REFERENCE                 & Name of the instrument (for VELOCE measurement) or source in the literature\\\hline
 BIBCODE                   & Bibcode of the reference (Set to NULL if reference is an instrument)\\\hline
 CLUSTER\_ID                     & Identification number of the data cluster that the measurement belongs to\\\hline
 BJD                         & Barycentric Julian Date of the observation\\\hline
 EPOCH\_RESIDUALS  & Residuals between the measurement and the \veloce\ template (shifted accordingly to the phase shift of the epoch and without offset) [\kms]\\\hline
 EPOCH\_RESIDUALS\_ERR & Uncertainty on EPOCH\_RESIDUALS [\kms] \\\hline

 EPOCH\_RESIDUALS\_ZPC  & Residuals between the measurement and the \veloce\ template (shifted accordingly to the phase shift of the epoch and without offset) after applying the zero-point correction to the measurement [\kms]\\\hline
 EPOCH\_RESIDUALS\_ZPC\_ERR & Uncertainty on EPOCH\_RESIDUALS\_ZPC [\kms] \\\hline
 ZP\_CORRECTION\_USED          & Zero-point offset correction of the instrument/source in the literature [\kms] \\\hline
ZP\_CORRECTION\_USED\_ERR     & Uncertainty on ZP\_CORRECTION\_USED [\kms] \\\hline

\end{tabular}
\end{center}
\tablefoot{Only present if BOOL\_TEMPLATE\_FITTING\_ANALYSIS is True, cf. \autoref{tab:data_structure_header}. Each line corresponds to a single RV measurement. Zero-point offset corrections could not be determined for some references. In such cases, the last four columns contain a \texttt{NULL} value.}

\end{table*}

\begin{table*}[h!]
\tiny
\begin{center}
\caption{Keplerian fit results based on both \veloce\ and literature data using the MCMC method\label{tab:KP_VL_structure_table}}
\begin{tabular}{|p{0.30\textwidth} | p{0.6\textwidth}|}
 \hline
 \multicolumn{2}{| c |}{\bf{\normalsize{VL\_KEPLER - Optional BinTableHDU}}}\\
 \hline
 \hline
 \bf{COLUMN}               & \bf{DESCRIPTION}\\                                                
\hline
VAR\_NAMES & Names of the variables of the Keplerian model \\ \hline
MEAN & Mean of the MCMC chain fitted parameters \\ \hline
STD & Standard deviation of the MCMC chain fitted parameters \\ \hline
Q50 & 50th percentile of the MCMC chain fitted parameters \\ \hline
Q16 & 16th percentile of the MCMC chain fitted parameters \\ \hline
Q84 & 84th percentile of the MCMC chain fitted parameters \\ \hline
\hline
\end{tabular}
\end{center}

\tablefoot{Only present if BOOL\_KEPLERIAN\_VL is True, cf. \autoref{tab:data_structure_header}).}
\end{table*}

\begin{table*}[h!]
\tiny
\begin{center}
\caption{Circular Keplerian fit results based on both \veloce\ and literature data using the MCMC method\label{tab:KC_VL_structure_table}}
\begin{tabular}{|p{0.30\textwidth} | p{0.6\textwidth}|}
 \hline
 \multicolumn{2}{| c |}{\bf{\normalsize{VL\_CIRCULAR - Optional BinTableHDU}}}\\
 \hline
 \hline
 \bf{COLUMN}               & \bf{DESCRIPTION}\\                                                
\hline
VAR\_NAMES & Name of the variables of the circular model \\ \hline
MEAN & Mean of the MCMC chain fitted parameters \\ \hline
STD & Standard deviation of the MCMC chain fitted parameters \\ \hline
Q50 & 50th percentile of the MCMC chain fitted parameters \\ \hline
Q16 & 16th percentile of the MCMC chain fitted parameters \\ \hline
Q84 & 84th percentile of the MCMC chain fitted parameters \\ \hline
\hline
\end{tabular}
\end{center}
\tablefoot{Only present if BOOL\_CIRCULAR\_VL is True, cf. \autoref{tab:data_structure_header}).}
\end{table*}

\end{appendix}

\end{document}